\documentclass[fleqn,usenatbib]{mnras}

\usepackage{newtxtext,newtxmath}

\usepackage{verbatim}
\usepackage{ae,aecompl}
\usepackage[T1]{fontenc}
\usepackage{threeparttable}

\usepackage{amsmath}	
\usepackage{amssymb}	
\usepackage{graphicx}	
\usepackage{pdflscape}	

\title[NGC 2527]{A multi-wavelength view of the open cluster NGC
	2527: Discovery of active stars}

\author[Shah, Singh $\&$ Subramaniam]{
	Nevil Shah,$^{1}$\thanks{E-mail: nevilshah235@gmail.com}
	K. P. Singh,$^{1}$ and Annapurni Subramaniam$^{2}$\\
	$^{1}$Indian Institute of Science Education and Research Mohali, Sector 81, SAS Nagar, Punjab, India \\
	$^{2}$Indian Institute of Astrophysics, Koramangala, Bangalore, India\\
}

\date{Accepted XXX. Received YYY; in original form ZZZ}

\pubyear{2019}

\begin{document}
	\label{firstpage}
	\pagerange{\pageref{firstpage}--\pageref{lastpage}}
	\maketitle
	
	\begin{abstract}
		Star clusters are ideal platforms for categorising X-ray emitting stars and to study X-ray emission as a function of stellar age and activity. We present a comprehensive study of an open star cluster, NGC2527, by combining data from XMM-UVOT-Gaia. 
		Cluster membership of stars and their photometry are taken from Gaia and cross-matched with XMM and UVOT detections. We estimate the age of NGC2527 as $\sim$630 Myr, reddening as E(B$-$V)=0.13 mag, and a distance of 642$\pm$30 pc using PARSEC isochrones. 
		We detect 5 sub-subgiants and 5 bandgap stars, which defy single star evolution. We estimate the temperature, mass, radius, and luminosity of 53 single stars and 10 potential binary stars using a python code which fits single and composite Kurucz spectra to broad-band Spectral Energy Distribution. Among the 12 X-ray emitting members, we find 5 are potential RS CVn type binaries, 2 are potential FK Comae type of RGB stars, and 5 are main sequence (MS) stars with high coronal activity. Members with strong UV emission comprise of 1 RGB star, and several MS stars with UV excess suggestive of chromospheric activity. 
		Based on comparison with other clusters, we tentatively suggest that X-ray luminosity of both RS CVn and contact binaries increases with age suggesting more active binaries are present in older clusters as compared to younger clusters. This study suggests possible presence of W UMa and FK Comae type stars in younger (age$\simeq$630 Myr) clusters.
		
	\end{abstract}
	
	\begin{keywords}
		X-rays: stars -- ultraviolet: stars -- (Galaxy:) open clusters and associations: individual: NGC 2527 -- (stars:) binaries (including multiple): close -- X-rays: binaries -- stars: activity
	\end{keywords}												
	
	
	
	\section{Introduction}
	
	
	X-ray emission from stars in young open clusters (OCs) is primarily from coronal activity driven by stellar rotation, magnetic field and convection. As clusters age, the X-ray luminosity of stars decreases as a result of spin down due to magnetic braking \citep{1989A&ARv...1..177P, 1997MmSAI..68..971R}. Thus study of X-rays from stars in old OCs reveals populations of stars spun-up in binary systems, or from systems undergoing accretion. X-ray observations of intermediate age open clusters are especially important for studying both the main sequence (MS) coronally active stars and rare interacting binaries. For example, previous X-ray studies of intermediate-age/old open clusters (NGC 6791 $\sim$8 Gyr, \citealt{2013ApJ...770...98V}; M67 $\sim$3 Gyr, \citealt{2015MNRAS.452.3394M}; NGC 6940 \citealt{1997A&A...326..608B}; NGC6633 and IC4756 $\sim$0.7 Gyr, \citealt{2000MNRAS.319..826B}) revealed a significant number of magnetically active - RS CVn, BY Dra, W UMa, FK Comae, Algol, and  sub-subgiant (SSG) type of stars and mass transfer systems - Cataclysmic Variables (CVs), Low/High Mass X-ray Binaries (L/HMXB). 
	
	UV emission from stars in open clusters, on the other hand,  is dominated by luminous hot stars and/or a white dwarfs (WDs) hidden in binary systems as WDs are less prominent in the optical bands wherein a large number of late-type MS and red giant branch (RGB) stars dominate. 
	Recent UV studies of OCs \citep{2014AJ....148..131S,2018MNRAS.481..226S,2019AJ....158...35S} have shown that UV Color-Magnitude Diagrams (CMDs) are essential tools for identifying and studying the properties of UV bright stellar population such as spatially unresolved MS stars with cool WD companions, post-interaction binaries with hot WD companions, MS and RGB stars with high chromospheric activity, and blue straggler stars (BSS). Though, there have been extensive studies of OCs in the optical and X-ray passbands, detailed studies in UV passbands have been rare, exceptions being old open clusters (M67 \citealt{2018MNRAS.481..226S}; NGC 188 \citealt{2014ApJ...783L...8G}). Another effort in this direction has been the Swift/UVOT Stars Survey \citet{2019AJ....158...35S} which led to a point-source photometry catalogue for 103 Galactic OC with isochrones fitted to the UV CMDs of stars. \\
	
	Here, we present first comprehensive study of UV and X-ray emission from an intermediate-age open cluster, NGC 2527. It is one of the nearest such cluster that lies in the southern hemisphere in the Puppis constellation. 
	It is located at $\alpha$ = $08^h$ $04^m$ $58.0^s$ and $\delta$ = -28\degr 08\arcmin 48\arcsec \citep{2009MNRAS.399.2146W}, at a distance of $\approx$ 600 pc and has a reddening of E(B-V) = 0.038 to 0.1 mag \citep{2002A&A...389..871D,1973A&AS....9..229L}. The estimation of cluster age shows a range between 450 Myr to 1.1 Gyr \citep{1973A&AS....9..229L,1977MNRAS.181..729D,2018A&A...615A..49C}. The cluster turn-off mass is about 2.8 M$_{\odot}$ \citep{2013MNRAS.431.3338R}.  Recently, \citet{2018A&A...618A..93C} have determined membership probability of stars using accurate Gaia astrometric measurement. The overall metallicity in the cluster is slightly lower than the solar metallicity with [Fe/H] = -0.10 \citep{2013MNRAS.431.3338R}.  The mean radial velocity is 42.4 $\pm$ 1.3 km/s 
	\citep{2014A&A...562A..54C}. The apparent diameter of NGC 2527 is 16\arcmin $ $ \citep{1973A&AS....9..229L}.\\
	

	We have used data from Gaia, XMM-Newton, and Swift UVOT observations to characterize the X-ray and UV emission of stars in NGC 2527. The paper is structured as follows: XMM-Newton: EPIC and OM, Swift UVOT, and Gaia observations are described in Section \ref{sec:observations}. In section \ref{sec:data_analysis} we describe reduction of XMM EPIC and OM data, finding of optical counterparts to X-ray and UV sources, analysis of optical and UV CMDs of cluster members, determinaton of X-ray luminosity of stars by using spectral fit parameters of one bright member, and analysis of Spectral Energy Distribution (SED) of X-ray and UV sources. In section \ref{sec:results} and \ref{sec:discussion} we present the results and discussion respectively, and summarise the results in section \ref{sec:summary of results}.\\
	
	\section{Observations}
	\label{sec:observations}
	\subsection{XMM-Newton}
	
\begin{figure*}
		\includegraphics[width=2\columnwidth]{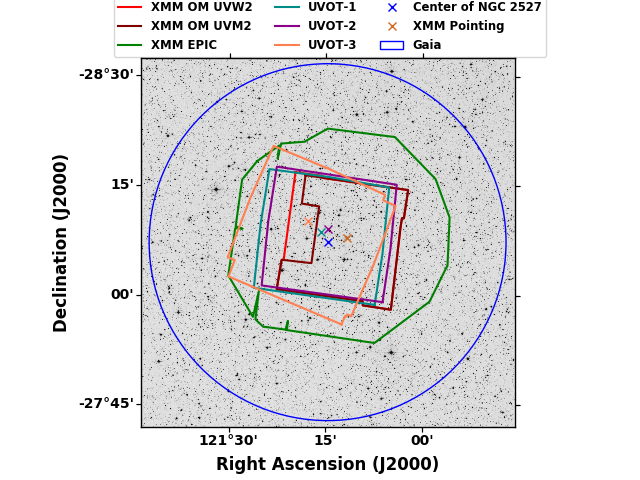}
		\caption{Digitized Sky Survey (DSS) image of NGC 2527. The footprints of XMM EPIC (PN, MOS1 and MOS2), and OM (UVW2 and UVW1) and Swift UVOT observations are overplotted over the image. A circle of radius 25$\arcmin$ is also shown to indicate the region containing half the cluster members \citep{2018A&A...618A..93C}. The crosses indicate the pointing direction and thus its offset from the center of NGC 2527.}
		\label{fig:Footprint}
	\end{figure*}

	
	We have analysed data from the XMM-Newton \citep{2001A&A...365L...1J} observations (ObsId = 0057570101) of NGC 2527 centered on $\alpha = 08^h 04^m 47.20^s$, $\delta$ = $-28\degr 07\arcmin 52.0\arcsec$ performed on March 18, 2002, and March 19, 2002 during orbit 432. During these observations, the PN camera \citep{2001A&A...365L..18S} was active in extended full frame mode, and the MOS cameras \citep{2001A&A...365L..27T} were active in the full frame mode, using the medium filter. We have analysed only the deepest XMM-Newton EPIC (PN, MOS1, and MOS2) exposures of NGC 2527. Several simultaneous short exposure OM \citep{2001A&A...365L..36M} observations in imaging mode with UVM2 (2310 \AA) and UVW2 (2120 \AA) filters were also used. In the OM observations with UVM2 filter, one of the mosaic windows needed to cover the full field of view was not taken. The observation log of XMM-Newton observations is given in Table ~\ref{tab:Log}. The footprints of EPIC (PN and MOS) and OM (UVM2 and UVW2) XMM-Newton observations of NGC 2527 are overplotted on the Digitized Sky Survey (DSS) image of NGC 2527 in Figure \ref{fig:Footprint}. The details of EPIC and OM data processing and analysis are given in Section ~\ref{sec:xraydataanalysis}.
	
	\begin{table*}
		\centering
		\caption{Log of XMM-Newton observations of NGC2527 analysed in this study.}
		\label{tab:Log}
		\begin{tabular}{lcccccr}
			\hline
			INSTRUMENT (MODE) & FILTER & EXPOSURE TIME (FILTERED) \\ 
			&        &        (sec) \\ \hline
			EPN (Extended Full Frame) & MEDIUM & 45998 (26600) \\
			EMOS1 (Full Frame) & MEDIUM & 33283 (25300) \\
			EMOS2 (Full Frame) & MEDIUM & 33265 (25600) \\
			OM (Image) & UVM2 & 4 $\times$ 4300 \\
			OM (Image) & UVW2 & 5 $\times$ 4400 \\	\hline
		\end{tabular}
	\end{table*}
	
	\subsection{Swift UVOT}

	We obtained the Swift UVOT photometric data of NGC 2527 from \citet{2019AJ....158...35S}. These data had been processed using the reduction pipeline outlined in \citet{2014AJ....148..131S}. The footprints of multiple UVOT observations in UVW2 (1928 \AA), UVM2 (2246 \AA), and UVW1 (2600 \AA) filters are overplotted on the DSS image of NGC 2527 (see Figure \ref{fig:Footprint}). Bright sources with high count rates that lead to significant coincidence losses were not included in the Swift UVOT catalogue. We converted the UVOT magnitudes to count rates using the eqn. \ref{eqn:crate} \citep{2008MNRAS.383..627P}. 
	
	\begin{equation}
	\label{eqn:crate}
	\centering
	C_{source} = 10^{\frac{Z_{pt}-m}{2.5}}
	\end{equation}
	
	where m and C$_{source}$ is the magnitude and count rate of a source in the UVOT filters, and Z$_{pt}$ is the zero point for each UVOT filter obtained from \citet{2011AIPC.1358..373B}. The flux in each UVOT filter was calculated by converting the estimated count rates to fluxes using the conversion factors for Vega magnitude scale. The conversion factors were obtained from \citet{2016AJ....152..102B}. The magnitudes and extinction corrected fluxes of cluster members in UVW1, UVW2 and UVM2 filters are given in online-only Table A.

	\subsection{Gaia}
	Gaia DR2 offers a precise and accurate 5-parameter astrometric solution i.e. positions ($\alpha$ and $\delta$), proper motions in right ascension ($\mu_\alpha$) and declination ($\mu$ $_\delta$), and parallaxes ($\varpi$), magnitudes in three photometric filters (G, $G_{BP}$, and $G_{RP}$) for more than 1.3 billion sources and radial velocities (RV) for more than 7 million stars \citep{2018A&A...616A...1G}. \citet{2018A&A...618A..93C} used precise and accurate astrometric (proper motion and parallax) measurements from Gaia DR2 catalogue to determine the cluster membership probability of stars for 1229 OCs in the Milky Way. They have excluded sources fainter than G = 18 mag because of high uncertainties in parallax and proper motion values for such sources. NGC 2527 is part of the Gaia DR2 open cluster population in Milky Way Catalogue \citep{2018A&A...618A..93C}. The catalogue gives membership probabilites of 400 potential members stars that lie around the center of NGC 2527. Out of these 400 stars, only 356 stars have membership probability $>50\%$ that are considered as highly probable members of NGC 2527. As per \citet{2018A&A...618A..93C}, half of the cluster members are
	present within a a circle of radius $\approx 25\arcmin$ from the cluster centre as shown on the DSS image of NGC
	2527 in Figure \ref{fig:Footprint}. 
	The Gaia DR2 catalogue gives radial velocity of 43 bright highly probable members of NGC 2527. 
	The mean radial velocity of these stars is 41 $\pm$ 9 km/s. It is consistent with the previous estimates of radial velocity (42.4 $\pm$ 1.3) by RAVE survey of OCs \citep{2014A&A...562A..54C}.
	
	\section{Data Analysis}
	\label{sec:data_analysis}
	\subsection{X-ray Data Reduction}
	\label{sec:xraydataanalysis}
	Raw data obtained from the XMM-Newton Science Archive were processed using the XMM-Newton Science Analysis Software (SAS 17.0.0). SAS tasks 'emchain' and 'epchain' were used to generate event files for the MOS and PN detectors, respectively. These tasks perform energy and astrometry calibration of events registered on each CCD chip and combine them into a single data file. The event lists of both MOS and PN were filtered using SAS task 'evselect' in the 0.3-7.0 keV energy band and only the single, double, triple, and quadruple pixel events were retained as genuine X-ray events. This removes the contamination due to low pulse height events. Furthermore, the data from all three cameras were individually screened for soft proton background flaring, by applying the criterion of total count rate for single events with energy $>$ 10 keV that exceeded 0.4 counts $s^{-1}$ and 1.0 counts $s^{-1}$ for the MOS and PN detectors respectively. 
	The time intervals affected by soft proton flares were thus excluded from the analysis.
	The observation log and useful exposure time for XMM-Newton observations are summarised in Table ~\ref{tab:Log}.\\
	
	\begin{figure*}
		\includegraphics[width=2\columnwidth]{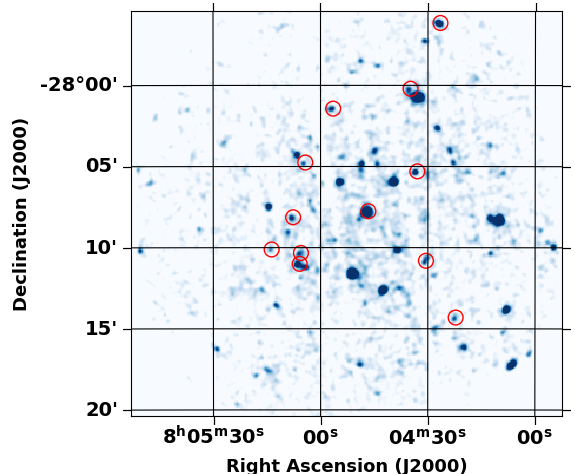}
		\caption{Mosaic of PN and MOS soft band (0.3-2.0 keV) X-ray image of NGC2527 smoothed with a 2D Gaussian $\sigma$ = 2 pixel. The circles indicate the position of the soft band X-ray sources which are members of NGC 2527 and have distinct optical counterparts.}
		\label{fig:X-ray_mosaic}
	\end{figure*}

	\begin{figure*}
		\includegraphics[width=\columnwidth]{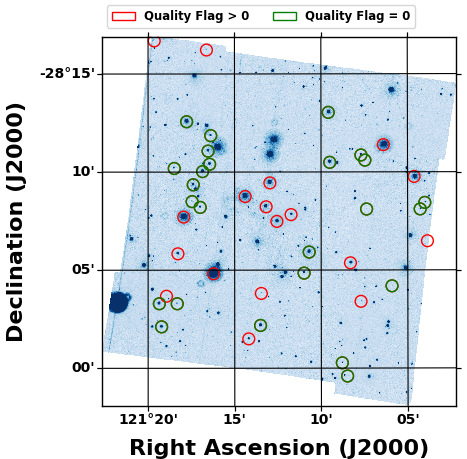}\includegraphics[width=\columnwidth]{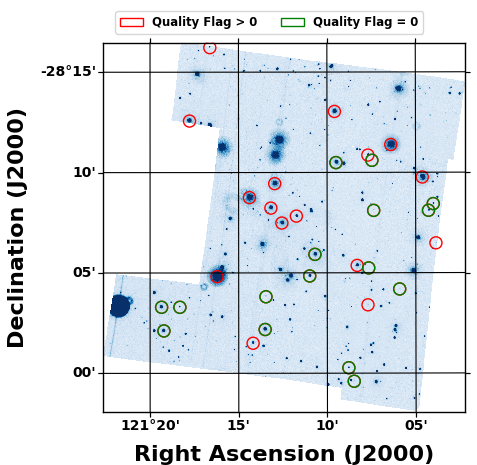}
		\caption{Mosaic of all the OM exposures in UVW2 filter (left) and UVM2 filter (right). The green circles represent the detected sources in each filter which are members of NGC 2527 and have distinct optical counterparts.}
		\label{fig:UV_mosaic}
	\end{figure*}
	
	\subsection{X-ray Images and Source Detection}
	We produced images in two energy bands, a soft band (0.3-2.0 keV), and a hard band (2.0-7.0 keV) for all three EPIC detectors using the 'evselect' and 'emosaic' SAS tasks with the filtered MOS and PN event lists. During the imaging process, only the events with FLAG = 0, PATTERN$\leq$4 (single and double pixel events) in the hard band and PATTERN == 0 (single pixel events) in the soft band were selected. These events were selected to reject noise at the extremities of the PN detector along Y-direction. Default image binning of 600 $\times$ 600 pixels was chosen, similar to the one used in the 3XMM catalogue. \\
	
	We used the SAS task 'edetect$\_$chain' to search for sources present in both the soft and hard bands in the XMM-Newton EPIC (PN, MOS1 and MOS2). The 'edetect$\_$chain' task is a script which runs a series of subtasks to detect point sources from the input filtered event files. It determines the source parameters such as coordinates, count, count rates, etc through simultaneous maximum likelihood PSF (point spread function) fitting to the source count distribution in the given energy bands of each EPIC instrument. A combined Maximum Likelihood (ML) value for all three instruments was taken to be greater than 10. We detect 64 sources in the soft band and 34 sources in the hard band. 
	The merged PN and MOS soft band X-ray image of NGC 2527 smoothed with a 2D Gaussian $\sigma$ = 2 pixel is shown in Figure \ref{fig:X-ray_mosaic}.
	
	\subsection{OM Data Reduction}
	The OM observations taken in Imaging mode were processed using the SAS task 'omichain' with default parameters. This task is a script which detects sources in the OM image and performs aperture photometry on the detected sources. It  processes the data from multiple exposures simultaneously and returns various data products. We obtained images and source lists for individual exposures in each filter, mosaiced images, and merged source list 
	from  all the observations. The merged source list contains information of source position, count rates, fluxes, magnitudes, and quality flags. The SAS pipeline corrects the count rates and fluxes for coincidence losses, dead time and time dependent sensivity of the detector. The combined multiple exposures of OM images in UVW2 and UVM2 filters is shown in Figure \ref{fig:UV_mosaic}. A total of 729 sources were thus detected in the merged OM catalogue. The extinction corrected fluxes and the quality flags of cluster members detected in OM are given in online-only Table A. The SAS pipeline provides a quality flag for each OM source. The sources with quality flag equal to zero are good detections. SAS task detects a significant number of sources which have quality flag $>$ 0 that lie on a readout streak, a bad pixel, near a bright source, near the edge or on a star-spike which are spurious detections and sources with high count rates (rate > 0.97 c/frame). Out of the 729 sources, only 316 detected in UVW2 filter and 217 sources in UVM2 are good detections and not artefacts. The OM sources with bad quality flags were excluded from the subsequent analysis for Spectral Energy Distributions.
	
	\begin{table}
		\centering
		\caption{Summary of total number of sources, and cluster members observed in Gaia, XMM and Swift fov of NGC 2527.}
		\label{tab:Sources}
		\resizebox{\columnwidth}{!}{
					\begin{tabular}{lcccr}
			\hline
			Energy Band 	   & No. of observed sources     & No. of cluster members\\
			\hline
			Gaia Optical       & 400        & 358\\
			OM UVM2            & 600        & 32\\
			OM UVW2            & 426        & 43\\
			UVOT UVW1          & 1292       & 51\\
			UVOT UVM2          & 1025       & 51\\
			UVOT UVW2          & 1279       & 51\\    
			X-ray soft band    & 64         & 12\\
			X-ray hard band    & 34         & 0 \\ \hline
		\end{tabular}}
	\end{table}
	
	\subsection{Optical Counterparts}
	In order to determine the nature of sources detected in Swift UVOT and XMM observations, we searched for their optical counterparts in the Gaia members catalogue of NGC 2527.\\
	
	We determined the optimum cross-correlation separation between soft band X-ray catalogue and optical counterparts using the method outlined by \citet{1997MNRAS.287..350J}. We find that the best cross-correlation radius of 3\arcsec $ $ optimizes the number of real identifications while minimizing the number of spurious detections due to background sources. The details of the best fit cross-correlation function is given in Appendix A. Within a radius of 3\arcsec, 12 soft band X-ray sources possess a unique optical counterpart that is a cluster member. None of the hard band X-ray sources has an optical counterpart that is a cluster member within a radius of $\le$ 30\arcsec $ $ from the hard band X-ray source position, suggesting that the sources in the hard X-ray band are not members of the cluster. X-ray sources that have a unique optical counterpart that is a proper motion member of the cluster have been considered as real sources, and other sources are considered as spurious detections. To identify optical counterparts to the UV (OM/UVOT) sources, we assumed a cross-correlation radius of 3\arcsec $ $. Within a 3\arcsec $ $ radius, we find that 44 XMM-OM, and 54 UVOT sources have unique optical counterparts (see Table \ref{tab:Sources}).\\
	
	The total Galactic column density along the line of sight of NGC 2527, but extending to the edge of the Galaxy as given by the (\citealt{2016A&A...594A.116H}; see HEASARC $N_H$ column density tool\footnote{https://heasarc.gsfc.nasa.gov/cgi-bin/Tools/w3nh/w3nh.pl}) is N$_H$ =  44.2 $\times$ 10$^{20}$ cm$^{-2}$. To estimate the number of extragalactic sources in the field of view (fov) of XMM-Newton, we assumed a power-law model with a photon index $\Gamma$= 2, and the quoted value of N$_H$. An upper limit on the background count rate in the hard band for EPIC MOS1 observations was found to be $\sim 0.002$ counts/s (see also Sect. 3.6). Since the log N $-$ log S distribution derived from the ROSAT deep sky survey \citep{1993A&A...275....1H} gives flux in the 0.5 $-$ 2.0 keV energy band, we determined the total flux for the power-law model in this energy band using the HEASARC WebSpec\footnote{https://heasarc.gsfc.nasa.gov/webspec/webspec.html} tool. For a count rate of $\sim$0.002 counts/s, exposure time of 25300 sec, the $\Gamma$ = 2 power law with N$_H$ = 44.2 $\times$ 10$^{20}$ cm$^{-2}$ and normalization = 1.63 $\times$ 10$^{-5}$ has a flux of 1.3 $\times$ 10$^{-14}$ ergs cm$^{-2}$ s$^{-1}$ and 3.1 $\times$ 10$^{-14}$ ergs cm$^{-2}$ s$^{-1}$ in 0.5-2.0 keV and 2.0-7.0 keV energy band respectively. A comparison of the background flux in the 0.5-2.0 keV energy band with the log N $-$ log S distribution suggests that approximately 30-35 extragalactic sources are likely to be detected above a maximum likelihood threshold of 10.
	The number of hard band X-ray sources detected in our XMM-Newton observations is consistent with the estimated number of background sources.\\

	X-ray properties of cluster members are given in Table \ref{tab:Gaia_XMM}. Positions of X-ray, OM - UVW2, and UVW1 sources which have unique optical counterparts that are cluster members are shown in Figure \ref{fig:X-ray_mosaic} and Figure \ref{fig:UV_mosaic}. In the OM images, sources with good and poor quality flags are marked with green and red colours respectively to distinguish between good and bad quality detection. A unique object id SwiftXMMOM (SXOM) has been assigned to the cluster members that are detected in the XMM and Swift UVOT observations. 
	Sources that have been detected only in the XMM-OM and not in the Swift UVOT observations are considered as spurious detections. SXOM30 and SXOM46 are two such sources which were excluded from further analysis.\\

	\begin{table*}
		\centering
		\caption{Properties of cluster members detected in the XMM observations. The superscript $^{\alpha}$ on the source ID denotes that the corresponding X-ray properties are from the PN detector as the source lies outside the MOS CCD. Column 1 gives the ID of stars detected in X-ray or UV, Columns 2, 3, 4, 5 \& 8 list the RA, Dec, distance, radial velocity from Gaia DR2, and membership probability of stars obtained from \citet{2018A&A...618A..93C}, columns 6 \& 7, give the G mag and the G$_{BP}$-G$_{RP}$ colour taken from Gaia DR2 catalogue, columns 9, 10 \& 11 give the countrate, flux and X-ray luminosity of sources that lie in the field of view of XMM detector, and column 12 gives the type of stars depending upon their position in the optical CMD.}
		\label{tab:Gaia_XMM}
		\resizebox{\textwidth}{!}{
			\begin{threeparttable}
				\begin{tabular}{lcccccccccccccccccr}
					\hline
					
					SXOM   &  RA  &  Dec  &    Distance   & Radial Velocity & G   &  G$_{BP}$-G$_{RP}$
					&   PMemb    &  Rate  & $F_X$ & $L_X$ & Type \\ 
					& ($^h$ $\arcmin$ $\arcsec$)  & ($\degr$ $\arcmin$ $\arcsec$) & (pc) & (km/s)& (mag) & (mag) &  & (10$^{-3}$ counts $s^{-1}$) & (10$^{-15}$ ergs s$^{-1}$ cm$^{-2}$ ) &  (10$^{29}$ ergs s$^{-1}$)   & \\ \hline
					1	&8:05:27.17	&-28:17:33.29	&645.0$\pm$18.6	&-	&	11.31	&	0.52	&	1.0	&	<1.867	&<3.22	&<1.6	&B\tnote{1}\\
					2	&8:05:33.73	&-28:12:43.42	&623.8$\pm$13.5	&-	&	12.6	&	0.65	&	1.0	&	<1.867	&<3.22	&<1.5	&MS\tnote{2}\\
					3	&8:05:35.17	&-28:13:12.38	&656.4$\pm$15.9	&41.77$\pm$2.28	&	12.17	&	0.58	&	1.0	&	<1.867	&<3.22	&<1.66	&MS\\
					4	&8:05:18.65	&-28:16:41.64	&639.9$\pm$9.2	&-	&	13.87	&	0.88	&	1.0	&	<1.867	&<3.22	&<1.58	&MS\\
					5	&8:05:16.3	&-28:16:52.21	&617.7$\pm$14.5	&-	&	11.52	&	0.43	&	1.0	&	<1.867	&<3.22	&<1.47	&MS\\
					6	&8:05:36.06	&-28:16:53.53	&670.3$\pm$16.2	&-	&	10.77	&	0.31	&	1.0	&	<1.867	&<3.22	&<1.73	&MS\\
					7	&8:04:30.86	&-28:10:52.97	&642.6$\pm$11.8	&33.52$\pm$13.54	&	12.86	&	0.69	&	1.0	&	1.44$\pm$0.4	&2.49	&1.23	&MS\\
					8	&8:04:29.89	&-28:10:35.99	&635.8$\pm$8.0	&-	&	13.67	&	1.0	&	1.0	&	<1.867	&<3.22	&<1.56	&B\\
					9	&8:04:38.37	&-28:13:02.64	&629.4$\pm$14.6	&-	&	11.29	&	0.39	&	1.0	&	<1.867	&<3.22	&<1.52	&MS\\
					10	&8:04:54.64	&-28:18:08.26	&628.3$\pm$7.0	&-	&	13.07	&	0.76	&	1.0	&	<1.867	&<3.22	&<1.52	&MS\\
					11	&8:05:06.52	&-28:16:13.63	&634.1$\pm$8.4	&-	&	13.9	&	0.87	&	1.0	&	<1.867	&<3.22	&<1.55	&MS\\
					12	&8:04:32.53	&-28:12:02.0	&631.1$\pm$12.4	&-	&	15.32	&	1.26	&	1.0	&	<1.867	&<3.22	&<1.53	&MS\\
					13	&8:04:15.48	&-28:06:29.1	&637.0$\pm$9.7	&-	&	14.19	&	0.94	&	1.0	&	<1.867	&<3.22	&<1.56	&MS\\
					14	&8:04:38.09	&-28:10:30.78	&638.2$\pm$15.0	&-	&	11.43	&	0.46	&	1.0	&	<1.867	&<3.22	&<1.57	&PB\tnote{3}\\
					15	&8:04:29.57	&-28:08:06.18	&633.6$\pm$9.6	&-	&	14.39	&	0.99	&	1.0	&	<1.867	&<3.22	&<1.54	&MS\\
					16	&8:04:16.1	&-28:08:26.46	&636.5$\pm$7.3	&-	&	13.26	&	0.9	&	1.0	&	<1.867	&<3.22	&<1.56	&B\\
					17	&8:04:35.46	&-28:10:42.69	&608.5$\pm$28.7	&-	&	16.35	&	1.87	&	1.0	&	<1.867	&<3.22	&<1.43	&B\\
					18	&8:04:18.54	&-28:09:46.17	&661.4$\pm$17.8	&-	&	10.8	&	0.27	&	1.0	&	<1.867	&<3.22	&<1.68	&MS\\
					19	&8:04:25.65	&-28:11:23.72	&674.0$\pm$16.4	&-	&	10.11	&	0.23	&	1.0	&	<1.867	&<3.22	&<1.75	&MS\\
					20	&8:04:17.16	&-28:08:06.61	&631.0$\pm$10.0	&-	&	12.97	&	0.82	&	1.0	&	<1.867	&<3.22	&<1.53	&B\\
					21	&8:04:44.06	&-28:04:52.13	&630.7$\pm$12.4	&-	&	12.34	&	0.6	&	1.0	&	<1.867	&<3.22	&<1.53	&MS\\
					22	&8:05:05.56	&-28:11:51.47	&633.1$\pm$13.7	&-	&	12.9	&	0.7	&	1.0	&	<1.867	&<3.22	&<1.54	&MS\\
					23	&8:04:35.19	&-28:00:16.71	&567.0$\pm$11.8	&-	&	13.31	&	0.95	&	0.9	&	1.65$\pm$0.49	&2.85	&1.1	&B\\
					24	&8:05:47.42	&-28:05:53.4	&661.8$\pm$15.0	&32.23$\pm$5.74	&	12.54	&	0.71	&	1.0	&	<1.867	&<3.22	&<1.69	&PB\\
					25	&8:05:48.68	&-28:07:07.07	&626.2$\pm$8.7	&-	&	14.14	&	0.94	&	1.0	&	<1.867	&<3.22	&<1.51	&MS\\
					26	&8:05:17.44	&-28:03:17.37	&621.0$\pm$13.2	&38.28$\pm$0.87	&	11.93	&	0.54	&	1.0	&	<1.867	&<3.22	&<1.48	&MS\\
					27	&8:05:29.77	&-28:09:03.37	&601.0$\pm$36.4	&-	&	11.09	&	0.35	&	0.5	&	<1.867	&<3.22	&<1.39	&PB\\
					28	&8:05:04.97	&-28:04:50.36	&678.0$\pm$23.5	&-	&	9.46	&	0.23	&	1.0	&	<1.867	&<3.22	&<1.77	&MS\\
					29	&8:05:09.61	&-28:09:21.13	&626.7$\pm$11.7	&-	&	11.95	&	0.52	&	1.0	&	<1.867	&<3.22	&<1.51	&MS\\
					30	&8:04:23.71	&-28:04:11.78	&628.2$\pm$9.4	&-	&	14.39	&	0.99	&	1.0	&	<1.867	&<3.22	&<1.52	&MS\\
					31	&8:05:05.82	&-28:10:24.09	&618.7$\pm$12.0	&42.66$\pm$6.36	&	11.69	&	0.6	&	0.8	&	1.72$\pm$0.43	&2.98	&1.36	&B\\
					32	&8:04:57.65	&-28:08:46.31	&652.8$\pm$16.1	&-	&	10.09	&	0.32	&	1.0	&	<1.867	&<3.22	&<1.64	&B\\
					33	&8:05:06.18	&-28:11:4.33	&612.7$\pm$11.6	&-	&	11.79	&	0.63	&	1.0	&	3.74$\pm$0.6	&6.47	&2.91	&B\\
					34	&8:04:56.79	&-28:01:30.34	&715.0$\pm$17.3	&43.32$\pm$2.86	&	12.34	&	0.71	&	1.0	&	2.83$\pm$0.55	&4.9	&2.99	&B\\
					35	&8:05:13.13	&-28:05:50.23	&630.7$\pm$14.2	&40.67$\pm$1.46	&	12.09	&	0.58	&	1.0	&	<1.867	&<3.22	&<1.53	&MS\\
					36	&8:05:11.14	&-28:12:34.01	&641.6$\pm$15.4	&-	&	11.05	&	0.38	&	1.0	&	<1.867	&<3.22	&<1.58	&PB\\
					37	&8:04:50.25	&-28:07:29.2	&653.4$\pm$12.9	&-	&	11.76	&	0.49	&	1.0	&	<1.867	&<3.22	&<1.64	&MS\\
					38	&8:05:11.83	&-28:07:42.43	&677.5$\pm$16.6	&-	&	9.57	&	0.34	&	0.7	&	<1.867	&<3.22	&<1.77	&MS\\
					39	&8:05:15.75	&-28:03:40.07	&651.2$\pm$13.0	&-	&	14.64	&	1.03	&	1.0	&	<1.867	&<3.22	&<1.63	&MS\\
					40	&8:05:07.45	&-28:10:01.78	&626.0$\pm$11.7	&27.4$\pm$11.79	&	11.66	&	0.48	&	1.0	&	<1.867	&<3.22	&<1.51	&MS\\
					41	&8:05:16.9	&-28:02:06.61	&607.8$\pm$12.9	&41.26$\pm$16.89	&	12.07	&	0.57	&	1.0	&	<1.867	&<3.22	&<1.42	&MS\\
					42	&8:05:33.91	&-28:08:58.2	&643.3$\pm$16.0	&40.26$\pm$0.11	&	9.3	&	1.13	&	1.0	&	<1.867	&<3.22	&<1.59	&RGB\tnote{4}\\
					43	&8:04:54.06	&-28:02:11.25	&654.5$\pm$15.7	&41.41$\pm$3.53	&	11.88	&	0.5	&	1.0	&	<1.867	&<3.22	&<1.65	&MS\\
					44	&8:05:05.48	&-27:59:44.17	&625.0$\pm$18.6	&-	&	10.84	&	0.31	&	1.0	&	<1.867	&<3.22	&<1.5	&B\\
					45	&8:05:40.68	&-28:02:25.15	&615.6$\pm$11.1	&40.59$\pm$2.9	&	12.12	&	0.57	&	1.0	&	<1.867	&<3.22	&<1.46	&MS\\
					46	&8:04:33.98	&-27:59:36.88	&633.8$\pm$19.1	&-	&	16.15	&	1.51	&	1.0	&	<1.867	&<3.22	&<1.55	&MS\\
					47	&8:04:51.94	&-28:09:27.85	&628.1$\pm$13.6	&-	&	10.51	&	0.4	&	0.6	&	<1.867	&<3.22	&<1.52	&SSG\tnote{5}\\
					48	&8:04:33.28	&-28:05:22.32	&616.0$\pm$12.1	&39.41$\pm$11.09	&	11.57	&	0.6	&	1.0	&	1.75$\pm$0.39	&3.03	&1.37	&B\\
					49	&8:05:22.42	&-28:10:18.65	&637.8$\pm$16.0	&-	&	15.65	&	1.34	&	1.0	&	<1.867	&<3.22	&<1.57	&MS\\
					50	&8:05:36.6	&-28:03:38.58	&593.3$\pm$17.2	&-	&	11.14	&	0.48	&	1.0	&	<1.867	&<3.22	&<1.35	&B\\
					51	&8:04:42.42	&-28:08:56.32	&643.3$\pm$14.8	&-	&	15.56	&	1.33	&	1.0	&	<1.867	&<3.22	&<1.59	&MS\\
					52	&8:04:30.69	&-28:05:14.74	&623.6$\pm$11.8	&-	&	14.62	&	1.04	&	1.0	&	<1.867	&<3.22	&<1.5	&MS\\
					53	&8:05:07.99	&-28:08:12.08	&619.5$\pm$10.4	&-	&	12.94	&	0.7	&	1.0	&	1.9$\pm$0.45	&3.29	&1.51	&MS\\
					54	&8:04:42.86	&-28:05:56.44	&615.8$\pm$11.9	&-	&	11.53	&	0.49	&	1.0	&	<1.867	&<3.22	&<1.46	&PB\\
					55	&8:04:30.88	&-28:03:24.34	&644.9$\pm$9.8	&-	&	14.36	&	0.97	&	1.0	&	<1.867	&<3.22	&<1.6	&MS\\
					56	&8:04:54.11	&-28:03:50.29	&638.0$\pm$11.9	&-	&	15.04	&	1.14	&	1.0	&	<1.867	&<3.22	&<1.57	&PB\\
					57	&8:05:13.31	&-28:03:17.53	&614.1$\pm$8.4	&-	&	13.68	&	0.84	&	1.0	&	<1.867	&<3.22	&<1.45	&MS\\
					58	&8:05:24.96	&-28:13:14.44	&619.6$\pm$10.8	&-	&	14.8	&	1.09	&	1.0	&	<1.867	&<3.22	&<1.48	&MS\\
					59	&8:05:14.0	&-28:10:11.19	&634.8$\pm$7.2	&-	&	13.32	&	0.77	&	1.0	&	1.17$\pm$0.41	&2.02	&0.98	&MS\\
					60	&8:04:52.8	&-28:08:14.82	&664.5$\pm$15.0	&46.43$\pm$3.2	&	11.42	&	0.43	&	0.9	&	<1.867	&<3.22	&<1.7	&PB\\
					61	&8:04:46.96	&-28:07:49.84	&644.3$\pm$12.7	&40.32$\pm$0.29	&	9.34	&	1.12	&	1.0	&	12.06$\pm$0.82	&20.86	&10.36	&RGB\\
					62	&8:05:09.88	&-28:08:29.82	&643.1$\pm$9.9	&-	&	13.9	&	0.88	&	1.0	&	<1.867	&<3.22	&<1.59	&MS\\
					63	&8:04:22.53	&-28:14:22.79	&659.7$\pm$48.1	&-	&	17.75	&	2.06	&	0.8	&	1.73$\pm$0.48	&2.99	&1.56	&MS\\
					64	&8:05:04.59	&-28:04:49.57	&626.4$\pm$9.7	&-	&	13.6	&	0.77	&	1.0	&	1.55$\pm$0.39	&2.68	&1.26	&MS\\
					65	&8:04:26.85	&-27:56:13.49	&646.0$\pm$14.4	&38.63$\pm$0.23	&	9.5	&	1.15	&	1.0	&	3.79$\pm$0.74	&6.56	&3.27	&RGB\\ \hline
				\end{tabular}
				\begin{tablenotes}
					\item[1] B : Binary
					\item[2] MS : Main Sequence
					\item[3] PB : Potential Binary
					\item[4] RGB : Red Giant Branch
					\item[5] SSG : Sub-subgiant
				\end{tablenotes}
		\end{threeparttable}}
	\end{table*}
	
	\subsection{Optical and UV Color Magnitude Diagram}
	The Hertzsprung-Russell (H-R) Diagram or the Color-Magnitude Diagram (CMD) of OCs can be used to distinguish between stars in various stellar evolutionary stages and identify peculiar stars that deviate from standard evolutionary models. 
	These are essential tools to determine the fundamental properties of a cluster like age and extinction. These parameters can be determined by fitting standard stellar evolutionary tracks or isochrones to the observed CMD.\\
	
\begin{figure}
		\includegraphics[width=\columnwidth]{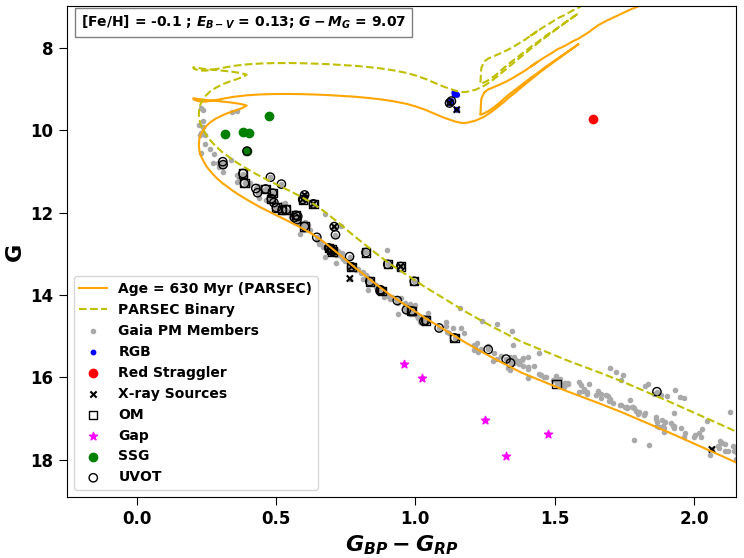}  
		\caption{Optical CMD of NGC 2527 based on Gaia magnitudes. The best fitting PARSEC isochrone along with an equal mass binary isochrone are overlaid on the observed CMD. }
		\label{fig:CMD}
	\end{figure}

	The optical CMD of NGC 2527 based on the Gaia magnitudes (G, $G_{BP}-G_{RP}$) is shown in Figure \ref{fig:CMD}. The main sequence, as well as the binary sequence, can be distinctly identified in the optical CMD. Previous ground-based optical photometric studies of NGC 2527 using CMD \citep{1973A&AS....9..229L, 1977MNRAS.181..729D} were limited only to a few cluster members and the distance to NGC 2527 was not well known. Therefore, the fundamental parameters of NGC 2527 given in the literature have high uncertainity. The distance to NGC 2527 can now be calculated fairly accurately from the average of Gaia DR2 parallaxes of cluster members. We have corrected the Gaia DR2 parallaxes for bias by adding +0.029 mas to the parallax values as prescribed by \citet{2018A&A...616A...2L} and inferred the mean distance to NGC 2527 to be $642\pm30$ pc, which corresponds to a distance modulus of $G-M_G$ =  9.0 $\pm$ 2.4 mag. The PARSEC-COLIBRI isochrone \citep{2017ApJ...835...77M} of log(age) = 8.8 or 630 million years with a reddening of E(B-V) = 0.13, fixed metallicity [Fe/H] = -0.1 \citep{2013MNRAS.431.3338R} and distance modulus of $G-M_G$ = 9.0 mag is found to match the observed MS and main sequence turn off (MSTO) very well. The extinction coefficients used for the Gaia and UVOT filters were obtained from the VOSA Filter Profile Service \citep{2012ivoa.rept.1015R}. 
	The PARSEC-Isochrone with these parameters along with an equal mass binary isochrone are superimposed over the observed optical Gaia CMD in Figure \ref{fig:CMD}. \\
	
	\begin{figure}
		\includegraphics[width=\columnwidth]{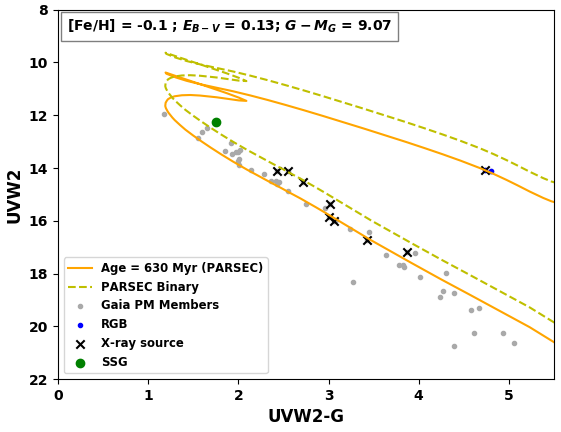}  
		\caption{UV-optical CMD of NGC 2527 based on UVW2 magnitude. The best fitting PARSEC isochrone along with an equal mass binary isochrone are overlaid on the observed CMD. }
		\label{fig:UVW2_CMD}
	\end{figure}    
	
	\begin{figure}
		\includegraphics[width=\columnwidth]{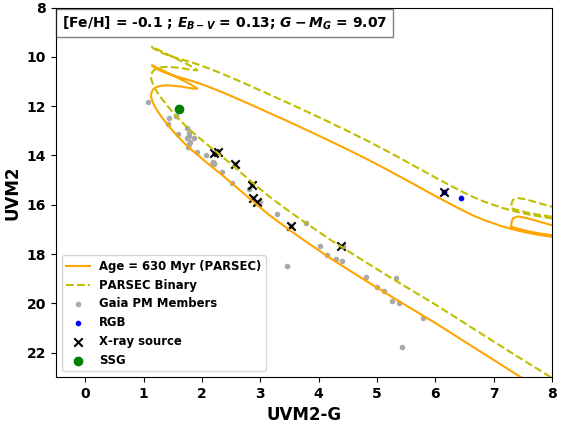}  
		\caption{UV-optical CMD of NGC 2527 based on UVM2 magnitude. The best fitting PARSEC isochrone along with an equal mass binary isochrone are overlaid on the observed CMD. }
		\label{fig:UVM2_CMD}
	\end{figure}    
	
	\begin{figure}
		\includegraphics[width=\columnwidth]{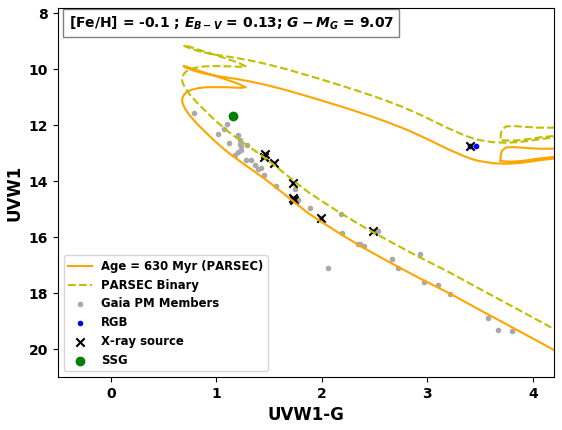}  
		\caption{UV-optical CMD of NGC 2527 based on UVW1 magnitude. The best fitting PARSEC isochrone along with an equal mass binary isochrone are overlaid on the observed CMD.}
		\label{fig:UVW1_CMD}
	\end{figure}    
	
	Of the 44 cluster members detected in the OM only 16 (and 25) have good quality flags in the UVM2 (in the UVW2) filter, as compared to 51 cluster members in each UVOT (UVM2, UVW2, and UVW1) filter. Since there are fewer sources in OM, it is difficult to distinctly visualize the main sequence in UV-optical CMDs with OM magnitudes. Therefore, we only use UVOT magnitudes for UV-optical CMDs. The UV-optical CMD based upon UVOT photometry in the UVW2, UVM2, and UVW1 filters is shown in Figures \ref{fig:UVW2_CMD}, \ref{fig:UVM2_CMD} and \ref{fig:UVW1_CMD}. 
	The PARSEC-COLIBRI isochrones with the same parameters as the optical isochrones are overlaid on UV-optical CMDs. The stars near the main sequence turn-off are not detected in the Swift UVOT because they are saturated or lie outside the fov of Swift UVOT. The PARSEC isochrones fit the MS very well in the UV-optical CMDs based on the UVW1-G and UVM2-G colours. 
	The faint MS stars in the UVW2-G CMD have an excess blue colour as compared to the PARSEC isochrones. \citet{2019AJ....158...35S} had also found an excess in the UVW2 - UVW1 colours of faint MS stars as compared to the isochrones.\\
	
	We checked the position of each star in the Gaia optical CMD and UV-optical CMD, and have classified those which lie close to the binary isochrone as Binary (B). Similarly those lying close to the single star isochrone have been classified as Main Sequence (MS), the ones near the red giant phase of the isochrone as RGB, and the ones lying between single star and binary isochrone in the optical CMD but closer to the binary isochrone in the UV CMD as Potential Binaries (PBs) (see Table \ref{tab:Gaia_XMM}). A few stars that deviate from the standard evolutionary models are classified as red straggler (RS) or sub-subgiant (SSG) depending upon whether they lie in the RS or the SSG region. As per \citet{2017ApJ...840...66G}, SSG refer to stars that are fainter than the normal subgiants and redder than the normal MS star, whereas the RS stars are brighter than the normal subgiants and redder than the normal RGB.
	The fundamental parameters of the peculiar stars that lie outside the fov of XMM and Swift UVOT detectors are given in the appendix Table D. Below, we focus only on the stars which are detected in either XMM-OM, Swift UVOT, and XMM-EPIC observations.  \\

	\subsection{Spectral Fitting and Luminosity of X-ray Sources}
	
	A bright X-ray source SXOM61 is the only source that is found to have a significant number of counts ($>$ 100) for spectral analysis. We extracted PN and MOS spectra for the source and background regions. 
	The source spectrum was extracted from a circular region of radius 30\arcsec centred on the X-ray source. For the MOS, the background spectrum was extracted from an annulus with inner radius of 40\arcsec and outer radius of 60\arcsec centered on the X-ray source. For the PN, such an annular region coincided with PN chip gap, so the background spectrum was extracted from a circle with a radius of 60\arcsec close to the source position. We used the SAS task 'especget' to extract these spectra from the filtered PN and MOS event files. We grouped the spectral (PHA) channels using the 'specgroup' task such that each grouped PHA channel had at least 16 counts, sufficient for the assumption of Gaussian distribution of uncertainties. This task also performs the subtraction of the background spectrum from the source spectrum. \\
	
	\begin{figure} 
		\includegraphics[width=\columnwidth]{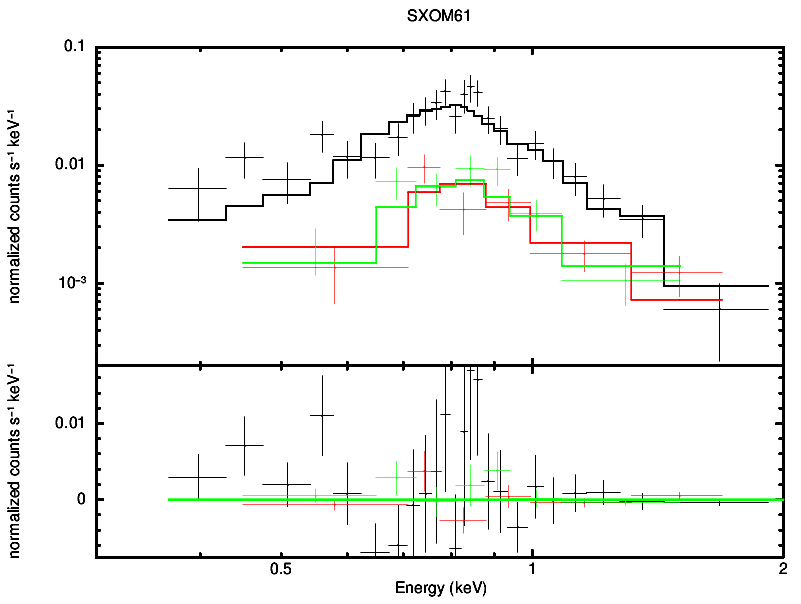} 
		\caption{X-ray spectrum of SXOM61. The black, red and green points are MOS1, MOS2 and PN observed flux data points with error bars respectively. The best fitting tbabs*apec model is indicated by black, green and red lines for MOS1, PN and MOS2 respectively.}
		\label{fig:spectra_SXOM61}
	\end{figure}

	\begin{table}
		\centering
		\caption{Results of fitting tbabs*apec model to the observed PN, MOS1 and MOS2 X-ray spectrum of SXOM61.  Column density of the absorber, N$_{H}$, in the tbabs model was fixed at 8.12 $\times$ 10$^{20}$  cm$^{-2}$. The errors quoted are with 90\% confidence. Column 1 gives the common elemental abundance value used by the model, Column 2 gives the plasma temperature of the best fitting apec model, Column 3 gives the total flux in the 0.3-2.0 keV energy band, and Column 4 gives the reduced $\chi^2$ of the fit and the degrees of freedom (dof: equal to the number of channel groups minus one).}
		\label{tab:Spectra_Best_Fit}
		\begin{tabular}{lccccr}
			\hline
			Abundance & kT & log$_{10}$ Flux  & $\chi^2_{red}$ (dof)\\ 
			&(keV)&  (ergs cm$^{-2}$ s$^{-1}$) & \\
			\hline
			
			1.0 (fixed) & 0.613$\pm$0.05 & -13.69$\pm$0.04 &1.123(39)\\

			
			0.33$^{+0.34}_{-0.13}$ & 0.62$\pm$0.06 & -13.67$\pm$0.04  &1.015(38) \\
			
			
			0.3 (fixed) & 0.62 $\pm$ 0.06 & -13.67 $\pm$ 0.4 &0.991(39) \\ \hline
			
		\end{tabular}
	\end{table}
	
	We performed the spectral fitting using the XSPEC \citep{1996ASPC..101...17A} software package. The PN, MOS1 and MOS2 background subtracted source spectra were loaded in XSPEC, and the spectral groups containing channels with energies $<$ 0.3 keV and $>$ 2.0 keV were ignored. The X-ray spectra of SXOM61 obtained from PN, MOS1, and MOS2 were fitted simultaneously 
	using collisionally-ionised plasma (apec) model with one characteristic plasma temperature multiplied with an absorber model (tbabs). 
	We estimated the hydrogen column density $N_{H}$ value from the standard relation for dust and extinction for our Milky Way Galaxy, $N_{H}/A_{V}$ = (2.08 $\pm$ 0.02) $\times$ 10$^{21}$ H cm$^{-2}$ mag$^{-1}$ \citep{2017MNRAS.471.3494Z}. For an E(B-V) = 0.13 along the line of sight of NGC 2527, we found that $N_{H}$ = (8.12 $\pm$ 0.08) $\times$ 10$^{20}$  cm$^{-2}$, and we fixed the column density to this value in the tbabs model.
	We used a common value for the metallicity of all the elements relative to the solar abundance values as given by \citet{2009ARA&A..47..481A}. We used the $\chi^2_{red}$ minimization method to determine the best fitting model.  At first we determined the best fit parameters for the models using the elemental abundance values fixed to 1 relative to the solar. The observed X-ray spectra of SXOM61, along with the best fitting model are shown in Figure \ref{fig:spectra_SXOM61}. The best fit model parameters are given in Table \ref{tab:Spectra_Best_Fit}. 
	We also tried a fixed value for the sub-solar metallicity, fixed to 0.3 solar usually seen in many X-ray active late type stars \citet{2003MNRAS.345..714B} and performed the spectral fitting again and obtaining the best fit temperature of the plasma and its normalization. There was little change in the $\chi^2_{red}$ value in between the sub-solar and solar fits (see Table \ref{tab:Spectra_Best_Fit}). We estimated the X-ray flux  coming from SXOM61 in the 0.3-2 keV energy band for both the solar and sub-solar fits using the task 'flux' in the XSPEC. The total flux in the 0.3-2 keV energy band of SXOM61 from the X-ray spectrum is (2.1 $\pm$ 0.2) $\times$ 10$^{-14}$. There is no significant difference in the estimated flux from a solar and sub-solar fits (see Table \ref{tab:Spectra_Best_Fit}).\\
	
	We obtained the energy conversion factor (ECF = Flux/Rate) using the estimated flux and MOS1 count rate of SXOM61 in the 0.3-2 keV energy band. The value of the ECF is found to be 1.73 $\times$ 10$^{-12}$ ergs cm$^{-2}$ counts$^{-1}$. We used this value of ECF to calculate the X-ray fluxes of the other 11 X-ray cluster members. The corresponding X-ray Luminosity was calculated using the formula L = 4$\pi$ $d^2$ $F_X$ where $F_X$ is the X-ray flux of sources and d is the distance of source obtained from Gaia. The estimated X-ray fluxes and luminosity are reported in Table \ref{tab:Gaia_XMM}. We used the 'esensmap' task to generate a background sensitivity map of MOS1 soft and hard band image for an ML of 10 and took it's median value to estimate the detection limit of rate (counts $s^{-1}$). 
	For a given XMM EPIC image, the 'esensmap' task  (part of the 'emchain' task) creates a sensitivity map giving point source detection upper limits for each image pixel. The median background count rate in the soft band and hard band in the XMM MOS1 image is 0.00187 counts $s^{-1}$. The detection limit for the X-ray luminosity of cluster members is thus found to lie between 1.35 $\times$ 10$^{29}$ $-$ 1.77 $\times$ 10$^{29}$ ergs s$^{-1}$ cm$^{-2}$.
	
	\subsection{Spectral Energy Distribution of UV sources}
	The optical CMD given in Figure \ref{fig:CMD} shows that many sources detected in the UV and X-rays are located along single and binary isochrones, while a few other stars deviate from these isochrones. To understand the physical properties of these stars, it is essential to estimate their fundamental parameters that include luminosity ($L_\odot$), radius ($R_\odot$), mass ($M_\odot$), and effective temperature ($T_{eff}$). These parameters can be derived by fitting UV to IR Spectral Energy Distribution (SED) of these stars with Kurucz models \citep{1997A&A...318..841C}. Kurucz model library consists of 3808 discrete stellar atmosphere models with a unique set of $T_{eff}$, metallicity and log $g$ parameters. The $T_{eff}$ of these models ranges from 3500 to 50000 K and log $g$ from 0.0 to 0.5 dex. The available Kurucz model metallicities are -2.5, -2.0, -1.5, -1.0, -0.5, 0.0, 0.2 and 0.5.\\
	
	The SEDs of X-ray and UV bright stars were constructed by combining observed photometric data from UV (XMM OM - UVW2 and UVM2; UVOT - UVM2, UVW2, and UVW1), optical (Gaia - G, G$_{BP}$ and G$_{RP}$; Paranal - u, g, r, Halpha, and i; MISC/APASS B and B; PAN-STARRS - g, r, z, y and i), near-infrared (2MASS - J, H, and K$_s$), and mid-infrared (WISE - W1, W2) observations. We used the Virtual Observatory SED Analyzer (VOSA) \citep{2008A&A...492..277B} tool to obtain the fluxes in optical (Gaia, APASS, Paranal, and PAN-STARRS) and IR (2MASS and WISE) bands within a search radius of 5\arcsec. We downloaded the Kurucz model synthetic flux from VOSA for these bands. VOSA corrects the observed flux for extinction in their respective wavebands using the extinction law of \citet{1999PASP..111...63F} improved by \citet{2005ApJ...619..931I} in the infrared. The dereddened fluxes in optical and infrared bands are given in online-only Tables B and C. \\
	
	Since VOSA only performs single spectrum SED fitting which is not suited for potential binary stars, we developed a python code for SED analysis that can perform both single and composite spectral fitting. The code constructs the observed SED and compares the observed flux with the synthetic Kurucz model flux using $\chi^2_{red}$ statistical analysis. The synthetic Kurucz model fluxes are scaled by a variable scaling parameter $M_d$ = ($\frac{R}{D}$)$^2$, which is optimized using Nelder-Mead optimization technique to get the least $\chi^2_{red}$ for each model. For each Kurucz model, the code finds the optimum scaling parameter to minimize the $\chi^2_{red}$. It compares the optimum value of $\chi^2_{red}$ for each model and returns the best fit model parameters whose $\chi^2_{red}$ is closest to 1. The code also estimates the residual flux, where Residual = $\frac{Observed Flux - Model Flux}{Observed Flux}$.\\
	
	For a single spectrum SED fitting, the $\chi^2_{red}$ function is given by eqn. \ref{eqn:chi}. In the case of composite spectrum fitting, the sum of fluxes of two Kurucz models is compared with the observed fluxes using eqn. \ref{eqn:chi_double}. The composite spectrum obtained by the addition of two Kurucz model spectra is referred to as Kurucz and Kurucz model.\\
	
	\begin{equation}
	\label{eqn:chi}
	\centering
	\chi^{2}_{red} = \frac{1}{N-N_{fit}} \sum_{i=1}^{N} \bigg\{\frac{ (F_{o,i} - M_d F_{m,i})^2 }{\sigma^2_{o,i}}\bigg\}
	\end{equation}
	
	where N is the number of photometric data points, N$_{fit}$ is the number of free parameters in the model, $F_{o,i}$ is the observed flux, $M_d F_{m,i}$ is the model flux of the star, and $\sigma_{o,i}$ is the error in the observed flux, $M_d$ = ($\frac{R}{D}$)$^2$ is the scaling parameter, R is the radius of the star, and D is the distance to the star. \\

	\begin{equation}
	\label{eqn:chi_double}
	\centering
	\chi^{2}_{red} = \frac{1}{N-N_{fit}} \sum_{i=1}^{N} \bigg\{\frac{ (F_{o,i} - M^1_{d} F^1_{m,i} - M^2_{d} F^2_{m,i})^2}{\sigma^2_{o,i}}\bigg\}
	\end{equation}
	
	where N is the number of photometric data points, N$_{fit}$ is the number of free parameters in the model, $F_{o,i}$ is the observed flux, $F^1_{m,i}$ and $F^2_{m,i}$ is the model flux of first and second component respectively, $\sigma_{o,i}$ is error in the observed flux, $M_d^1$ and $M_d^2$ are the scaling parameters for the first and second component. \\
	
	We obtained $T_{eff}$, log $g$, and metallicity from the best fitting Kurucz model parameters. We estimated the radius, R$_\odot$, from the scaling parameter $M_d$, and bolometric luminosity (L$_\odot$) of the star using eqn \ref{eqn:luminosity}. We found the mass of the star by comparing the estimated luminosity and temperature with the best fitting PARSEC isochrone. The uncertainties in R$_\odot$ and L$_\odot$ were estimated using eqn. \ref{eqn:R} and eqn. \ref{eqn:L} respectively.\\

	\begin{equation}
	\label{eqn:luminosity}
	L = 4 \pi R^2 \sigma T^4
	\end{equation}
	
	where R is the radius of star, $\sigma$ the Stefan-Boltzmann constant and T is the effective temperature of star estimated from SED fitting.

	\begin{equation}
	\label{eqn:R}
	\bigtriangleup R = \frac{R\bigtriangleup D}{D}
	\end{equation}
	
	where $\bigtriangleup D$ is the uncertainty in distance for each star given in Table \ref{tab:Gaia_XMM}. 
	
	\begin{equation}
	\label{eqn:L}
	\bigtriangleup L = L\sqrt{\bigg(\frac{2\bigtriangleup R}{R}\bigg)^2 + \bigg(\frac{4\bigtriangleup T}{T}\bigg)^2}
	\end{equation}	 
	
	where $\bigtriangleup$T is the uncertainty in T.\\

	We estimated the fundamental parameters of 53 cluster members that were detected by XMM-OM and Swift UVOT. The metallicity of the NGC2527 cluster is near solar ([Fe/H] = -0.1, \citealt{2013MNRAS.431.3338R}). The metallicities of Kurucz models that bracket this value are [Fe/H] = 0.0 and -0.5. In our analysis, we thus adopted the [Fe/H] = 0.0 models. Given the near-solar metallicity of the cluster, this simplification is unlikely to have a significant impact on the SED fits.\\
	
	
	If a star shows residual flux in excess of more than 50\% in UV band, it is classified as having UV excess. 23 out of 53 stars are found to show an excess in the UV band. 10 stars are found to show a prominent dip feature near 2250 \AA $ $. The fundamental parameters of these stars are given in Table \ref{tab:SED_Single_Component}. A closer inspection of PAN-STARRS, 2MASS, and WISE images shows that three stars SXOM12, SXOM17, and SXOM56 are multiple stars that are resolved in the PAN-STARRS image but are unresolved in the WISE and 2MASS images. So the results obtained from SED fitting of these stars are deemed spurious. \\
	
	Of the remaining 20 stars that show UV excess, 14 are MS stars, 4 are binary stars, and 2 are RGB stars according to their position in the CMD. In order to understand the nature of excess UV residual, and determine the fundamental parameters of both the components in binaries, we performed composite spectrum SED fitting for 4 binaries and 2 RGB stars that show UV excess. In addition, we also performed composite spectrum SED fitting for 4 more binary stars which emit in X-rays.
	A comparison of single and composite spectrum SED fit of these stars is summarized in Table \ref{tab:SED_Double_Component}. We present the results below.
	
	\begin{table*}
		\centering
		\caption{Fundamental parameters of cluster members detected in X-ray and UV, estimated from single Kurucz spectrum SED fitting. Column 1 gives the ID of stars detected in X-ray or UV, columns 2, 3 and 6 list the T$_{eff}$, log g and reduced $\chi^2$ of the best fitting Kurucz spectrum, columns 4 \& 5 list the estimated luminosity and radius, columns 7 and 8 indicate whether a Excess and Dip feature is observed in the residual plots of the single spectrum SED fit.}
		\label{tab:SED_Single_Component}
		\begin{tabular}{cccccccccc}
			\hline
			SXOM   &$T_{eff}$    &log g   &Radius &Luminosity &Mass   &$\chi^2_{red}$ &Excess &Dip \\
			& (K)            &        & ($R_\odot$) &($L_\odot$)   &($M_\odot$)  &     & \\
			\hline
			1	&7750$\pm$250	&4.5$\pm$0.5	&2.05$\pm$0.06	&13.7$\pm$1.94	&1.78	&28.1	&N	&N	&\\
			2	&7250$\pm$250	&3.5$\pm$0.5	&1.24$\pm$0.03	&3.8$\pm$0.55	&1.35	&48.81	&N	&N	&\\
			3	&7500$\pm$250	&4.0$\pm$0.5	&1.49$\pm$0.04	&6.32$\pm$0.9	&1.51	&26.52	&N	&N	&\\
			4	&6250$\pm$250	&3.5$\pm$0.5	&0.93$\pm$0.01	&1.19$\pm$0.19	&1.09	&70.31	&Y	&Y	&\\
			5	&8000$\pm$250	&4.0$\pm$0.5	&1.7$\pm$0.04	&10.64$\pm$1.42	&1.68	&28.87	&N	&N	&\\
			6	&8750$\pm$250	&4.5$\pm$0.5	&2.25$\pm$0.05	&26.66$\pm$3.31	&2.03	&14.47	&N	&N	&\\
			7	&7000$\pm$250	&3.5$\pm$0.5	&1.21$\pm$0.02	&3.15$\pm$0.47	&1.32	&73.48	&N	&N	&\\
			8	&5750$\pm$250	&3.0$\pm$0.5	&1.2$\pm$0.02	&1.41$\pm$0.25	&1.1	&65.94	&Y	&N	&\\
			9	&8500$\pm$250	&4.0$\pm$0.5	&1.73$\pm$0.04	&14.13$\pm$1.79	&1.78	&29.83	&N	&N	&\\
			10	&6500$\pm$250	&5.0$\pm$0.5	&1.23$\pm$0.01	&2.43$\pm$0.38	&1.24	&16.48	&Y	&N	&\\
			11	&6250$\pm$250	&5.0$\pm$0.5	&0.92$\pm$0.01	&1.16$\pm$0.19	&1.04	&102.64	&Y	&Y	&\\
			12	&5250$\pm$250	&0.5$\pm$0.5	&0.68$\pm$0.01	&0.32$\pm$0.06	&0.82	&165.27	&Y	&N	&\\
			14	&8000$\pm$250	&4.0$\pm$0.5	&1.8$\pm$0.04	&11.91$\pm$1.59	&1.73	&31.54	&N	&N	&\\
			15	&5750$\pm$250	&3.5$\pm$0.5	&0.87$\pm$0.01	&0.74$\pm$0.13	&0.99	&41.51	&Y	&N	&\\
			16	&6250$\pm$250	&4.0$\pm$0.5	&1.23$\pm$0.01	&2.08$\pm$0.34	&1.19	&70.46	&Y	&Y	&\\
			17	&3500$\pm$250	&3.0$\pm$0.5	&1.75$\pm$0.08	&0.41$\pm$0.12	&0.85	&2308.31	&Y	&N	&\\
			20	&6500$\pm$250	&4.0$\pm$0.5	&1.29$\pm$0.02	&2.67$\pm$0.42	&1.24	&59.05	&Y	&Y	&\\
			21	&7500$\pm$250	&3.5$\pm$0.5	&1.33$\pm$0.03	&5.0$\pm$0.7	&1.44	&58.19	&N	&N	&\\
			22	&7000$\pm$250	&3.5$\pm$0.5	&1.15$\pm$0.02	&2.88$\pm$0.43	&1.29	&76.85	&N	&Y	&\\
			23	&6000$\pm$250	&4.0$\pm$0.5	&1.17$\pm$0.02	&1.61$\pm$0.28	&1.14	&43.91	&Y	&N	&\\
			24	&7000$\pm$250	&3.5$\pm$0.5	&1.45$\pm$0.03	&4.54$\pm$0.68	&1.42	&36.64	&N	&N	&\\
			25	&6000$\pm$250	&3.0$\pm$0.5	&0.89$\pm$0.01	&0.91$\pm$0.15	&1.0	&37.52	&Y	&Y	&\\
			26	&7750$\pm$250	&3.5$\pm$0.5	&1.49$\pm$0.03	&7.21$\pm$0.98	&1.56	&29.36	&N	&N	&\\
			27	&8500$\pm$250	&4.5$\pm$0.5	&1.82$\pm$0.11	&15.53$\pm$2.62	&1.83	&19.76	&N	&N	&\\
			29	&7750$\pm$250	&4.0$\pm$0.5	&1.49$\pm$0.03	&7.19$\pm$0.97	&1.56	&30.29	&N	&N	&\\
			31	&7500$\pm$250	&3.5$\pm$0.5	&1.78$\pm$0.03	&9.03$\pm$1.25	&1.63	&49.17	&N	&N	&\\
			33	&7250$\pm$250	&3.5$\pm$0.5	&1.78$\pm$0.03	&7.91$\pm$1.13	&1.58	&67.17	&N	&N	&\\
			34	&7000$\pm$250	&4.0$\pm$0.5	&1.71$\pm$0.04	&6.34$\pm$0.96	&1.51	&84.42	&N	&Y	&\\
			35	&7500$\pm$250	&4.0$\pm$0.5	&1.48$\pm$0.03	&6.27$\pm$0.88	&1.51	&27.38	&N	&N	&\\
			36	&8250$\pm$250	&4.5$\pm$0.5	&2.06$\pm$0.05	&17.74$\pm$2.31	&1.88	&18.19	&N	&N	&\\
			37	&7750$\pm$250	&4.5$\pm$0.5	&1.7$\pm$0.03	&9.44$\pm$1.27	&1.63	&36.24	&N	&N	&\\
			39	&5750$\pm$250	&3.0$\pm$0.5	&0.79$\pm$0.02	&0.62$\pm$0.11	&0.95	&62.21	&Y	&N	&\\
			40	&8000$\pm$250	&3.5$\pm$0.5	&1.62$\pm$0.03	&9.69$\pm$1.26	&1.63	&32.32	&N	&N	&\\
			41	&7500$\pm$250	&4.0$\pm$0.5	&1.45$\pm$0.03	&5.97$\pm$0.83	&1.49	&26.99	&N	&N	&\\
			42	&5500$\pm$250	&2.5$\pm$0.5	&10.07$\pm$0.25	&83.5$\pm$15.74	&2.55	&528.64	&Y	&N	&\\
			43	&8000$\pm$250	&4.0$\pm$0.5	&1.51$\pm$0.04	&8.39$\pm$1.12	&1.58	&35.79	&N	&N	&\\
			44	&8750$\pm$250	&2.0$\pm$0.5	&1.99$\pm$0.06	&20.92$\pm$2.69	&1.95	&11.07	&N	&N	&\\
			45	&7500$\pm$250	&5.0$\pm$0.5	&1.44$\pm$0.03	&5.92$\pm$0.82	&1.49	&6.95	&N	&N	&\\
			47	&8000$\pm$250	&4.5$\pm$0.5	&2.71$\pm$0.06	&27.0$\pm$3.57	&2.03	&14.05	&N	&N	&\\
			48	&7250$\pm$250	&4.5$\pm$0.5	&1.96$\pm$0.04	&9.58$\pm$1.37	&1.63	&28.41	&N	&N	&\\
			49	&5000$\pm$250	&0.5$\pm$0.5	&0.67$\pm$0.02	&0.25$\pm$0.05	&0.8	&43.69	&Y	&N	&\\
			50	&8000$\pm$250	&4.0$\pm$0.5	&1.93$\pm$0.06	&13.8$\pm$1.9	&1.78	&27.81	&N	&N	&\\
			51	&5000$\pm$250	&1.0$\pm$0.5	&0.7$\pm$0.02	&0.28$\pm$0.06	&0.8	&22.05	&Y	&N	&\\
			52	&5750$\pm$250	&2.0$\pm$0.5	&0.76$\pm$0.01	&0.57$\pm$0.1	&0.95	&39.22	&Y	&N	&\\
			53	&7000$\pm$250	&3.5$\pm$0.5	&1.12$\pm$0.02	&2.71$\pm$0.4	&1.29	&62.12	&N	&Y	&\\
			54	&8000$\pm$250	&4.0$\pm$0.5	&1.67$\pm$0.03	&10.3$\pm$1.35	&1.67	&39.78	&N	&N	&\\
			55	&6000$\pm$250	&2.5$\pm$0.5	&0.82$\pm$0.01	&0.78$\pm$0.13	&0.99	&38.47	&Y	&N	&\\
			56	&5500$\pm$250	&5.0$\pm$0.5	&0.72$\pm$0.01	&0.43$\pm$0.08	&0.85	&339.48	&Y	&N	&\\
			57	&6500$\pm$250	&3.0$\pm$0.5	&0.92$\pm$0.01	&1.35$\pm$0.21	&1.1	&94.06	&Y	&Y	&\\
			58	&5500$\pm$250	&2.5$\pm$0.5	&0.76$\pm$0.01	&0.47$\pm$0.09	&0.9	&35.65	&Y	&N	&\\
			59	&6500$\pm$250	&3.5$\pm$0.5	&1.09$\pm$0.01	&1.91$\pm$0.3	&1.19	&108.31	&Y	&N	&\\
			60	&8000$\pm$250	&3.5$\pm$0.5	&1.89$\pm$0.04	&13.19$\pm$1.75	&1.78	&22.31	&N	&N	&\\
			61	&5500$\pm$250	&2.5$\pm$0.5	&9.68$\pm$0.19	&77.21$\pm$14.37	&2.45	&228.28	&Y	&N	&\\
			62	&6250$\pm$250	&3.5$\pm$0.5	&0.92$\pm$0.01	&1.17$\pm$0.19	&1.09	&72.79	&Y	&Y	&\\
			65	&5250$\pm$250	&2.5$\pm$0.5	&10.17$\pm$0.23	&70.69$\pm$13.83	&2.49	&5.12	&N	&N	&\\ \hline
		\end{tabular}
	\end{table*}
	
	\begin{table*}
		\centering
		\caption{Comparison of fundamental parameters of cluster members detected in X-ray and UV, estimated from single and composite Kurucz spectrum SED fitting. Column 1 gives the ID of stars detected in X-ray or UV. Columns 2 \& 3 are T$_{eff}$ and reduced $\chi^2$ of the best fitting single Kurucz spectrum. Columns 4, 7, 9, and 11 list the estimated parameters of the first component in the Kurucz and Kurucz composite spectrum fit. Columns 5, 8, 10, and 12 list the estimated parameters of the second component in the Kurucz and Kurucz composite spectrum fit. Column 6 gives the reduced $\chi^2$ of the best fitting composite spectrum.}
		\label{tab:SED_Double_Component}
		\resizebox{2\columnwidth}{!}{
			\begin{tabular}{ccccccccccccc}
				\hline
				SXOM& T & $\chi^2_{red}$ & $T_1$ & $T_2$  & $\chi^2_{red}$ & $R_1$  & $R_2$  & $L_1$ & $L_2$ & $M_1$ & $M_2$   \\
				& (K) &  &  (K)  & (K)   &   &  ($R_\odot$)  & ($R_\odot$)  & ($L_\odot$) & ($L_\odot$) & ($M_\odot$) & ($M_\odot$)  &  \\ \hline
				
				8	&5750$\pm$250	&65.9	&7000$\pm$250	&5500$\pm$250	&37.0	&0.37$\pm$0.0	&1.18$\pm$0.01	&0.3$\pm$0.04	&1.14$\pm$0.21	&0.82	&1.04	&\\
				16	&6250$\pm$250	&70.5	&6000$\pm$250	&8500$\pm$250	&16.5	&1.28$\pm$0.01	&0.17$\pm$0.0	&1.93$\pm$0.32	&0.14$\pm$0.02	&1.19	&0.7	&\\
				20	&6500$\pm$250	&59.0	&7750$\pm$250	&6000$\pm$250	&21.1	&0.46$\pm$0.01	&1.31$\pm$0.02	&0.67$\pm$0.09	&1.99$\pm$0.34	&0.95	&1.19	&\\
				23	&6000$\pm$250	&43.9	&5750$\pm$250	&8000$\pm$250	&19.6	&1.2$\pm$0.02	&0.23$\pm$0.0	&1.41$\pm$0.25	&0.2$\pm$0.03	&1.1	&0.75	&\\
				31	&7500$\pm$250	&49.2	&8250$\pm$250	&6500$\pm$250	&4.5	&1.09$\pm$0.02	&1.56$\pm$0.03	&4.96$\pm$0.63	&3.91$\pm$0.62	&1.44	&1.34	&\\
				33	&7250$\pm$250	&67.2	&6500$\pm$250	&8250$\pm$250	&16.0	&1.63$\pm$0.03	&0.92$\pm$0.02	&4.28$\pm$0.68	&3.51$\pm$0.45	&1.39	&1.34	&\\
				34	&7000$\pm$250	&84.4	&6500$\pm$250	&8500$\pm$250	&25.7	&1.73$\pm$0.04	&0.55$\pm$0.01	&4.79$\pm$0.77	&1.42$\pm$0.18	&1.42	&1.1	&\\
				42	&5500$\pm$250	&528.6	&49000$\pm$1000	&5250$\pm$250	&496.6	&0.01$\pm$0.0	&11.17$\pm$0.28	&0.65$\pm$0.06	&85.38$\pm$16.81	&-	&2.55	&\\
				48	&7250$\pm$250	&28.4	&6750$\pm$250	&8250$\pm$250	&2.0	&1.71$\pm$0.03	&0.99$\pm$0.02	&5.48$\pm$0.84	&4.08$\pm$0.52	&1.46	&1.39	&\\
				61	&5500$\pm$250	&228.3	&5250$\pm$250	&49000$\pm$1000	&135.5	&10.87$\pm$0.21	&0.01$\pm$0.0	&80.8$\pm$15.72	&0.86$\pm$0.08	&2.52	&-	&\\ \hline
		\end{tabular}}
	\end{table*}
	
	\begin{figure*}
		\centering
		\includegraphics[width=0.5\linewidth]{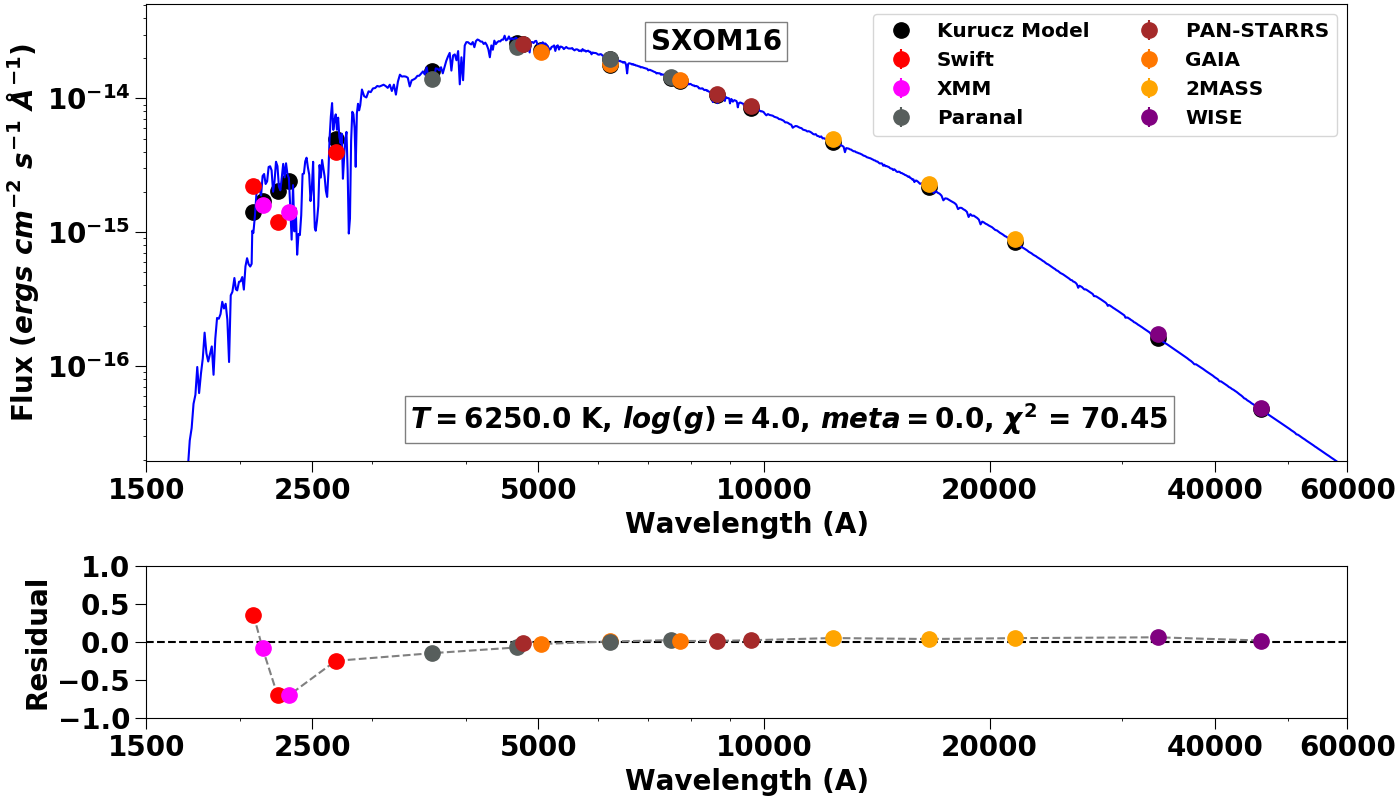}\includegraphics[width=0.5\linewidth]{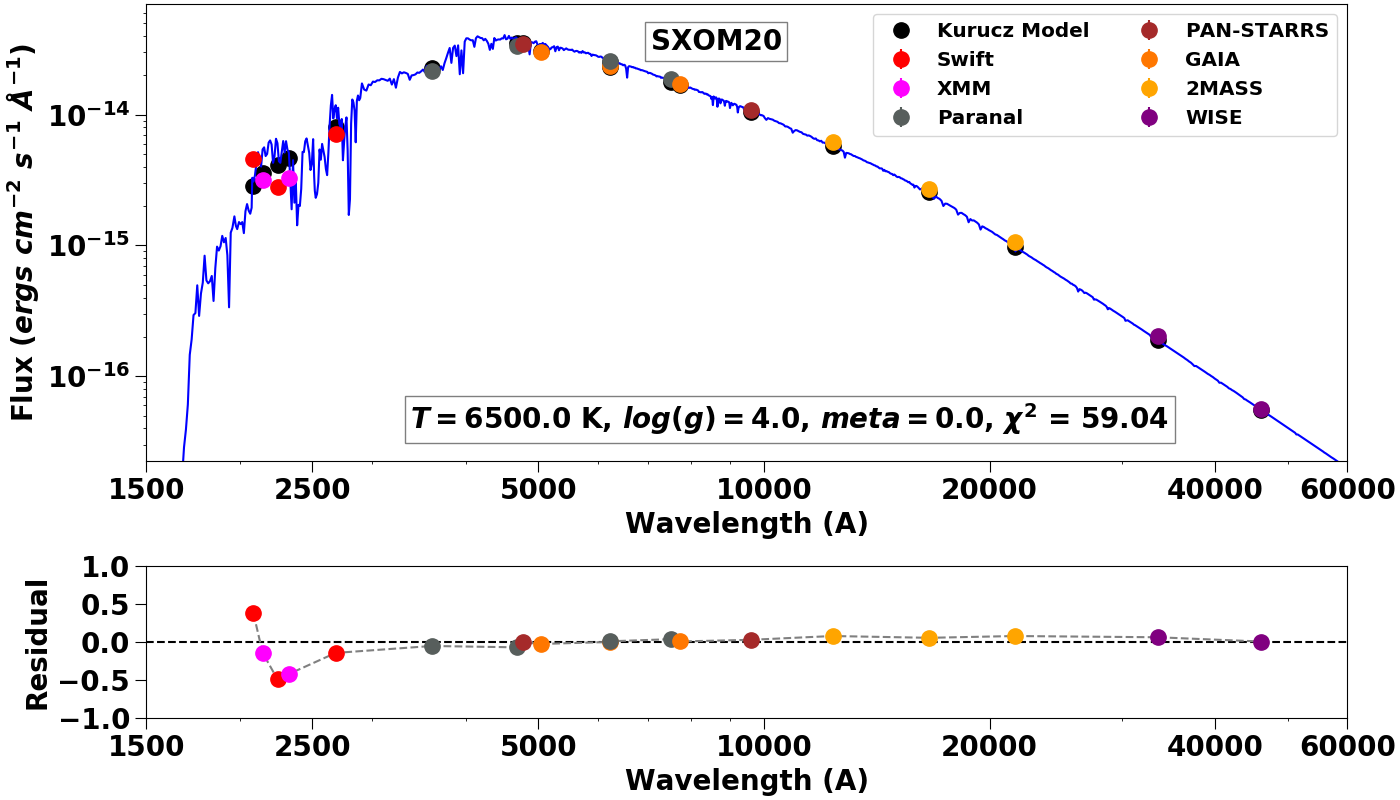}
		\includegraphics[width=0.5\linewidth]{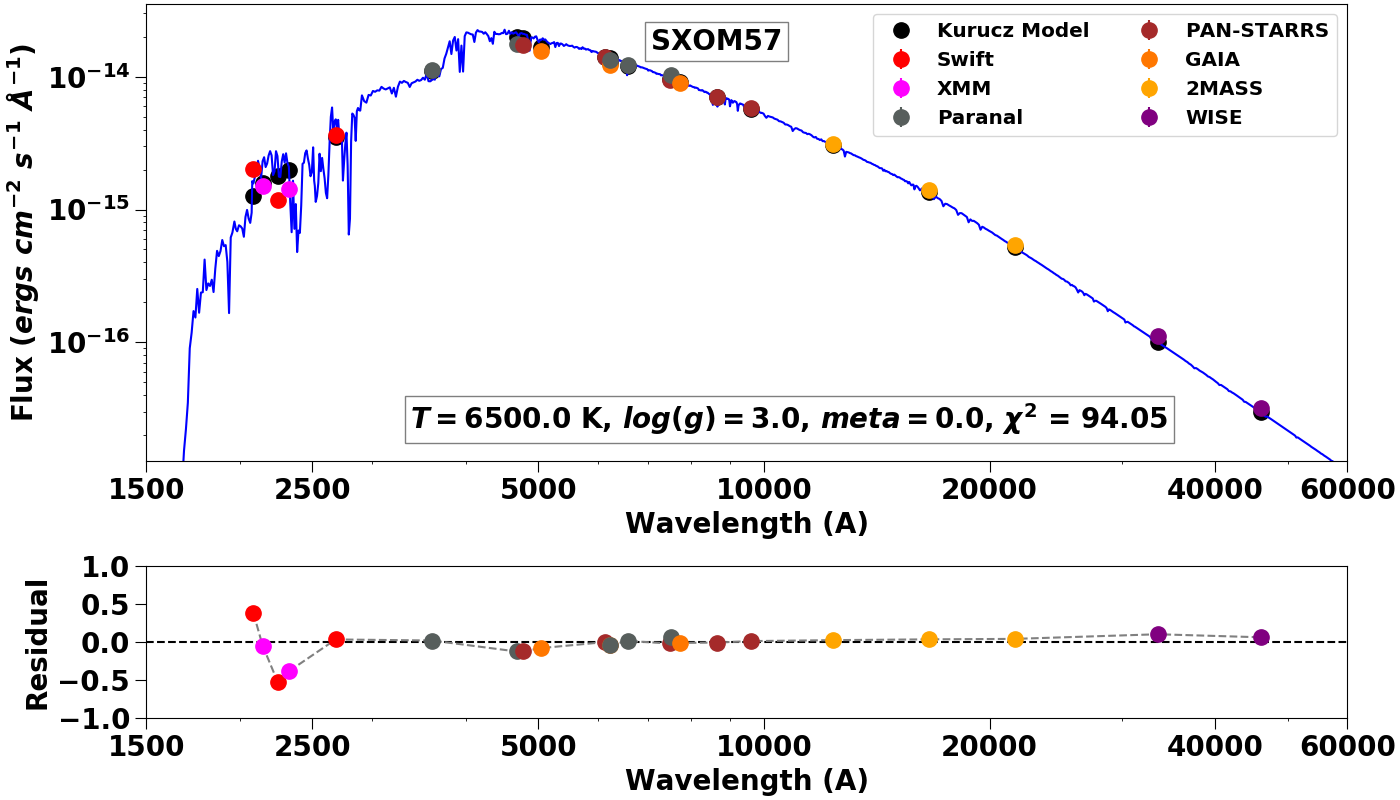}
		\caption{SED of stars that show a prominent dip around 2250 \AA $ $.}
		\label{fig:dip}
	\end{figure*}
	
	\section{Results}
	\label{sec:results}
	
	\subsection{Fundamental properties of NGC 2527}
	In the very early studies of NGC 2527 by \citet{1973A&AS....9..229L} and \citet{1977MNRAS.181..729D}, they had estimated the age to be around 500 Myr and 1 Gyr, with reddening E(B-V) = 0.077 $\pm$ 0.016, and a distance modulus of 8.70 mags. Since these studies had limited coverage in area and photometric depth, the sample of cluster members is incomplete, and estimated parameters were not reliable. \citet{2018A&A...615A..49C} determined the membership of cluster stars using Gaia-TYCHO astrometric solution and found the age of NGC 2527 to be around 1 Gyr, metallicity [Fe/H] = -0.26 $\pm$ 0.11, E(B$-$V) = 0.077 $\pm$ 0.016 and V$-$M$_V$ = 8.37 $\pm$ 0.10 mag by fitting isochrones to CMD based on 2MASS magnitudes. 
	They used an automated code BASE-9 to determine these parameters, but the estimated metallicity is relatively lower as compared to the average value of [Fe/H] = -0.1 \citep{2013MNRAS.431.3338R} obtained from spectroscopic studies of RGB stars.
	Also, the distance modulus was lower than that obtained from Gaia DR2. \citet{2019AJ....158...35S} fitted PARSEC isochrones to UV CMDs and found an age of 450 Myr, with a distance modulus of V$-$M$_V$ = 8.7 and a reddening E(B$-$V) = 0.04 mag. They assumed a lower distance modulus obtained from early studies of the cluster by \citep{1977MNRAS.181..729D}, which led them to estimate a younger age for the cluster. \\
	
	In this study, we determined the fundamental parameters of NGC 2527 by fitting PARSEC isochrones to the CMD based on Gaia magnitudes. The unprecedented coverage and depth in the photometry offered by Gaia DR2, allows us to cover a complete sample of cluster members.  We find that the age of NGC 2527 is $\sim$ 630 Myr, reddening as E(B$-$V) = 0.13 mag, and distance modulus as $G-M_G$ =  9.0 $\pm$ 2.4 mag from the isochrone fitting. The average radial velocity of cluster members from Gaia DR2 radial velocity measurements is 41 $\pm$ 9 km/s. \\
	
	\subsection{Red Giants}
	We detect 3 RGB stars in either X-ray or UV that lie in between single and binary isochrones in the optical CMD. SXOM61 emits in both X-ray and UV. SXOM42 lies in the field of view of XMM-EPIC detectors but is only detected in the UV. SXOM65 emits X-rays but is not detected in the UV observations because it lies outside the field of view of Swift UVOT and XMM-OM detectors. 
	The RGB stars have high proper motion membership probability (given in Table \ref{tab:Gaia_XMM}), and their Gaia radial velocities are close to the mean radial velocity of the cluster. Therefore the RGB stars are kinematic members of the cluster. \\
	
\begin{figure*}
		\centering
		\includegraphics[width=0.5\linewidth]{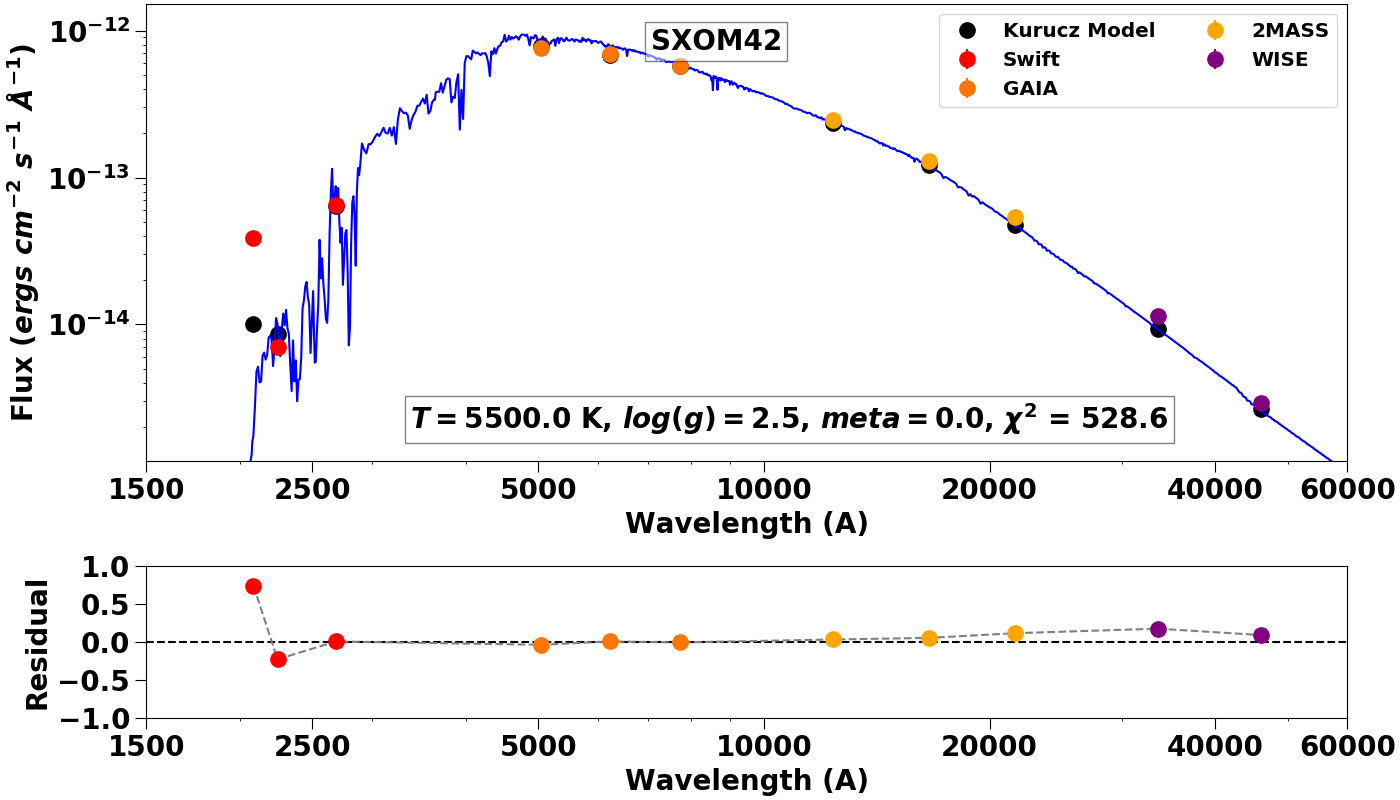}\includegraphics[width=0.5\linewidth]{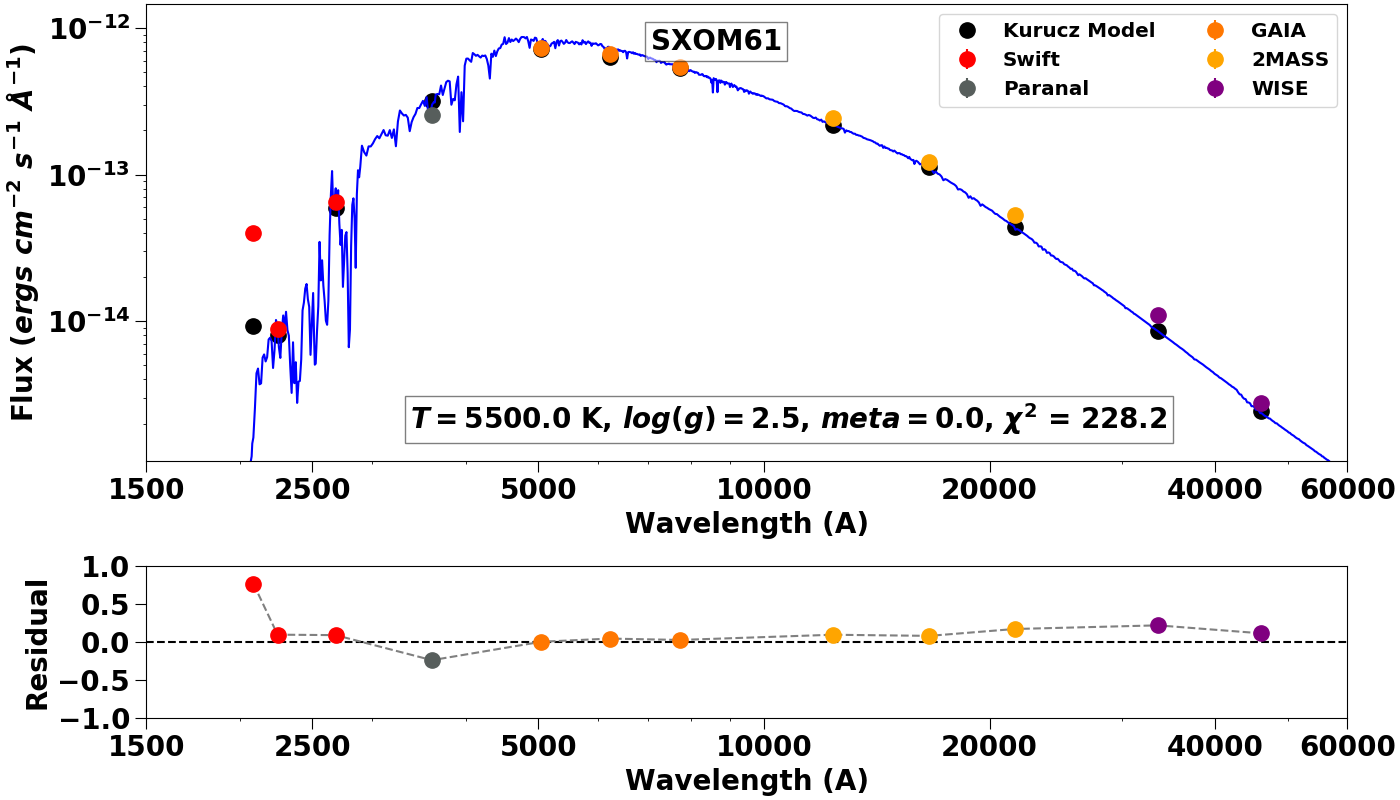}
		\caption{Broad band SED of two red giants that show excess UV flux. The best fit Kurucz spectrum is plotted over the observed SED and the fundamental parameters are given in each plot.}
		\label{fig:redgiants}
	\end{figure*}

	SXOM61 and SXOM42 occupy a similar position in between the single and binary isochrone in UV CMDs, except in UVW2-G CMD. In UVW2-G CMD, they show an excess UVW2-G colour and lie close to the single star isochrone. The single spectrum SED fits to these 3 stars estimate R$_\odot$ $\approx$ 10 $R_\odot$ and T$_{eff}$ $\approx$ 5500 K respectively, which corresponds to a spectral type G2 III. SED residual plots of SXOM42 and SXOM61 (see Figure \ref{fig:redgiants}) show a 75\% excess in UVW2 flux as compared to the Kurucz model flux of a G2 III star. The origin of excess flux in the UVW2 band in these stars could be due to chromospheric activity, presence of a hotter companion, or red leakage in the UVW1 filter. The composite spectrum fitting using a combination of Kurucz and Kurucz models does not significantly decrease the excess flux, and the best fit temperature of individual components is exceptionally high, making the fit unreliable. The flux of these RGB stars in UV filters is at least $10^4$ times higher as compared to the detection limits suggests that the high flux is unlikely due to red leakage. Hence, the excess in UV is most likely due to high chromospheric activity or a hotter companion. \\
	
	
	The X-ray spectrum of SXOM61 shown in Figure \ref{fig:spectra_SXOM61} is fitted by a single component plasma model that has a temperature of 0.62 $\pm$ 0.06 keV or (7.1 $\pm$ 0.7) $\times$ 10$^6$ K. The high plasma temperature is suggestive of a highly active corona. The X-ray luminosity (Lx) of SXOM61 and SXOM65 is 10.2 $\times$ 10$^{29}$ ergs s$^{-1}$ and 3.8 $\times$ 10$^{29}$ ergs s$^{-1}$ suggecting that both the stars have an active corona. As stars evolve off the main sequence, they undergo rapid loss of angular momentum as a result of magnetic breaking through coronal winds. Therefore, the RGB stars are slowly rotating stars with relatively less active corona. It is highly unlikely for a RGB star to have an active corona, unless it is tidally locked in a binary system or a fast rotating merger remnant of a contact binary, like FK Comae. The high coronal activity and excess UV flux hints that these stars are potential FK Comae candidates. It is a rare class of rapidly rotating late-type giants or subgiants, which show signs of active chromospheres and X-ray emission \citep{1981ApJ...247L.131B}. 
	
	\subsection{Main sequence stars}
	\subsubsection{Coronally active}
	We detect X-ray emission from 5 main sequence stars. Two of these stars, SXOM63 and SXOM64, lie outside the field of view of XMM OM and Swift UVOT detectors; therefore, we could not detect UV emission from them. We determined the fundamental parameters of other three stars (SXOM7, SXOM53, and SXOM59) by performing single spectrum SED fitting. The temperature of these stars suggests that they are F type stars and their X-ray luminosity is $\sim$ 10$^{29}$ $ergs$ $s^{-1}$, suggesting that they are coronally active. The origin of the active corona is directly linked to the coupling of convection and rotation in these stars, which produces a solar-type dynamo. The magnetic field produced by the solar dynamo heats the plasma in stellar corona, which produces X-rays \citep{1978ApJ...220..643R}.
	
	\subsubsection{UV excess}
	The residuals of single spectrum SED fit of 10 stars (see Table \ref{tab:SED_Single_Component}) shows a decrease in flux around 2250 \AA $ $ by $\sim$50\% and then rise in flux around 1930 \AA $ $ by $\sim$50\%, as compared to the model flux (see Figure \ref{fig:dip}). The estimated temperature of these stars suggests that they are solar-type stars with temperature $\le$ 6500 K. The prominent saddle dip in the observed SED of SXOM16, SXOM20, and SXOM57, hints towards the inefficiency of atmospheric models to characterize the absorption lines of cool stars. The combined spectro-photometric coverage of XMM-OM and Swift UVOT allows us to visualize the poorly characterized absorption line feature. SXOM16 and SXOM20 lie close to the binary isochrone, but their composite spectrum fit are ineffective because of the poor model fit in the UV band.\\
	
	The single spectrum SED fitting of many MS stars show an excess in the UV band suggestive of chromospheric activity or the presence of a hot WD companion. The absence of a far-UV (FUV) flux data point makes it difficult to determine the nature of excess UV flux. The broad-band SEDs of these stars are shown in Figure \ref{fig:chromospheric_active}. \citet{2018MNRAS.481..226S} also found several stars near and below the MSTO in M67 to show an excess in NUV band. They suggested that these stars could have high chromospheric activity. \citet{2009ApJ...707..852R} studied 15 solar-type stars in M67 and concluded that rapid rotation is the reason for the high chromospheric activity in these stars.\\
	
\begin{figure*}
		\centering
		\includegraphics[width=0.5\linewidth]{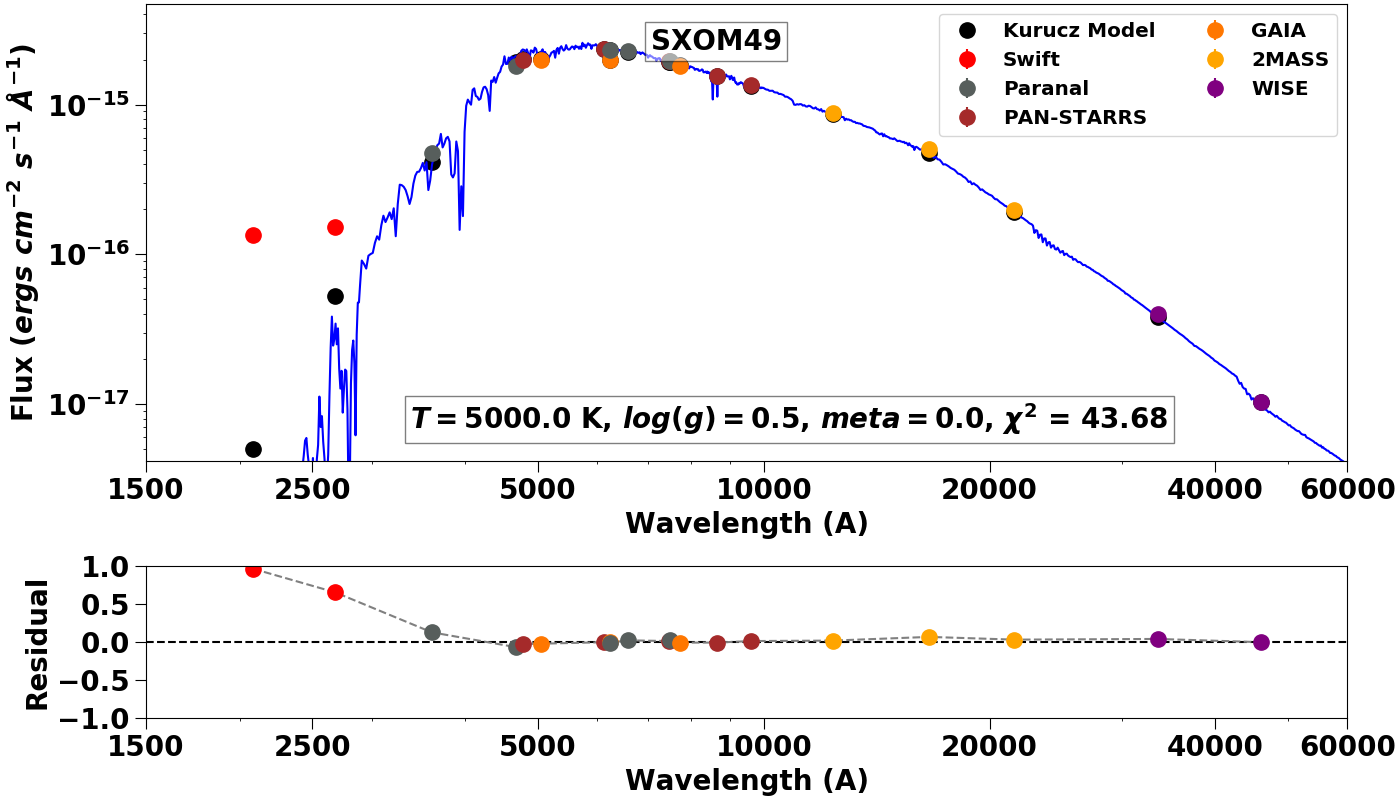}\includegraphics[width=0.5\linewidth]{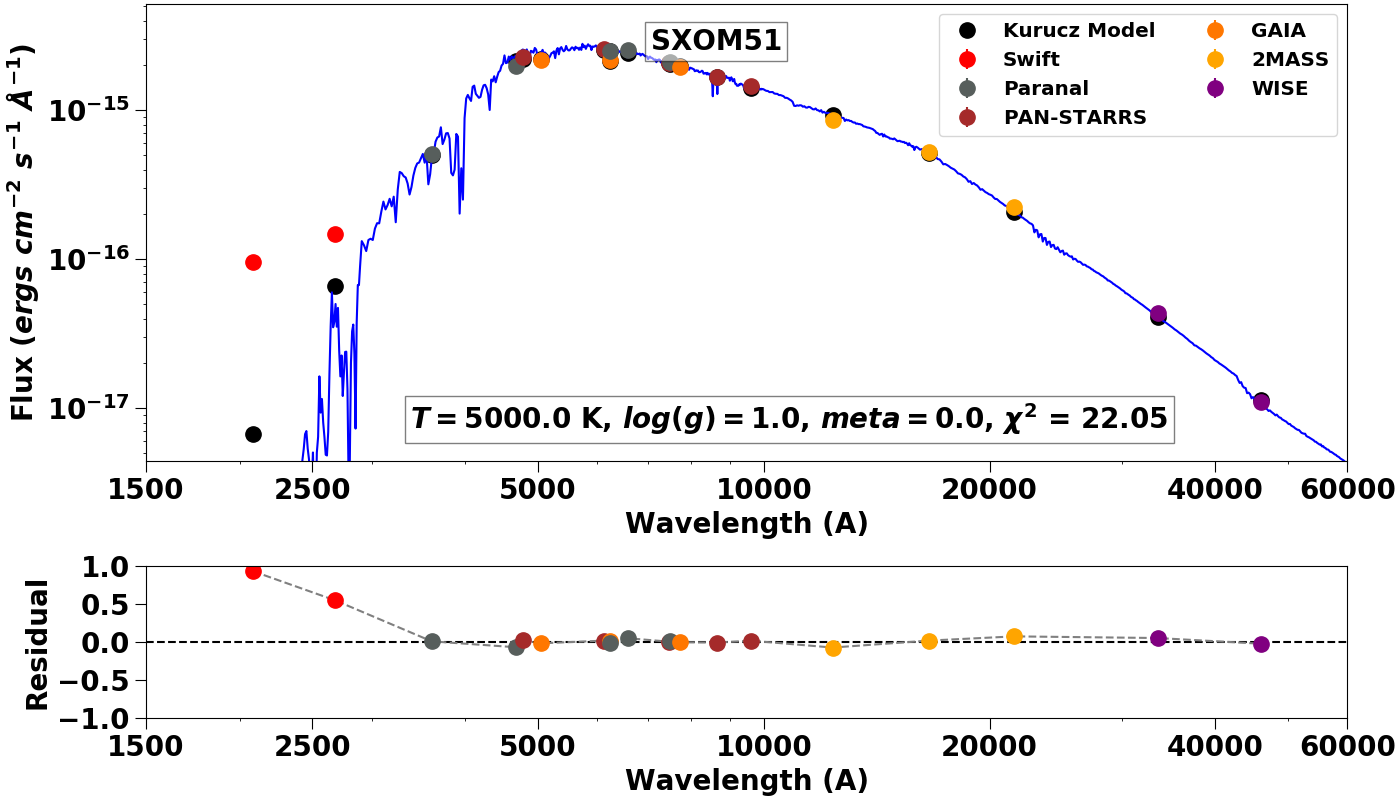}
		\caption{SED of two faint subgiant stars that show excess UV flux mostly due to red leak in UVW1 and UVW2 filters.}
		\label{fig:redleak}
	\end{figure*}

	The broad-band SED of two of these stars, SXOM49 and SXOM51 shows huge excess flux in the Swift UVW1 and UVW2 filters, but are not detected in Swift UVM2 filter (see Figure \ref{fig:redleak}). However, since these stars are not detected in UVM2, the excess flux in UVW1 and UVW2 filters could be due to red leak usually observed in Swift UVW1 and UVW2 filters, or the presence of a hot WD companion. \citet{2012AJ....144...65S} have shown that if a cool K type star is detected and shows an excess in all three UVOT filters, only then it has a potential hot WD companion. Since SXOM49 and SXOM51 are not detected in UVM2 filter, the excess UV flux could be an artefact. A comparison of the best fitting model flux and observed flux of SXOM49 and SXOM51 with the detection limits (see Appendix B) of UVOT filters is given in Table \ref{tab:red_leak}. The observed flux in UVW2 and UVW1 bands are at least 10 times higher than the model flux and the detection limits. For the given detection limits, if the UV excess is due to the presence of a hot WD companion, we should have detected an excess flux in the UVM2 band as well. \\
	
	The stellar parmaters of stars (given in Table 5) that suffer from significant red leakage (SXOM49 and SXOM51) corresponds to a spectral type of G5 VI. From the spectral models available in \citet{2010ApJ...721.1608B}, the spectral type closest to G5 VI is G5 V. To estimate the contribution from red leakage, we corrected the magnitudes of SXOM49 and SXOM51 using the red-leak magnitude correction factors for a spectral type G5 V given in Table 12 of \citet{2010ApJ...721.1608B}. The red leak corrected magnitudes were converted to fluxes using the conversion factors given in Table 13 of \citet{2010ApJ...721.1608B} and the fluxes were corrected for extinction. The extinction and red-leak corrected observed fluxes, and the percentage contribution from red-leak is given in Table 7. For these stars, the red-leak contributes approximately 97\% and 65\% of the observed flux in the UVW2 and UVW1 filters, which is extremely high. 
	Therefore, the UV excess in SXOM49 and SXOM51 is most likely due to red leakage.
	\begin{table*}
		\centering
		\caption{Red-leak corrected observed flux of two cool subgiant stars which show UV excess. The units of flux is 10$^{-18}$ ergs cm$^{-2}$ s$^{-1}$. Column 2 gives the detection limit for each UVOT filter in the given observations, Column 3 gives the extinction corrected observed flux, Column 4 gives the red-leak and extinction corrected observed flux, Column 5 gives the best fitting Kurucz model flux, and Column 6 gives the percentage contribution from red leak ($\frac{Observed - Observed_{RC}}{Observed}$) in the Swift UVOT filters.}
		\label{tab:red_leak}	
		\begin{tabular}{cccccccccc}
			\hline 
			Filter	& 	{Background}  & 	  	\multicolumn{2}{c}{Observed} & \multicolumn{2}{c}{Observed$_{RC}$} &  
			\multicolumn{2}{c}{Model} 	  		
			& \multicolumn{2}{c}{\% Red Leak in Observed Flux}\\
			&		& SXOM49 & SXOM51 & SXOM49 & SXOM51 & SXOM49 & SXOM51 & SXOM49 & SXOM51 \\ \hline
			UVW2	& 7  & 134 & 96 & 4 & 3 & 5 &  7 & 97 & 97\\ 
			UVM2	& 8  & - & - & - & - & 2 & 3 & - & - \\ 
			UVW1	& 4 & 152 & 147 & 54 & 52 & 53 & 66 & 65 & 65 \\ 
			\hline 
		\end{tabular}
	\end{table*}

	\subsection{Potential Binaries}
	\subsubsection{RS CVn Binaries}
\begin{figure*}
		\centering
		\includegraphics[width=0.5\linewidth]{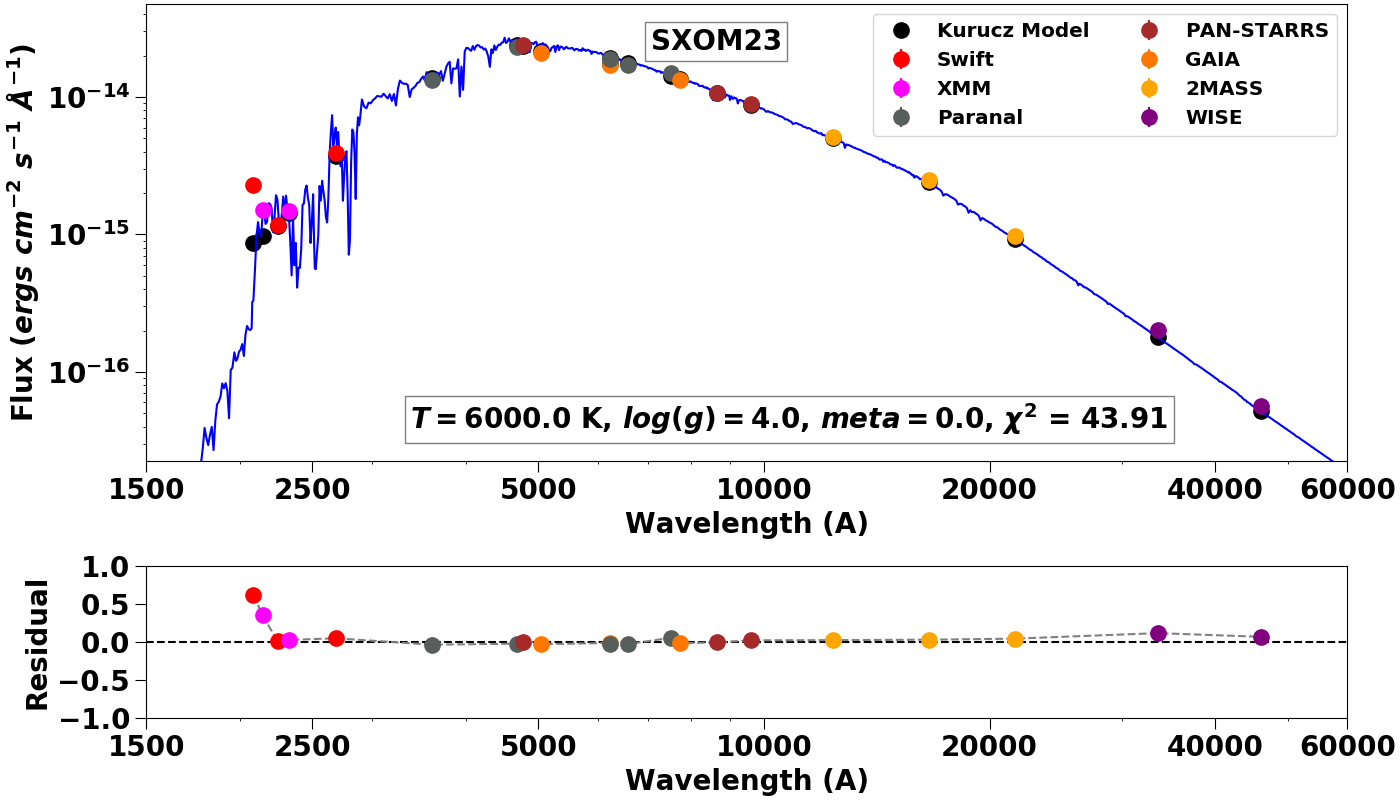}\includegraphics[width=0.5\linewidth]{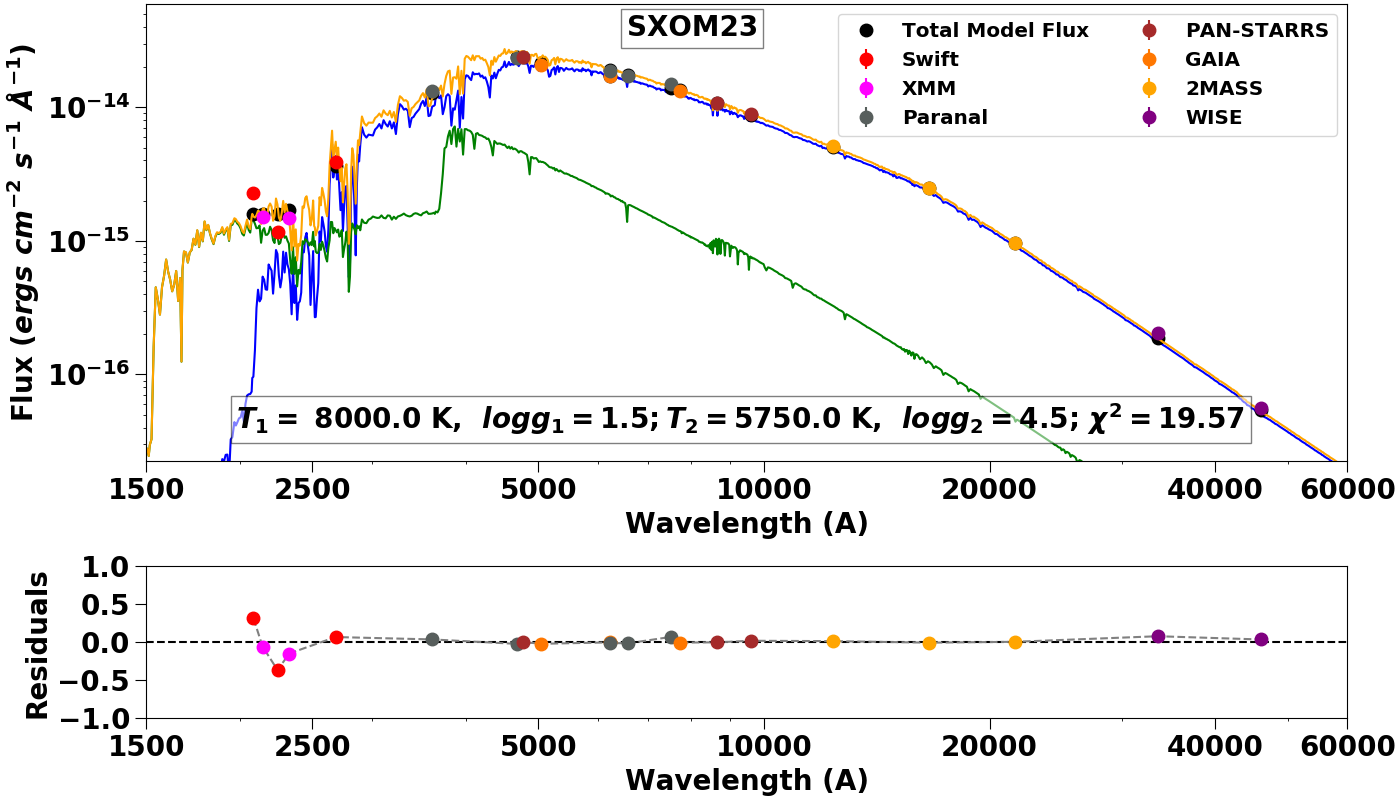}
		\includegraphics[width=0.5\linewidth]{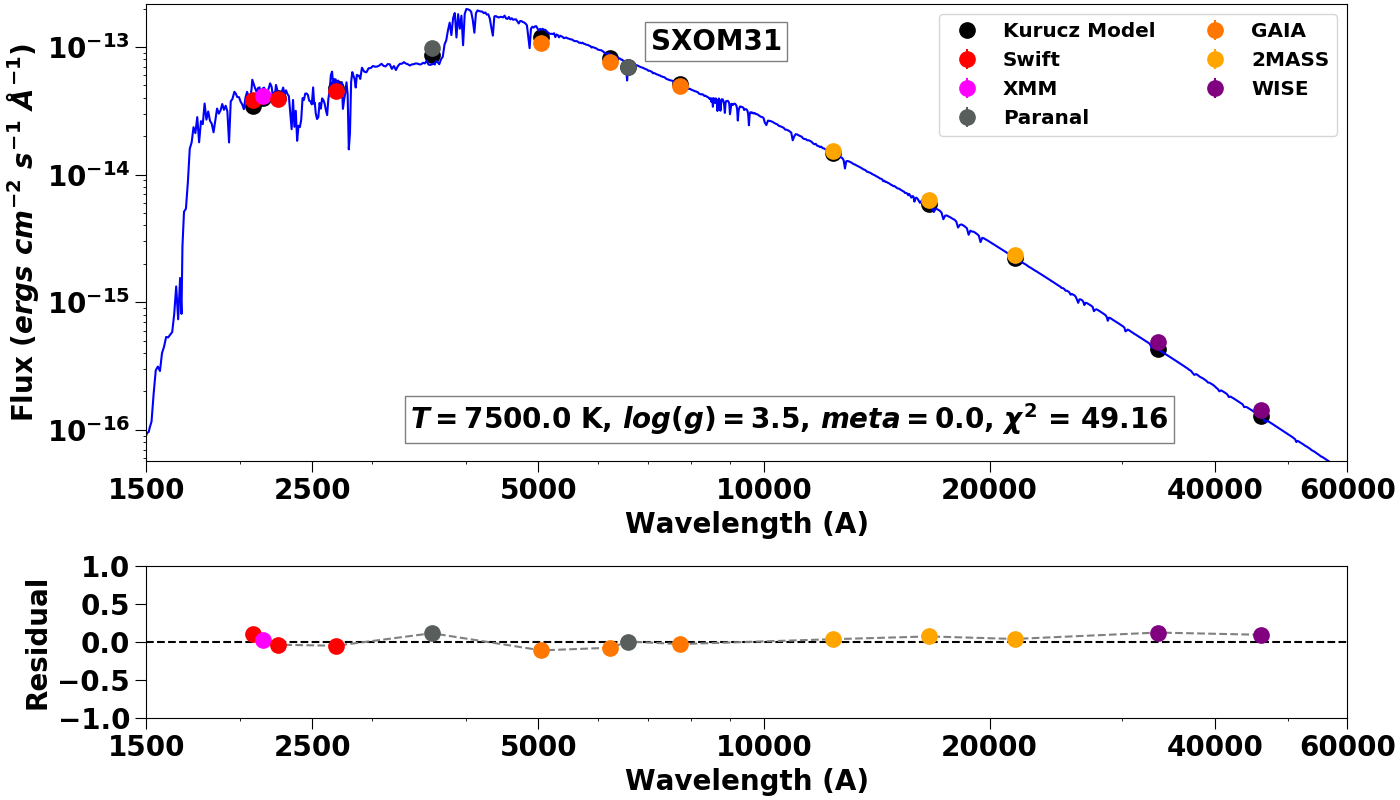}\includegraphics[width=0.5\linewidth]{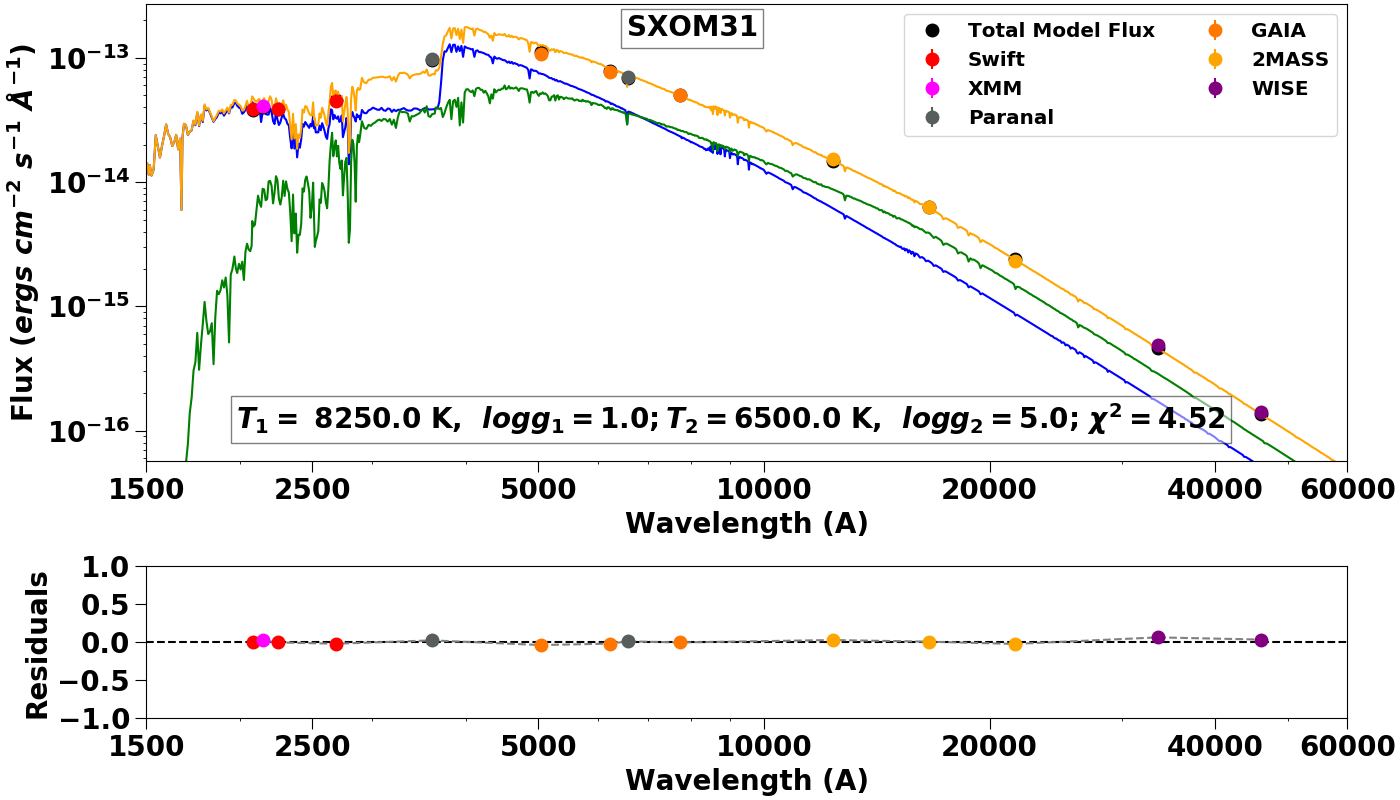}
		\caption{Multi-wavelength SED of 2 potential RS CVn candidates. The single spectrum SED fit is shown in the left panel and the double spectrum fit is shown in the right panel.}
		\label{fig:RSCVn_Binaries}
	\end{figure*}

	A few stars (SXOM23, SXOM31, SXOM33, SXOM34, and SXOM48) that lie close to the binary isochrone in the optical CMD are found to emit X-rays. SXOM23 is the only star which shows excess UV flux in single spectrum SED fit (see Figure \ref{fig:RSCVn_Binaries}). The composite spectrum fit of SXOM23 based on both Kurucz models gives reasonable parameters and also reduces the excess UV residual from 70\% to 30\%. The spectral type of hotter, and cooler star based on the estimated $T_{eff}$ and radius (see Table \ref{tab:SED_Double_Component}) from composite Kurucz model fit is A8 V, and G4 V respectively. SXOM23 has an X-ray luminosity of 1.08 $\times$ 10$^{29}$ $ergs$ $s^{-1}$, typical $L_{X}$ of RS CVn binaries. The $\chi^2_{red}$ of other stars decreases significantly on performing composite spectrum SED fit, suggesting that these stars are also potential binaries. Their X-ray luminosity is in the range 1.3 $-$ 3.0 $\times$ $10^{29}$ $ergs$ $s^{-1}$, typical of  RS CVn type binaries. The X-ray observations and broad-band SEDs of these stars indicate that they are potential RS CVn binaries. \\
	
	\subsubsection{Sub-subgiant and Red Straggler}
	Sub-subgiants (SSGs) and red stragglers (RS) are peculiar stars that defy standard evolutionary models. We identify 5 stars in the optical CMD that are cluster members and lie in the SSG region and 1 star which lies in the RS region as defined by \citet{2017ApJ...840...66G}. Of these stars, only 1 SSG (SXOM47) is detected in UV. SXOM47 lies in the SSG region in the optical CMD. The other 5 stars are too bright and saturated or lie outside the fov of Swift UVOT and XMM OM detectors. \citet{2017ApJ...840...66G} identified 43 SSG in previous studies of open and globular star clusters. These SSG belong to star clusters which are older than 2 Gyr. Of these 43 SSG stars, 25 are X-ray sources with typical 0.5-2.5 keV luminosities of order $10^{30-31}$ $ergs$ $s^{-1}$. SXOM47 lies in the fov of XMM detectors but is not detected, so we can estimate the upper limit of X-ray luminosity. SXOM47 is the first detection of a SSG in an intermediate-age open cluster with an upper limit of X-ray luminosity $\sim$ $10^{29}$ $ergs$ $s^{-1}$. SXOM47 lies in the region of optical CMD where SSGs are mostly formed by mass transfer mechanism as proposed by \citet{2017ApJ...840...67L}. 
	
	\subsubsection{Gap Objects}
	We identify 5 stars in the optical CMD which lie in the gap region similar to that found by \citet{2002ApJ...579..752K} in their FUV-optical CMD and \citet{2019arXiv190713152P} in the optical CMD based on Gaia filters. The stellar types which occupy these regions are Cataclysmic Variables (CV), MS+WD binaries, and He WDs. These stars lie outside the fov of XMM and Swift detectors, so are not detected in X-ray or UV. These stars are potential CV candidates which need further study.
	
	\subsubsection{Active UV binary}
\begin{figure*}
		\centering
		\includegraphics[width=0.5\linewidth]{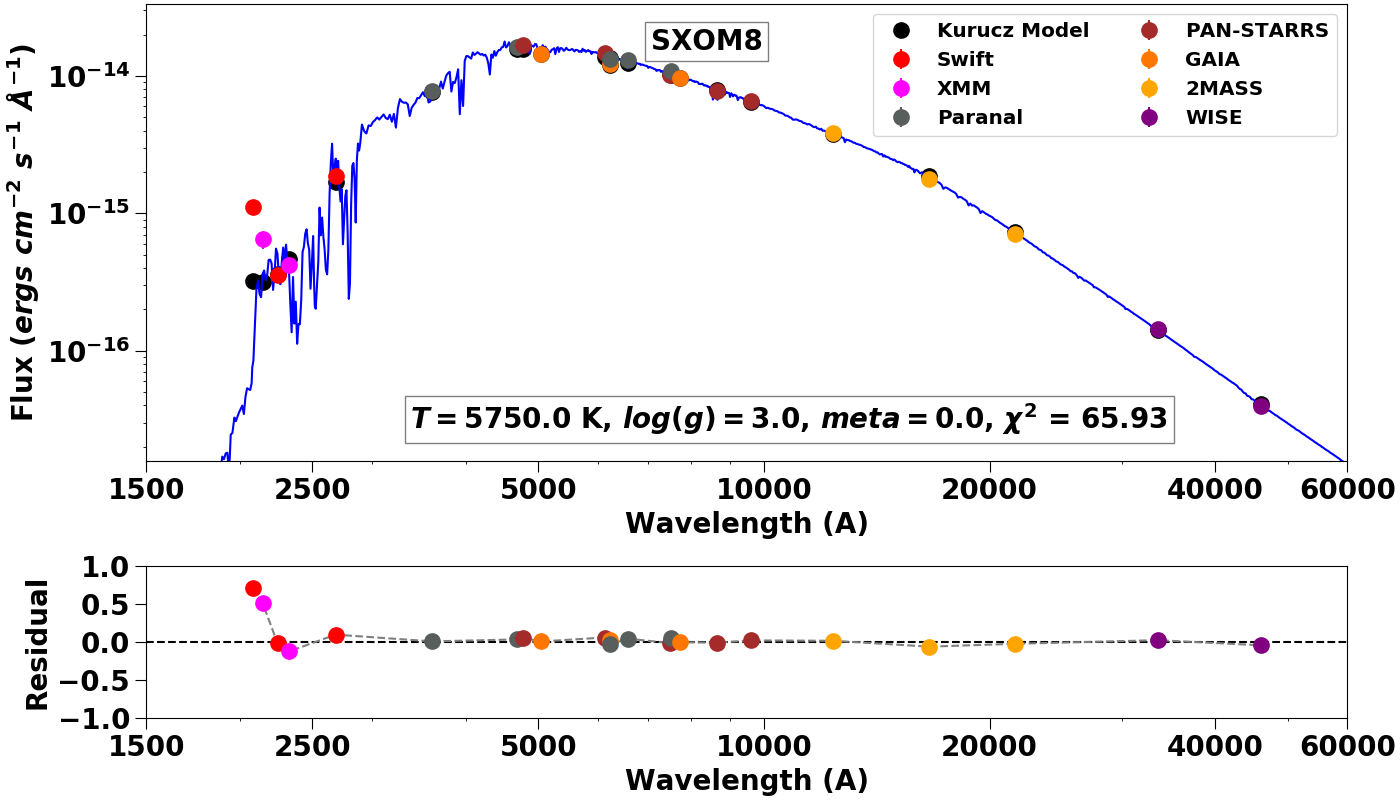}\includegraphics[width=0.5\linewidth]{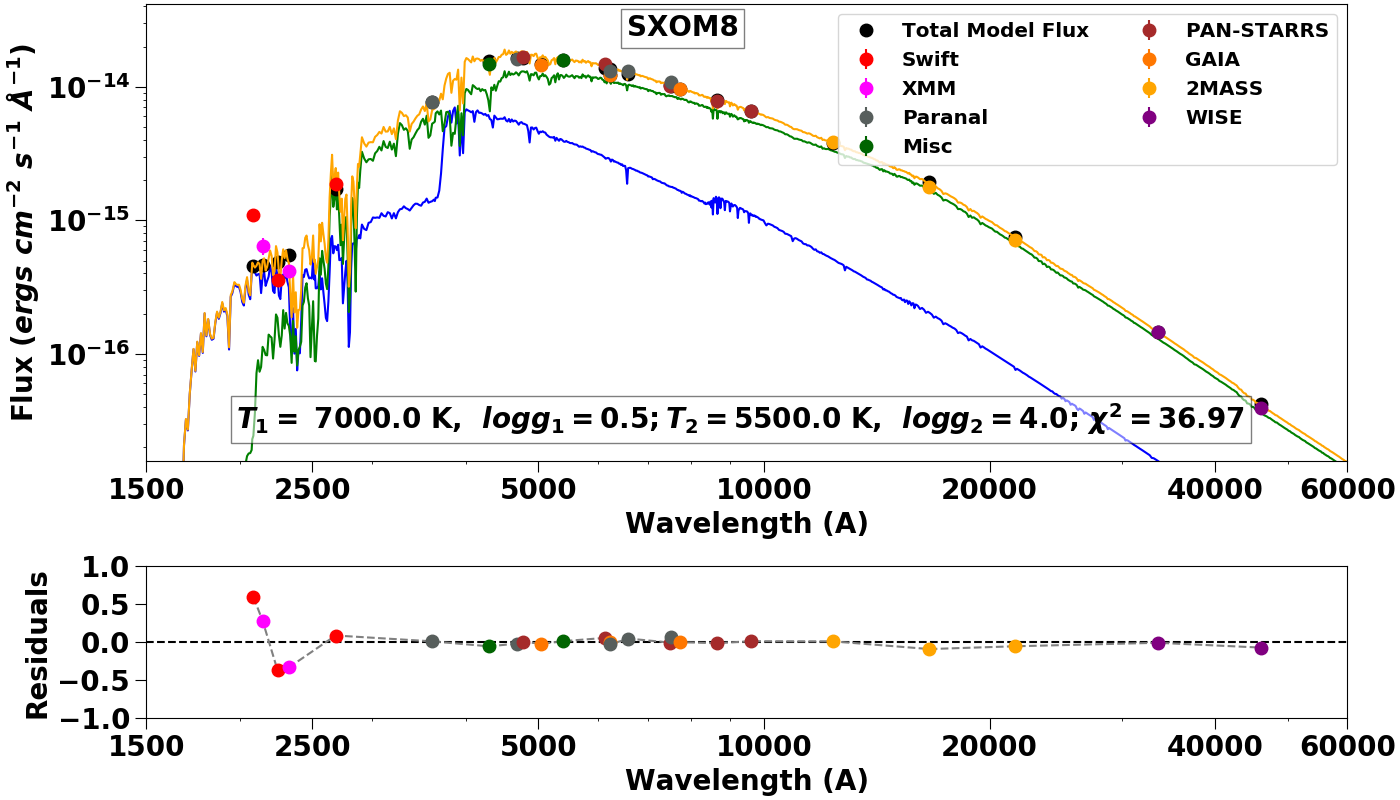}
		\includegraphics[width=0.5\linewidth]{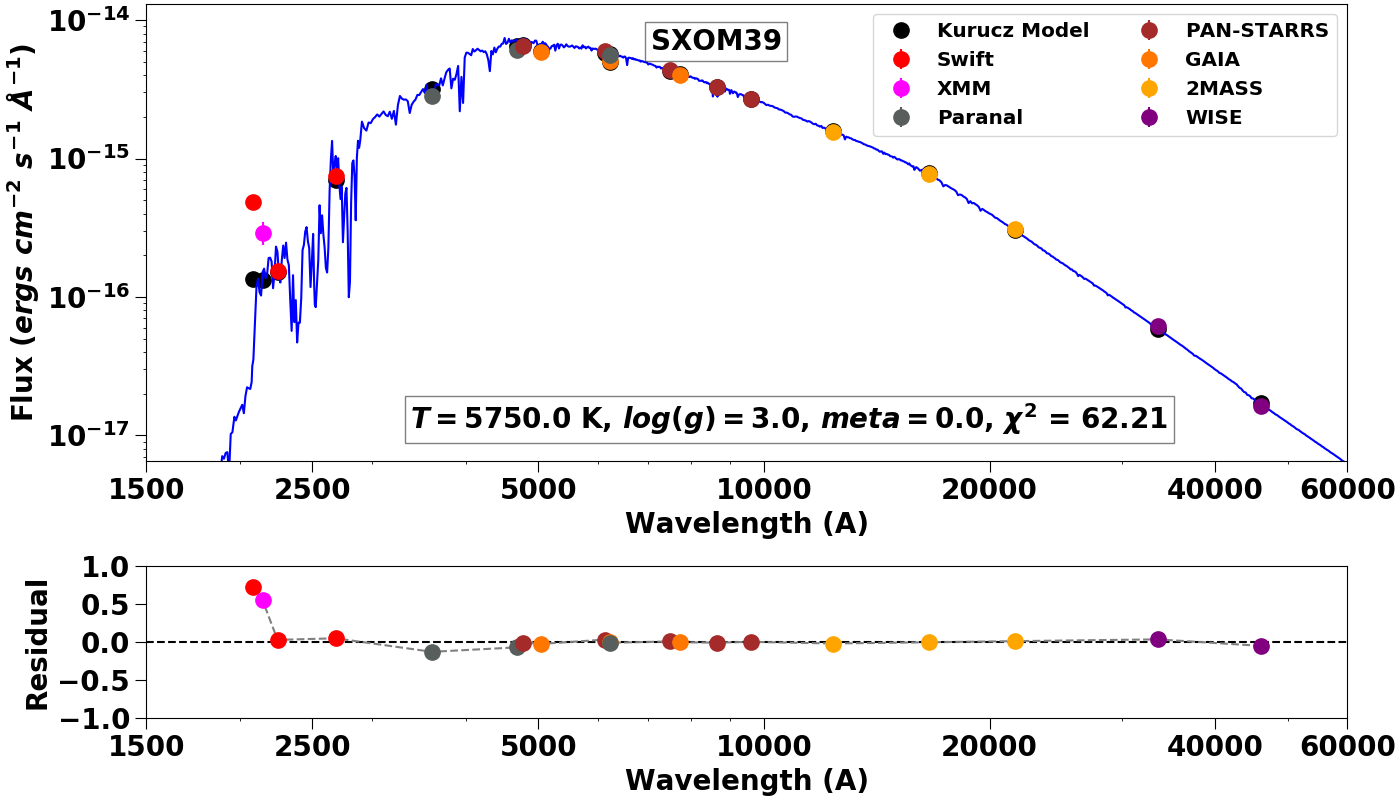}\includegraphics[width=0.5\linewidth]{SXOM42_oneStar_full}
		\caption{Spectral Energy Distribution of a few stars that show excess UV flux mostly due to high chromospheric activity. A comparison of single and double spectrum fitting of an active binary SXOM8 is shown.}
		\label{fig:chromospheric_active}
	\end{figure*}

	SXOM8 lies close to the binary isochrone and shows significant excess in the UV band. Though the composite spectrum fit of SXOM8 reduces the $\chi^2_{red}$ as compared to the single spectrum fit (see Table \ref{tab:SED_Double_Component}), it does not reduce the excess flux significantly (see Figure \ref{fig:chromospheric_active}). The excess UV flux and the position of SXOM8 in the CMD, suggests that it is a potential chromospherically active binary star.

	\begin{figure*}
		\centering
		\includegraphics[width=0.45\linewidth]{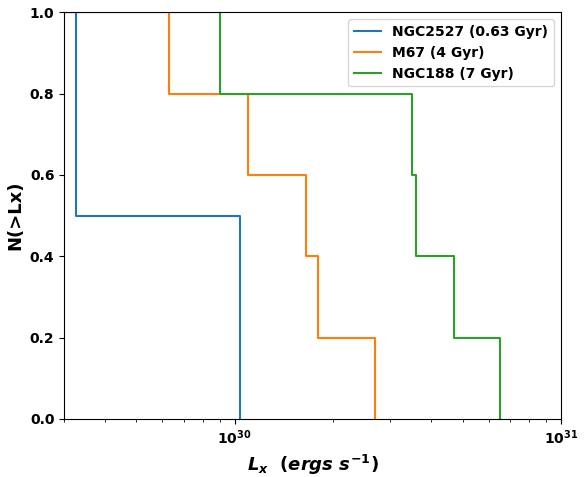}\hfill\includegraphics[width=0.46\linewidth]{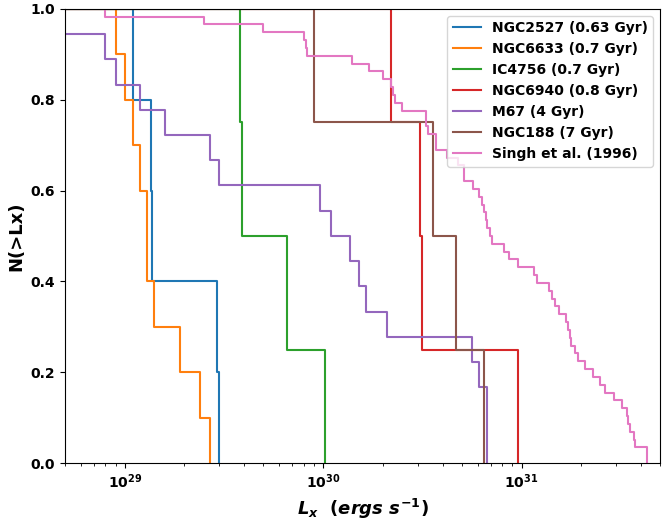}
		\caption{Comparison of the X-ray luminosity function of W UMa and FK Comae stars (left), and RS CVn binaries (right) observed in various open star clusters.}
		\label{fig:lx}
	\end{figure*}
	
	\begin{figure}
		\centering
		\includegraphics[width=\linewidth]{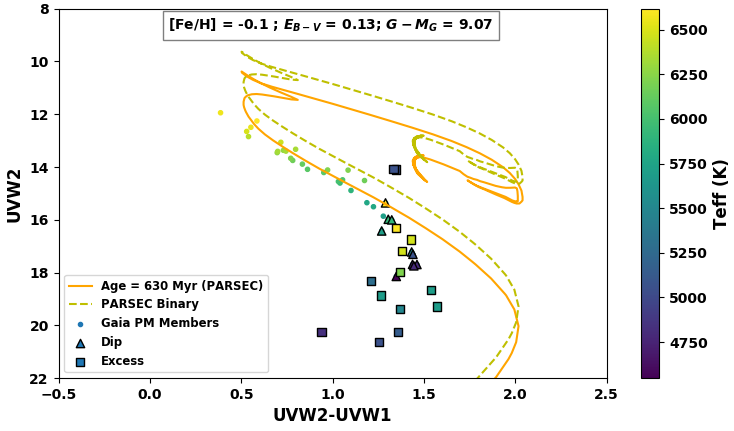}
		\caption{UV-optical CMD of NGC 2527 based on UVW2 magnitude. The best fitting PARSEC isochrone along with an equal mass binary isochrone is overlaid on the observed CMD. The stars which show a prominent dip feature in the SED and excess NUV flux are marked as triangles and squares respectively.}
		\label{fig:tempcmd}
	\end{figure}
	
	\section{Discussion}
	\label{sec:discussion}
	Though NGC 2527 is a nearby open cluster, it has hardly been studied in detail. The  optical CMD of NGC 2527, with only definite members allows us to identify and study various stages of stellar evolution distinctly. Several non-standard stellar evolutionary products such as red stragglers, sub-subgiants, band gap stars, and active binaries are identified. Several open clusters in the $\sim$ a few Gyr age range have been extensively studied in X-ray. This study is an attempt towards characterization of UV and X-ray emission from a relatively younger cluster ($\sim$ 630 Myr) to throw light on the type of population which dominates emission in these wavelengths. We have combined UV observations from Swift and XMM-OM, along with X-ray observations from XMM-EPIC to study the behavior of NGC 2527 members.

	\subsection{X-ray emission}
	We detect strong X-ray emission from 12 stars, out of which 2 are potential FK Comae candidates, 5 are RS CVn candidates, and 5 are MS stars which are coronally active. A comparison of X-ray luminosity function (XLF) of FK Comae and RS CVn stars in intermediate to old open clusters is shown in Figure \ref{fig:lx}. The star clusters used for comparison with NGC 2527 (0.63 Gyr) are : NGC 6633, IC 4756 \citep[0.7 Gyr; ][]{2000MNRAS.319..826B}, M67 \citep[4 Gyr; ][]{2015MNRAS.452.3394M}, and NGC 188 \citep[7 Gyr; ][]{1998A&A...339..431B, 2005A&A...438..291G}. We have shown a comparison with the XLF of field RS CVn binaries \citep{1996AJ....111.2415S}. NGC 6633 and IC 4756 do not have any significant detection of FK Comae type or contact binaries. The XLF suggest that the X-ray luminosity of both RS CVn and contact binaries increases with age i.e., more active binaries are present in older clusters as compared to younger clusters. \citet{2015MNRAS.452.3394M} have also compared the XLF of OCs of various ages, and they also find a similar trend. \\
	
	Main sequence stars as they evolve off the main sequence undergo rapid loss of angular momentum as a result of magnetic braking through coronal winds. These massive RGB stars can emit X-rays only if they are fast rotating stars. \citet{1976ApJ...209..829W} modeled the formation of FK Comae stars by the coalescence of W UMa contact binaries. During the merger, the orbital angular momentum of the binary is converted to rotational momentum of the single star which spins it up. The high X-ray luminosity is most likely due to dynamo action induced by the rapid rotation. Recently \citet{2016ApJ...831...27H} found 18 rapidly rotating FK Comae type of stars, in a sample of bright X-ray sources in the Kepler field. The detection of 2 potential FK Comae candidates suggests that there are many other hidden contact binaries or post-interaction mass transfer products lying along the main sequence in intermediate-age OCs.\\

	\subsection{UV emission}
	We detect UV excess from several stars lying close to the main sequence and binary isochrone, or the red giant branch in the optical CMD. The UV excess in these stars could be due to high chromospheric activity, presence of a hot WD companion or red leakage in UVW2 and UVW1 filters. The absence of a FUV flux data makes it difficult to distinguish between chromospheric activity and the presence of a WD companion.\\
	
	We find that cool subgiant stars (T $\le$ 5000 K) whose flux is close to detection limits could suffer from significant red leakage. These stars are detected only in UVW1 and UVW2 filters that suffer from red leak at optical wavelengths, suggesting that the UV excess in these stars is an artefact. Though RGB stars have similar temperatures (T $\sim$ 5000 K), their intrinsic flux is $\sim$ 10$^3$ larger than the cool subgiant stars. Therefore, the origin of UV excess in the RGB stars is a reliable feature.\\
	
	In a recent study of UV population in M67 by \citet{2018MNRAS.481..226S}, they detected 2 RGB stars which show excess in the GALEX NUV and FUV bands and the presence of Mg II h+k emission lines in the UV spectra suggestive of chromospheric activity. The models explaining the origin of chromospheric activity are not yet well-established \citep{2011MNRAS.414..418P}. The chromospheric activity could be due to acoustic heating of the atmosphere \citep{1995A&ARv...6..181S} or magnetic activity. Since SXOM42 is not observed in X-ray, the origin of the excess UV flux can be attributed to acoustic heating of the atmosphere leading to high chromospheric activity. \\ 
	
	
	
	Several cool stars deviate from the single star isochrones in the UVW1-UVW2 CMD as shown in the Figure \ref{fig:tempcmd}. \citet{2019AJ....158...35S} had also identified this strange feature in the UVW2-UVW1 CMDs of a few star clusters. They had suggested that this feature could be due to red leak in UVW1 and UVW2 filters or deficiencies in the underlying atmosphere models. According to the broad-band SED, most of these stars show a prominent saddle dip feature around 2250 \AA $ $ indicating the presence of strong absorption lines. These UV absorption features are not well characterized by atmosphere models of cool stars. The UVBLUE project by \citet{2005ApJ...626..411R, 2009ASSP....7..239R} and \citet{2007ApJ...657.1046C} also highlights this problem of incomplete characterization of absorption features in cool stars. They had retained the models longward of 2200 \AA $ $ for intermediate-type F, and G stars as the lower wavelengths do not properly fit the observed spectrum. In Figure \ref{fig:tempcmd} the stars have been differentially coloured based on their Teff obtained by comparing the optical magnitudes with the best fitting PARSEC isochrone. We notice that the deviation of these cool stars from isochrones begins around 6500 K, which is similar to that identified by \citet{2019AJ....158...35S}. These deviations from isochrones have also been noted in NUV photometry of cool stars by \citet{2018PASP..130c4204B}. The conspicuous dip feature in most of these stars suggests that the deviations are likely due to the inefficient characterization of absorption lines in atmospheric models of cool stars.
	
	\section{Summary of Results}
	\label{sec:summary of results}
	The main results of our study are as follows:
	\begin{enumerate}
		\item We have re-estimated the fundamental parameters of NGC 2527 based on the recent Gaia DR2 data, and find an age of 630 Myr,  distance of 642$\pm$30 pc, and a mean radial velocity of 41 $\pm$ 9 km/s. We estimate a reddening of 0.13 mag in the direction of this OC.\\
		
		
		
		
		\item We have discovered 5 MS stars that show high coronal activity and are potentially fast-rotating stars of F spectral type. \\
		
		\item We have discovered 2 potential FK Comae stars. One of these RGB stars shows strong X-ray and UV emission. \\
		
		\item We have characterized the fundamental parameters of 5 active binaries that emit X-rays and are potential candidates for being RS CVn type binaries. \\
		
		\item We have identified a sub-subgiant star with an upper limit of X-ray luminosity as $\sim$ $10^{29}$ $ergs$ $s^{-1}$ $-$ first such detection in an intermediate age OC. \\
		
		
		
		
		\item We confirm that the X-ray luminosity of both RS CVn and contact binaries increases with age suggesting that more active binaries may be present in the old and intermediate age open clusters.\\
		
		\item This study also suggests possible presence of W UMa and FK Comae type stars in younger (age $\simeq$ 630 Myr) clusters.
	\end{enumerate}
	
	\section*{Acknowledgements}
	We thank the referee for insightful and incisive suggestions that helped in improving the quality of the manuscript. We thank Michael Siegel for providing the Swift UVOT photometry catalogue of NGC 2527. Nevil thanks Vishal Joshi for useful discussion and guidance in creating the SED fitting code, COSPAR for giving him the opportunity to learn X-ray data analysis, Lynne Valencic for guiding him to process and analyze XMM-Newton data, and Sindhu and Vikrant for fruitful discussion on the results of paper. This research has made use of data and/or software provided by the High Energy Astrophysics Science Archive Research Center (HEASARC), which is a service of the Astrophysics Science Division at NASA/GSFC and the High Energy Astrophysics Division of the Smithsonian Astrophysical Observatory. This work is based on observations obtained with XMM-Newton, an ESA science mission with instruments and contributions directly funded by ESA Member States and NASA. This publication makes use of VOSA, developed under the Spanish Virtual Observatory project supported by the Spanish MINECO through grant AyA2017-84089. This research has made use of the SVO Filter Profile Service supported from the Spanish MINECO through grant AYA2017-84089. This research also made use of Topcat, Matplotlib, and Astropy, a community-developed core Python package for Astronomy.
	
	
	


	\appendix
	
	\section{Cross-correlation}
	
	To determine the optimal cross-correlation radius in between the soft band X-ray catalogue and the Gaia cluster members catalogue, we followed a technique used by \citet{1997MNRAS.287..350J}. We generated the cumulative distribution of X-ray sources, $\phi (r)$, for a correlation radius within range 0.1\arcsec $ $ to 30\arcsec $ $ with a step of 0.1\arcsec. We fitted the cumulative distribution function assuming it is formed from the sum of cumulative distribution of the true and spurious correlations as given in Eqn. \ref{eq:corr}.
	
	\begin{equation}
	\label{eq:corr}
	\phi (r) = A[1 - exp(\frac{-r^2}{2\sigma^2})] + (N-A)[1-exp(-\pi r^2 B)]
	\end{equation}
	
	where N is the number of soft band X-ray sources, A is the number of true correlations with the Gaia cluster members catalogue, and B is the number density of objects in the Gaia cluster members catalogue.\\
		
\begin{figure}
		\centering
		\includegraphics[width=\linewidth]{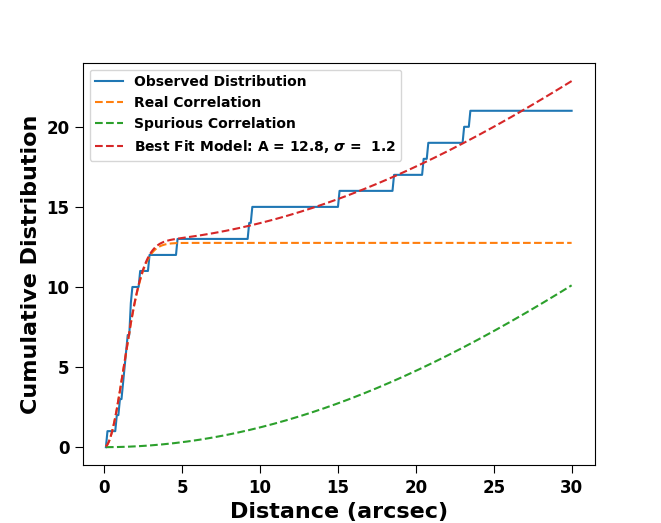}
		\caption{Cumulative distribution function of the distances in between the soft band X-ray catalogue and the Gaia cluster members catalogue. }
		\label{fig:corrsgaia}
	\end{figure}

	For the soft band X-ray and Gaia cluster members catalogue, we find that the cross-corelation function gives an extremely good fit with parameters A = 12.8, $\sigma$ = 1.2\arcsec, and B = 7.8 $\times$ 10$^{-5}$ arcsec$^{-2}$ as shown in Fig. \ref{fig:corrsgaia}. Using this fit, we choose a maximum cross-correlation radius of 3\arcsec, within which we should be able to detect approximately 12 true correlations and 0 spurious correlations. 
	
	\section{UVOT Data - backgound flux}
	We reprocessed the Swift UVOT observations of NGC 2527 to estimate the background flux limits. The UVOT images and exposure maps of NGC 2527 in each filter were added using HEASOFT image analysis package ximage. We processed the images using 'uvotdetect' task, which detects all the sources in the given image, and provides the background flux or threshold flux. The background flux limits attained in the Swift UVOT observations of NGC 2527 are given in Table \ref{tab:red_leak}.\\
	
	\section{Effect of red leak on the dip feature}
\begin{figure}
		\centering
		\includegraphics[width=\linewidth]{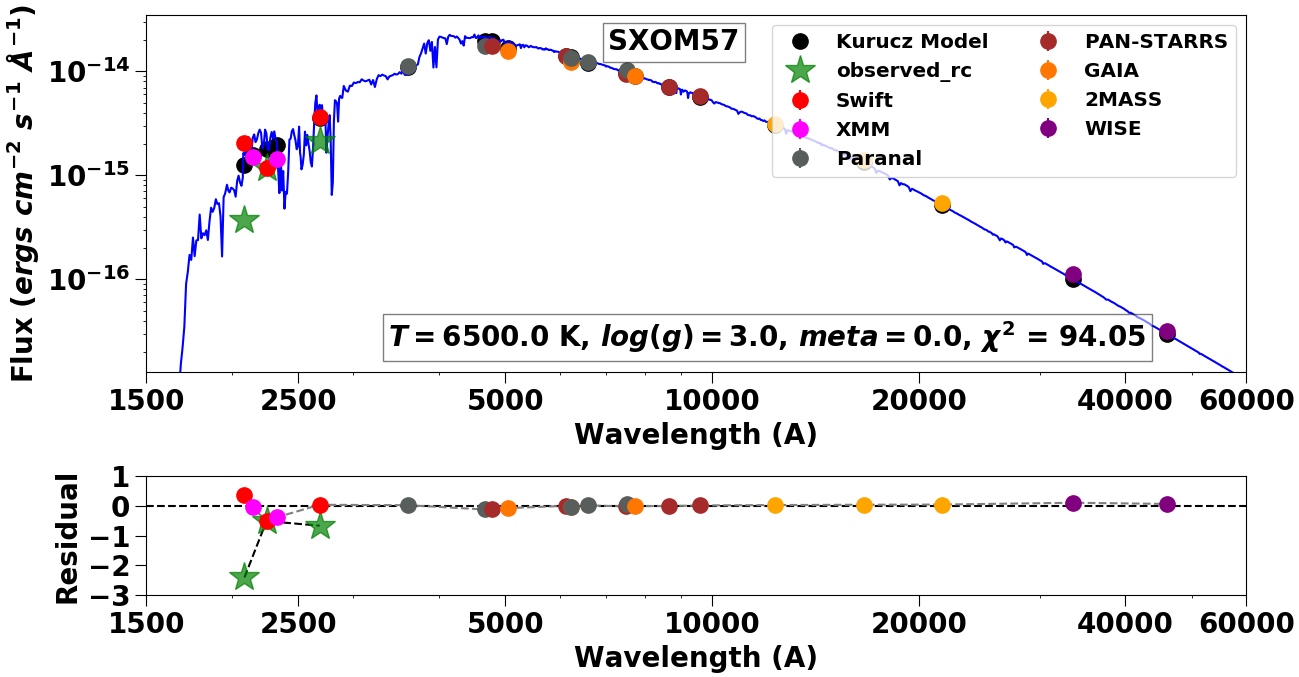}
		\caption{Red-leak corrected observed SED of a star that shows prominent dip feature. The green stars represent the red-leak corrected observed flux and residual.}
		\label{fig:SED_RC}
	\end{figure}

	To check whether the dip feature could be due to red leak, we corrected the UVOT observed magnitudes of SXOM57 (shows a prominent dip feature) using the red-leak magnitude correction factors for a spectral type F5 V given in Table 12 of \citet{2010ApJ...721.1608B}. These red leak corrected magnitudes were converted to fluxes using the conversion factors given in Table 13 of \citet{2010ApJ...721.1608B} and the fluxes were corrected for extinction. We re-estimated the residuals for UVW2 and UVW1 filters using the red-leak and extinction corrected observed fluxes as shown in Fig. \ref{fig:SED_RC}. After correcting the fluxes for red-leak, we observe a more prominent dip feature in the NUV band. Therefore, we conclude that the saddle dip feature is most likely due to inefficient characterization of absorption lines in atmospheric models of cool stars.
	
	\bsp	
	\label{lastpage}
\end{document}


\begin{table*}
		\centering
		\renewcommand\thetable{A} 
		\caption{Magnitudes and fluxes of cluster members that are detected in XMM OM and Swift UVOT observations. The units on the XMM OM and Swift UVOT fluxes are $10^{-14}$ ergs cm$^{-2}$ s$^{-1}$ \AA$^{-1}$.}
		\label{tab:Gaia_Swift}
		\resizebox{\textwidth}{!}{
			\begin{tabular}{lcccccccccccr}
				\hline
				SXOM  & \multicolumn{3}{c}{UVOT Magnitude} & \multicolumn{3}{c}{UVOT Flux} & \multicolumn{2}{c}{OM Flux}  & \multicolumn{2}{c}{OM Quality Flag} \\
				&UVW2	&UVM2	&UVW1 &UVW2	&UVM2	&UVW1	&UVW2	&UVM2	&UVW2	&UVM2 \\ 
				&(mag)	&(mag)	&(mag)  \\
				\hline1	&13.33$\pm$0.02	&13.1$\pm$0.01	&12.53$\pm$0.01	&7.92$\pm$0.15	&8.08$\pm$0.07	&7.93$\pm$0.09	&-	&-	&-	&-	\\
				2	&15.35$\pm$0.01	&15.11$\pm$0.01	&14.16$\pm$0.01	&1.23$\pm$0.01	&1.26$\pm$0.01	&1.76$\pm$0.01	&-	&-	&-	&-	\\
				3	&14.61$\pm$0.01	&14.34$\pm$0.01	&13.57$\pm$0.01	&2.44$\pm$0.02	&2.56$\pm$0.03	&3.05$\pm$0.04	&-	&-	&-	&-	\\
				4	&17.69$\pm$0.02	&18.27$\pm$0.03	&16.23$\pm$0.01	&0.14$\pm$0.0	&0.07$\pm$0.0	&0.26$\pm$0.0	&0.1$\pm$0.01	&-	&130	&-	\\
				5	&13.37$\pm$0.01	&13.12$\pm$0.02	&12.64$\pm$0.01	&7.64$\pm$0.09	&7.92$\pm$0.11	&7.19$\pm$0.08	&-	&-	&-	&-	\\
				6	&11.94$\pm$0.03	&11.85$\pm$0.02	&11.56$\pm$0.03	&28.33$\pm$0.78	&25.54$\pm$0.44	&19.42$\pm$0.5	&-	&-	&-	&-	\\
				7	&15.86$\pm$0.01	&15.73$\pm$0.01	&14.58$\pm$0.01	&0.77$\pm$0.01	&0.72$\pm$0.01	&1.2$\pm$0.01	&0.74$\pm$0.02	&0.8$\pm$0.01	&0	&128	&\\
				8	&17.97$\pm$0.03	&18.99$\pm$0.05	&16.6$\pm$0.03	&0.11$\pm$0.0	&0.04$\pm$0.0	&0.19$\pm$0.0	&0.06$\pm$0.01	&0.04$\pm$0.0	&0	&0	&\\
				9	&12.85$\pm$0.01	&12.71$\pm$0.01	&12.31$\pm$0.02	&12.33$\pm$0.14	&11.48$\pm$0.12	&9.73$\pm$0.16	&13.06$\pm$0.08	&11.13$\pm$0.05	&0	&256	&\\
				10	&16.31$\pm$0.05	&16.36$\pm$0.03	&14.96$\pm$0.02	&0.51$\pm$0.02	&0.4$\pm$0.01	&0.85$\pm$0.02	&-	&-	&-	&-	\\
				11	&17.69$\pm$0.02	&18.05$\pm$0.03	&16.25$\pm$0.01	&0.14$\pm$0.0	&0.08$\pm$0.0	&0.26$\pm$0.0	&0.11$\pm$0.01	&0.11$\pm$0.01	&130	&128	&\\
				12	&20.25$\pm$0.08	&-	&18.9$\pm$0.05	&0.01$\pm$0.0	&-	&0.02$\pm$0.0	&-	&-	&-	&-	\\
				13	&-	&-	&-	&-	&-	&-	&0.03$\pm$0.01	&0.05$\pm$0.0	&128	&128	&\\
				14	&13.4$\pm$0.01	&13.21$\pm$0.01	&12.65$\pm$0.01	&7.42$\pm$0.08	&7.3$\pm$0.06	&7.08$\pm$0.08	&8.01$\pm$0.06	&7.19$\pm$0.04	&0	&0	&\\
				15	&18.65$\pm$0.04	&19.51$\pm$0.09	&17.11$\pm$0.04	&0.06$\pm$0.0	&0.02$\pm$0.0	&0.12$\pm$0.0	&0.03$\pm$0.01	&0.03$\pm$0.0	&0	&0	&\\
				16	&17.22$\pm$0.03	&17.68$\pm$0.06	&15.78$\pm$0.03	&0.22$\pm$0.01	&0.12$\pm$0.01	&0.4$\pm$0.01	&0.16$\pm$0.02	&0.14$\pm$0.01	&0	&0	&\\
				17	&20.74$\pm$0.15	&21.78$\pm$0.49	&-	&0.01$\pm$0.0	&0.0$\pm$0.0	&-	&-	&-	&-	&-	\\
				18	&-	&-	&-	&-	&-	&-	&26.9$\pm$0.12	&21.79$\pm$0.08	&256	&274	&\\
				19	&-	&-	&-	&-	&-	&-	&56.42$\pm$0.14	&45.65$\pm$0.09	&274	&274	&\\
				20	&16.42$\pm$0.03	&16.76$\pm$0.04	&15.15$\pm$0.03	&0.46$\pm$0.01	&0.28$\pm$0.01	&0.71$\pm$0.02	&0.32$\pm$0.02	&0.33$\pm$0.01	&0	&0	&\\
				21	&14.89$\pm$0.01	&14.68$\pm$0.01	&13.79$\pm$0.01	&1.88$\pm$0.01	&1.88$\pm$0.02	&2.49$\pm$0.02	&2.05$\pm$0.03	&2.02$\pm$0.02	&0	&0	&\\
				22	&15.98$\pm$0.01	&15.9$\pm$0.01	&14.68$\pm$0.01	&0.69$\pm$0.01	&0.61$\pm$0.01	&1.1$\pm$0.01	&0.63$\pm$0.02	&-	&0	&-	\\
				23	&17.18$\pm$0.03	&17.69$\pm$0.05	&15.8$\pm$0.06	&0.23$\pm$0.01	&0.12$\pm$0.01	&0.39$\pm$0.02	&0.15$\pm$0.01	&0.15$\pm$0.01	&0	&0	&\\
				24	&15.5$\pm$0.02	&15.35$\pm$0.02	&14.28$\pm$0.02	&1.07$\pm$0.02	&1.02$\pm$0.02	&1.58$\pm$0.02	&-	&-	&-	&-	\\
				25	&18.14$\pm$0.03	&18.94$\pm$0.06	&16.8$\pm$0.02	&0.09$\pm$0.0	&0.04$\pm$0.0	&0.16$\pm$0.0	&-	&-	&-	&-	\\
				26	&14.21$\pm$0.01	&13.99$\pm$0.02	&13.26$\pm$0.02	&3.52$\pm$0.02	&3.54$\pm$0.05	&4.06$\pm$0.06	&3.98$\pm$0.03	&3.77$\pm$0.03	&0	&0	&\\
				27	&12.5$\pm$0.01	&12.34$\pm$0.01	&12.06$\pm$0.03	&17.07$\pm$0.18	&16.21$\pm$0.19	&12.22$\pm$0.32	&-	&-	&-	&-	\\
				28	&-	&-	&-	&-	&-	&-	&117.76$\pm$0.31	&94.47$\pm$0.24	&274	&274	&\\
				29	&14.09$\pm$0.01	&13.86$\pm$0.01	&13.22$\pm$0.01	&3.94$\pm$0.03	&4.0$\pm$0.03	&4.18$\pm$0.06	&4.21$\pm$0.05	&-	&0	&-	\\
				30	&-	&-	&-	&-	&-	&-	&0.04$\pm$0.01	&0.03$\pm$0.0	&0	&0	&\\
				31	&14.11$\pm$0.01	&13.89$\pm$0.01	&13.14$\pm$0.02	&3.85$\pm$0.04	&3.88$\pm$0.03	&4.52$\pm$0.07	&4.1$\pm$0.05	&-	&0	&-	\\
				32	&-	&-	&-	&-	&-	&-	&39.59$\pm$0.15	&32.55$\pm$0.09	&274	&274	&\\
				33	&14.51$\pm$0.01	&14.35$\pm$0.01	&13.34$\pm$0.02	&2.66$\pm$0.02	&2.54$\pm$0.03	&3.77$\pm$0.05	&2.84$\pm$0.04	&-	&0	&-	\\
				34	&15.36$\pm$0.01	&15.2$\pm$0.01	&14.07$\pm$0.01	&1.22$\pm$0.01	&1.16$\pm$0.01	&1.92$\pm$0.02	&1.22$\pm$0.03	&1.29$\pm$0.02	&2	&2	&\\
				35	&14.55$\pm$0.01	&14.3$\pm$0.01	&13.52$\pm$0.02	&2.57$\pm$0.02	&2.67$\pm$0.03	&3.2$\pm$0.05	&2.85$\pm$0.04	&-	&2	&-	\\
				36	&12.65$\pm$0.02	&12.49$\pm$0.01	&12.12$\pm$0.02	&14.76$\pm$0.23	&14.12$\pm$0.14	&11.54$\pm$0.26	&15.6$\pm$0.06	&13.57$\pm$0.05	&0	&256	&\\
				37	&13.75$\pm$0.01	&13.54$\pm$0.01	&12.97$\pm$0.01	&5.37$\pm$0.05	&5.37$\pm$0.04	&5.3$\pm$0.07	&5.85$\pm$0.02	&5.37$\pm$0.01	&32	&288	&\\
				38	&-	&-	&-	&-	&-	&-	&57.43$\pm$0.2	&-	&274	&-	\\
				39	&18.87$\pm$0.03	&19.89$\pm$0.09	&17.6$\pm$0.02	&0.05$\pm$0.0	&0.02$\pm$0.0	&0.07$\pm$0.0	&0.03$\pm$0.01	&-	&2	&-	\\
				40	&13.67$\pm$0.01	&13.46$\pm$0.01	&12.9$\pm$0.02	&5.78$\pm$0.06	&5.8$\pm$0.05	&5.65$\pm$0.09	&6.49$\pm$0.06	&-	&0	&-	\\
				41	&14.49$\pm$0.01	&14.26$\pm$0.02	&13.43$\pm$0.02	&2.73$\pm$0.01	&2.75$\pm$0.05	&3.46$\pm$0.06	&3.11$\pm$0.03	&2.95$\pm$0.02	&0	&0	&\\
				42	&14.1$\pm$0.01	&15.74$\pm$0.01	&12.75$\pm$0.02	&3.89$\pm$0.05	&0.71$\pm$0.01	&6.47$\pm$0.14	&-	&-	&-	&-	\\
				43	&13.89$\pm$0.01	&13.64$\pm$0.01	&13.06$\pm$0.01	&4.72$\pm$0.03	&4.89$\pm$0.05	&4.89$\pm$0.05	&5.28$\pm$0.05	&4.95$\pm$0.03	&0	&0	&\\
				44	&12.5$\pm$0.02	&12.39$\pm$0.02	&11.94$\pm$0.04	&17.06$\pm$0.31	&15.42$\pm$0.34	&13.62$\pm$0.47	&-	&-	&-	&-	\\
				45	&14.49$\pm$0.02	&14.33$\pm$0.05	&-	&2.72$\pm$0.06	&2.6$\pm$0.11	&-	&-	&-	&-	&-	\\
				46	&-	&-	&-	&-	&-	&-	&0.03$\pm$0.01	&0.02$\pm$0.0	&0	&0	&\\
				47	&12.26$\pm$0.02	&12.11$\pm$0.01	&11.67$\pm$0.03	&21.26$\pm$0.48	&20.01$\pm$0.27	&17.5$\pm$0.51	&22.01$\pm$0.11	&18.94$\pm$0.06	&2	&274	&\\
				48	&14.12$\pm$0.01	&13.85$\pm$0.01	&13.03$\pm$0.02	&3.83$\pm$0.04	&4.03$\pm$0.04	&4.99$\pm$0.1	&4.06$\pm$0.05	&4.12$\pm$0.03	&128	&128	&\\
				49	&20.26$\pm$0.07	&-	&19.32$\pm$0.07	&0.01$\pm$0.0	&-	&0.02$\pm$0.0	&-	&-	&-	&-	\\
				50	&13.06$\pm$0.01	&12.88$\pm$0.02	&12.35$\pm$0.02	&10.11$\pm$0.08	&9.82$\pm$0.17	&9.39$\pm$0.17	&-	&-	&-	&-	\\
				51	&20.62$\pm$0.1	&-	&19.36$\pm$0.05	&0.01$\pm$0.0	&-	&0.01$\pm$0.0	&-	&-	&-	&-	\\
				52	&19.29$\pm$0.08	&20.0$\pm$0.14	&17.72$\pm$0.06	&0.03$\pm$0.0	&0.01$\pm$0.0	&0.07$\pm$0.0	&-	&0.02$\pm$0.0	&-	&0	&\\
				53	&16.0$\pm$0.01	&15.88$\pm$0.01	&14.68$\pm$0.01	&0.67$\pm$0.01	&0.62$\pm$0.01	&1.1$\pm$0.01	&0.67$\pm$0.02	&-	&0	&-	\\
				54	&13.46$\pm$0.01	&13.27$\pm$0.01	&12.76$\pm$0.01	&7.05$\pm$0.06	&6.87$\pm$0.07	&6.43$\pm$0.09	&7.54$\pm$0.06	&6.6$\pm$0.04	&0	&0	&\\
				55	&18.75$\pm$0.06	&19.36$\pm$0.1	&-	&0.05$\pm$0.0	&0.03$\pm$0.0	&-	&0.04$\pm$0.01	&0.03$\pm$0.0	&2	&2	&\\
				56	&18.31$\pm$0.04	&18.5$\pm$0.05	&17.1$\pm$0.04	&0.08$\pm$0.0	&0.06$\pm$0.0	&0.12$\pm$0.0	&0.05$\pm$0.01	&0.06$\pm$0.0	&128	&0	&\\
				57	&17.31$\pm$0.01	&17.69$\pm$0.03	&15.87$\pm$0.01	&0.2$\pm$0.0	&0.12$\pm$0.0	&0.37$\pm$0.0	&0.15$\pm$0.01	&0.14$\pm$0.01	&0	&0	&\\
				58	&19.38$\pm$0.05	&20.59$\pm$0.15	&18.01$\pm$0.03	&0.03$\pm$0.0	&0.01$\pm$0.0	&0.05$\pm$0.0	&-	&-	&-	&-	\\
				59	&16.74$\pm$0.01	&16.85$\pm$0.01	&15.31$\pm$0.01	&0.34$\pm$0.0	&0.26$\pm$0.0	&0.61$\pm$0.01	&0.26$\pm$0.02	&-	&0	&-	\\
				60	&13.41$\pm$0.01	&13.28$\pm$0.01	&12.71$\pm$0.02	&7.36$\pm$0.08	&6.8$\pm$0.05	&6.73$\pm$0.13	&7.7$\pm$0.06	&6.58$\pm$0.04	&34	&34	&\\
				61	&14.08$\pm$0.02	&15.5$\pm$0.01	&12.74$\pm$0.02	&3.97$\pm$0.09	&0.89$\pm$0.01	&6.51$\pm$0.13	&1.8$\pm$0.01	&1.2$\pm$0.01	&33	&33	&\\
				62	&17.74$\pm$0.02	&18.21$\pm$0.03	&16.3$\pm$0.01	&0.14$\pm$0.0	&0.07$\pm$0.0	&0.25$\pm$0.0	&0.07$\pm$0.01	&-	&0	&-	\\
				\hline 
		\end{tabular}}
	\end{table*}

	\begin{table*}
		\centering
		\renewcommand\thetable{B} 
		\caption{Extinction corrected fluxes obtained from VO photometry. The units on APASS, PAN-STARRS, and Gaia fluxes are 10$^{-14}$ ergs s$^{-1}$ cm$^{-2}$ \AA$^{-1}$.}
		\label{tab:Flux_1}
		\resizebox{\textwidth}{!}{
			\begin{tabular}{lcccccccccccr}
				\hline
				SXOM    &\multicolumn{2}{c}{APASS} & \multicolumn{5}{c}{PAN-STARRS} & \multicolumn{3}{c}{Gaia} \\ 
				&B	&V	&g	&r	&z	&y	&i	&Gbp	&G	&Grp	& \\ \hline
				1	&21.59$\pm$0.54	&15.85$\pm$0.16	&-	&-	&-	&-	&-	&15.67$\pm$0.16	&10.78$\pm$0.11	&6.73$\pm$0.07	&\\
				2	&6.23$\pm$0.2	&5.02$\pm$0.05	&-	&-	&-	&1.32$\pm$0.0	&-	&4.57$\pm$0.05	&3.28$\pm$0.03	&2.21$\pm$0.02	&\\
				3	&9.24$\pm$0.05	&7.11$\pm$0.11	&-	&-	&-	&1.85$\pm$0.0	&-	&6.97$\pm$0.07	&4.86$\pm$0.05	&3.15$\pm$0.03	&\\
				4	&1.4$\pm$0.07	&1.36$\pm$0.05	&1.44$\pm$0.01	&1.16$\pm$0.0	&0.59$\pm$0.0	&0.48$\pm$0.0	&0.79$\pm$0.01	&1.28$\pm$0.01	&1.02$\pm$0.01	&0.76$\pm$0.01	&\\
				5	&19.3$\pm$0.43	&13.23$\pm$0.15	&-	&-	&-	&-	&-	&13.43$\pm$0.13	&8.89$\pm$0.09	&5.33$\pm$0.05	&\\
				6	&43.16$\pm$0.6	&27.16$\pm$0.6	&-	&-	&-	&-	&-	&27.97$\pm$0.28	&17.77$\pm$0.18	&9.89$\pm$0.1	&\\
				7	&4.2$\pm$0.04	&3.5$\pm$0.03	&-	&-	&-	&1.09$\pm$0.0	&-	&3.55$\pm$0.04	&2.59$\pm$0.03	&1.78$\pm$0.02	&\\
				8	&1.48$\pm$0.05	&1.6$\pm$0.02	&1.67$\pm$0.01	&1.47$\pm$0.03	&0.78$\pm$0.01	&0.66$\pm$0.0	&1.02$\pm$0.0	&1.46$\pm$0.01	&1.23$\pm$0.01	&0.97$\pm$0.01	&\\
				9	&24.11$\pm$0.82	&16.22$\pm$0.27	&-	&-	&-	&-	&-	&16.78$\pm$0.17	&10.96$\pm$0.11	&6.38$\pm$0.06	&\\
				10	&3.49$\pm$0.09	&3.02$\pm$0.08	&-	&-	&-	&0.95$\pm$0.0	&-	&2.83$\pm$0.03	&2.13$\pm$0.02	&1.52$\pm$0.02	&\\
				11	&1.35$\pm$0.06	&1.3$\pm$0.03	&1.4$\pm$0.0	&1.12$\pm$0.0	&0.58$\pm$0.0	&0.47$\pm$0.0	&0.82$\pm$0.0	&1.24$\pm$0.01	&0.99$\pm$0.01	&0.74$\pm$0.01	&\\
				12	&-	&-	&0.29$\pm$0.0	&0.32$\pm$0.0	&0.2$\pm$0.0	&0.17$\pm$0.0	&0.25$\pm$0.0	&0.28$\pm$0.0	&0.27$\pm$0.0	&0.24$\pm$0.0	&\\
				13	&0.97$\pm$0.05	&1.02$\pm$0.03	&1.07$\pm$0.0	&0.88$\pm$0.0	&0.46$\pm$0.0	&0.39$\pm$0.0	&0.63$\pm$0.0	&0.93$\pm$0.01	&0.76$\pm$0.01	&0.59$\pm$0.01	&\\
				14	&20.13$\pm$0.74	&14.14$\pm$0.21	&-	&-	&-	&-	&-	&14.43$\pm$0.14	&9.67$\pm$0.1	&5.88$\pm$0.06	&\\
				15	&0.71$\pm$0.04	&0.92$\pm$0.01	&0.85$\pm$0.0	&0.74$\pm$0.0	&0.4$\pm$0.0	&0.33$\pm$0.0	&0.54$\pm$0.0	&0.75$\pm$0.01	&0.63$\pm$0.01	&0.5$\pm$0.0	&\\
				16	&2.32$\pm$0.09	&2.32$\pm$0.07	&2.53$\pm$0.0	&-	&1.07$\pm$0.0	&0.87$\pm$0.0	&-	&2.23$\pm$0.02	&1.79$\pm$0.02	&1.36$\pm$0.01	&\\
				17	&-	&-	&0.06$\pm$0.0	&0.08$\pm$0.0	&-	&-	&0.12$\pm$0.01	&0.11$\pm$0.0	&0.1$\pm$0.0	&0.16$\pm$0.0	&\\
				18	&43.12$\pm$1.15	&26.91$\pm$0.57	&26.96$\pm$1.17	&-	&0.32$\pm$0.0	&0.94$\pm$0.0	&1.84$\pm$0.11	&27.5$\pm$0.27	&17.3$\pm$0.17	&9.42$\pm$0.09	&\\
				19	&84.0$\pm$2.32	&49.51$\pm$1.5	&-	&-	&-	&-	&-	&52.47$\pm$0.52	&32.53$\pm$0.33	&17.21$\pm$0.17	&\\
				20	&3.42$\pm$0.11	&3.21$\pm$0.03	&3.5$\pm$0.0	&-	&-	&1.09$\pm$0.0	&-	&3.01$\pm$0.03	&2.34$\pm$0.02	&1.71$\pm$0.02	&\\
				21	&7.67$\pm$0.23	&5.99$\pm$0.08	&-	&-	&-	&1.63$\pm$0.0	&-	&5.94$\pm$0.06	&4.19$\pm$0.04	&2.75$\pm$0.03	&\\
				22	&4.21$\pm$0.04	&3.54$\pm$0.02	&-	&-	&-	&1.04$\pm$0.0	&-	&3.39$\pm$0.03	&2.49$\pm$0.02	&1.72$\pm$0.02	&\\
				23	&2.02$\pm$0.04	&1.78$\pm$0.02	&2.37$\pm$0.0	&-	&1.08$\pm$0.0	&0.9$\pm$0.0	&-	&2.09$\pm$0.02	&1.71$\pm$0.02	&1.33$\pm$0.01	&\\
				24	&-	&-	&-	&-	&-	&1.48$\pm$0.0	&-	&4.7$\pm$0.05	&3.47$\pm$0.03	&2.41$\pm$0.02	&\\
				25	&1.07$\pm$0.01	&1.09$\pm$0.02	&1.15$\pm$0.0	&0.95$\pm$0.0	&0.49$\pm$0.0	&0.41$\pm$0.0	&0.67$\pm$0.0	&0.98$\pm$0.01	&0.8$\pm$0.01	&0.61$\pm$0.01	&\\
				26	&12.03$\pm$0.21	&8.89$\pm$0.11	&-	&-	&-	&2.26$\pm$0.0	&-	&8.86$\pm$0.09	&6.09$\pm$0.06	&3.87$\pm$0.04	&\\
				27	&30.72$\pm$0.62	&19.88$\pm$0.51	&-	&-	&-	&-	&-	&20.5$\pm$0.21	&13.25$\pm$0.13	&7.52$\pm$0.08	&\\
				28	&106.04$\pm$1.06	&-	&-	&-	&-	&-	&-	&95.85$\pm$0.96	&59.49$\pm$0.59	&31.54$\pm$0.32	&\\
				29	&-	&-	&-	&-	&-	&2.18$\pm$0.01	&-	&8.8$\pm$0.09	&6.01$\pm$0.06	&3.79$\pm$0.04	&\\
				30	&0.8$\pm$0.01	&0.95$\pm$0.01	&0.85$\pm$0.0	&0.74$\pm$0.0	&0.4$\pm$0.0	&0.33$\pm$0.0	&0.54$\pm$0.0	&0.76$\pm$0.01	&0.63$\pm$0.01	&0.5$\pm$0.0	&\\
				31	&14.81$\pm$0.37	&11.37$\pm$0.27	&-	&-	&-	&-	&-	&10.82$\pm$0.11	&7.63$\pm$0.08	&4.99$\pm$0.05	&\\
				32	&79.48$\pm$2.64	&50.2$\pm$2.31	&-	&-	&-	&-	&-	&51.89$\pm$0.52	&33.26$\pm$0.33	&18.48$\pm$0.18	&\\
				33	&12.73$\pm$0.28	&10.0$\pm$0.1	&-	&-	&-	&-	&-	&9.68$\pm$0.1	&6.93$\pm$0.07	&4.62$\pm$0.05	&\\
				34	&6.95$\pm$0.15	&5.89$\pm$0.04	&-	&-	&-	&1.75$\pm$0.0	&-	&5.64$\pm$0.06	&4.17$\pm$0.04	&2.88$\pm$0.03	&\\
				35	&10.02$\pm$0.24	&7.56$\pm$0.1	&-	&-	&-	&2.0$\pm$0.01	&-	&7.5$\pm$0.08	&5.24$\pm$0.05	&3.4$\pm$0.03	&\\
				36	&31.18$\pm$0.8	&20.41$\pm$0.45	&-	&-	&-	&-	&-	&20.91$\pm$0.21	&13.65$\pm$0.14	&7.9$\pm$0.08	&\\
				37	&14.65$\pm$0.49	&10.47$\pm$0.23	&-	&-	&-	&-	&-	&10.55$\pm$0.11	&7.13$\pm$0.07	&4.42$\pm$0.04	&\\
				38	&110.63$\pm$16.51	&80.82$\pm$0.74	&-	&-	&-	&-	&-	&82.84$\pm$0.83	&53.44$\pm$0.53	&30.2$\pm$0.3	&\\
				39	&-	&-	&0.65$\pm$0.0	&0.6$\pm$0.0	&0.33$\pm$0.0	&0.27$\pm$0.0	&0.43$\pm$0.0	&0.59$\pm$0.01	&0.5$\pm$0.01	&0.4$\pm$0.0	&\\
				40	&16.27$\pm$0.24	&11.54$\pm$0.2	&-	&-	&-	&-	&-	&11.54$\pm$0.12	&7.77$\pm$0.08	&4.79$\pm$0.05	&\\
				41	&10.33$\pm$0.22	&7.78$\pm$0.06	&-	&-	&-	&2.03$\pm$0.0	&-	&7.7$\pm$0.08	&5.37$\pm$0.05	&3.47$\pm$0.03	&\\
				42	&67.65$\pm$1.87	&87.16$\pm$2.09	&-	&-	&-	&-	&-	&75.95$\pm$0.76	&68.91$\pm$0.69	&57.19$\pm$0.57	&\\
				43	&12.9$\pm$0.46	&9.32$\pm$0.26	&-	&-	&-	&-	&-	&9.4$\pm$0.09	&6.37$\pm$0.06	&3.97$\pm$0.04	&\\
				44	&39.61$\pm$0.91	&25.56$\pm$0.24	&-	&-	&-	&-	&-	&26.08$\pm$0.26	&16.64$\pm$0.17	&9.23$\pm$0.09	&\\
				45	&9.7$\pm$0.18	&7.35$\pm$0.14	&-	&-	&-	&1.93$\pm$0.0	&-	&7.33$\pm$0.07	&5.1$\pm$0.05	&3.29$\pm$0.03	&\\
				46	&-	&-	&0.1$\pm$0.0	&0.15$\pm$0.0	&0.11$\pm$0.0	&0.1$\pm$0.0	&0.13$\pm$0.0	&0.12$\pm$0.0	&0.12$\pm$0.0	&0.13$\pm$0.0	&\\
				47	&50.89$\pm$1.92	&33.66$\pm$0.74	&-	&-	&-	&-	&-	&34.35$\pm$0.34	&22.52$\pm$0.23	&13.16$\pm$0.13	&\\
				48	&15.27$\pm$0.14	&12.22$\pm$0.42	&-	&-	&-	&-	&-	&11.96$\pm$0.12	&8.46$\pm$0.08	&5.54$\pm$0.06	&\\
				49	&-	&-	&0.2$\pm$0.0	&0.23$\pm$0.0	&0.15$\pm$0.0	&0.13$\pm$0.0	&0.2$\pm$0.0	&0.2$\pm$0.0	&0.2$\pm$0.0	&0.18$\pm$0.0	&\\
				50	&26.25$\pm$0.58	&18.36$\pm$0.19	&-	&-	&-	&-	&-	&18.54$\pm$0.19	&12.56$\pm$0.13	&7.68$\pm$0.08	&\\
				51	&0.18$\pm$0.01	&0.31$\pm$0.03	&0.23$\pm$0.0	&0.26$\pm$0.0	&0.17$\pm$0.0	&0.14$\pm$0.0	&0.21$\pm$0.0	&0.22$\pm$0.0	&0.22$\pm$0.0	&0.2$\pm$0.0	&\\
				52	&-	&-	&0.66$\pm$0.0	&0.6$\pm$0.0	&0.33$\pm$0.0	&0.27$\pm$0.0	&0.44$\pm$0.0	&0.6$\pm$0.01	&0.51$\pm$0.01	&0.41$\pm$0.0	&\\
				53	&3.99$\pm$0.15	&3.36$\pm$0.02	&3.72$\pm$0.1	&-	&-	&1.0$\pm$0.0	&-	&3.25$\pm$0.03	&2.39$\pm$0.02	&1.65$\pm$0.02	&\\
				54	&17.94$\pm$0.61	&12.74$\pm$0.27	&-	&-	&-	&-	&-	&13.02$\pm$0.13	&8.81$\pm$0.09	&5.43$\pm$0.05	&\\
				55	&0.82$\pm$0.06	&0.98$\pm$0.1	&0.88$\pm$0.0	&0.75$\pm$0.0	&0.41$\pm$0.0	&0.34$\pm$0.0	&0.55$\pm$0.0	&0.78$\pm$0.01	&0.65$\pm$0.01	&0.51$\pm$0.01	&\\
				56	&0.84$\pm$0.05	&0.89$\pm$0.02	&0.43$\pm$0.0	&0.42$\pm$0.0	&0.24$\pm$0.0	&0.21$\pm$0.0	&0.31$\pm$0.0	&0.39$\pm$0.0	&0.35$\pm$0.0	&0.29$\pm$0.0	&\\
				57	&1.68$\pm$0.05	&1.64$\pm$0.02	&1.74$\pm$0.04	&1.42$\pm$0.0	&0.7$\pm$0.0	&0.58$\pm$0.0	&0.95$\pm$0.01	&1.57$\pm$0.02	&1.22$\pm$0.01	&0.9$\pm$0.01	&\\
				58	&0.43$\pm$0.07	&0.54$\pm$0.01	&0.52$\pm$0.0	&0.51$\pm$0.0	&0.29$\pm$0.0	&0.24$\pm$0.0	&0.38$\pm$0.0	&0.49$\pm$0.0	&0.43$\pm$0.0	&0.36$\pm$0.0	&\\
				59	&2.63$\pm$0.08	&2.35$\pm$0.03	&2.55$\pm$0.0	&-	&0.93$\pm$0.0	&0.75$\pm$0.0	&-	&2.24$\pm$0.02	&1.7$\pm$0.02	&1.21$\pm$0.01	&\\
				60	&21.02$\pm$0.64	&14.51$\pm$0.21	&-	&-	&-	&-	&-	&14.75$\pm$0.15	&9.77$\pm$0.1	&5.82$\pm$0.06	&\\
				61	&64.96$\pm$0.78	&83.78$\pm$1.77	&-	&-	&-	&-	&-	&72.82$\pm$0.73	&66.01$\pm$0.66	&54.49$\pm$0.54	&\\
				62	&1.37$\pm$0.09	&1.41$\pm$0.05	&1.42$\pm$0.0	&1.14$\pm$0.0	&0.58$\pm$0.0	&0.47$\pm$0.0	&0.8$\pm$0.0	&1.24$\pm$0.01	&0.99$\pm$0.01	&0.74$\pm$0.01	&\\ \hline
		\end{tabular}}
	\end{table*}
	
	\begin{table*}
		\centering
		\renewcommand\thetable{C} 
		\caption{The units on Paranal, 2MASS and WISE fluxes are 10$^{-14}$, 10$^{-15}$ and 10$^{-17}$ ergs s$^{-1}$ cm$^{-2}$ \AA$^{-1}$ respectively.}
		\label{tab:Flux_2}
		\resizebox{\textwidth}{!}{\begin{tabular}{lccccccccccr}
				\hline
				SXOM    &\multicolumn{5}{c}{Paranal/OmegaCAM} &\multicolumn{3}{c}{2MASS} & \multicolumn{2}{c}{WISE} \\
				&u       	&Halpha		&g	&r	&i	&J	        &H	            &Ks	            &W1                  &W2		\\ \hline
				1	&13.78$\pm$0.14	&9.84$\pm$0.1	&-	&-	&-	&1.97$\pm$0.04	&0.78$\pm$0.02	&0.31$\pm$0.01	&0.06$\pm$0.0	&0.02$\pm$0.0	&\\
				2	&4.2$\pm$0.04	&3.11$\pm$0.03	&5.58$\pm$0.06	&3.38$\pm$0.03	&2.5$\pm$0.03	&0.69$\pm$0.02	&0.28$\pm$0.01	&0.11$\pm$0.0	&0.02$\pm$0.0	&0.01$\pm$0.0	&\\
				3	&6.3$\pm$0.06	&4.54$\pm$0.05	&-	&-	&3.45$\pm$0.03	&0.96$\pm$0.02	&0.38$\pm$0.01	&0.15$\pm$0.0	&0.03$\pm$0.0	&0.01$\pm$0.0	&\\
				4	&0.85$\pm$0.01	&1.03$\pm$0.01	&1.44$\pm$0.01	&1.13$\pm$0.01	&0.85$\pm$0.01	&0.27$\pm$0.01	&0.12$\pm$0.0	&0.05$\pm$0.0	&0.01$\pm$0.0	&0.0$\pm$0.0	&\\
				5	&11.9$\pm$0.12	&7.67$\pm$0.08	&-	&-	&-	&1.5$\pm$0.03	&0.57$\pm$0.01	&0.22$\pm$0.0	&0.04$\pm$0.0	&0.01$\pm$0.0	&\\
				6	&-	&14.62$\pm$0.15	&-	&-	&-	&2.68$\pm$0.06	&0.92$\pm$0.02	&0.37$\pm$0.01	&0.07$\pm$0.0	&0.02$\pm$0.0	&\\
				7	&3.1$\pm$0.03	&2.49$\pm$0.02	&4.2$\pm$0.04	&2.66$\pm$0.03	&1.99$\pm$0.02	&0.57$\pm$0.02	&0.24$\pm$0.01	&0.09$\pm$0.0	&0.02$\pm$0.0	&0.01$\pm$0.0	&\\
				8	&0.77$\pm$0.01	&1.31$\pm$0.01	&1.62$\pm$0.02	&1.32$\pm$0.01	&1.08$\pm$0.01	&0.38$\pm$0.01	&0.18$\pm$0.0	&0.07$\pm$0.0	&0.01$\pm$0.0	&0.0$\pm$0.0	&\\
				9	&14.3$\pm$0.14	&9.31$\pm$0.09	&-	&-	&-	&1.75$\pm$0.04	&0.66$\pm$0.01	&0.25$\pm$0.0	&0.05$\pm$0.0	&0.01$\pm$0.0	&\\
				10	&2.33$\pm$0.02	&2.15$\pm$0.02	&3.18$\pm$0.03	&2.31$\pm$0.02	&1.71$\pm$0.02	&0.53$\pm$0.01	&0.23$\pm$0.01	&0.09$\pm$0.0	&0.02$\pm$0.0	&0.0$\pm$0.0	&\\
				11	&0.85$\pm$0.01	&1.0$\pm$0.01	&1.43$\pm$0.01	&1.09$\pm$0.01	&0.83$\pm$0.01	&0.27$\pm$0.01	&0.12$\pm$0.0	&0.05$\pm$0.0	&0.01$\pm$0.0	&0.0$\pm$0.0	&\\
				12	&0.08$\pm$0.0	&0.32$\pm$0.0	&0.26$\pm$0.0	&0.31$\pm$0.0	&0.26$\pm$0.0	&0.05$\pm$0.0	&0.03$\pm$0.0	&0.01$\pm$0.0	&0.01$\pm$0.0	&0.0$\pm$0.0	&\\
				13	&0.54$\pm$0.01	&0.79$\pm$0.01	&1.03$\pm$0.01	&0.83$\pm$0.01	&0.66$\pm$0.01	&0.22$\pm$0.0	&0.1$\pm$0.0	&0.04$\pm$0.0	&0.01$\pm$0.0	&0.0$\pm$0.0	&\\
				14	&12.46$\pm$0.12	&8.34$\pm$0.08	&-	&-	&-	&1.66$\pm$0.04	&0.64$\pm$0.01	&0.24$\pm$0.0	&0.05$\pm$0.0	&0.01$\pm$0.0	&\\
				15	&0.4$\pm$0.0	&-	&0.81$\pm$0.01	&0.73$\pm$0.01	&0.54$\pm$0.01	&0.19$\pm$0.0	&0.09$\pm$0.0	&0.04$\pm$0.0	&0.01$\pm$0.0	&0.0$\pm$0.0	&\\
				16	&1.39$\pm$0.01	&-	&2.41$\pm$0.02	&1.98$\pm$0.02	&1.45$\pm$0.01	&0.5$\pm$0.01	&0.23$\pm$0.0	&0.09$\pm$0.0	&0.02$\pm$0.0	&0.0$\pm$0.0	&\\
				17	&0.03$\pm$0.0	&0.62$\pm$0.01	&0.28$\pm$0.0	&0.56$\pm$0.01	&0.71$\pm$0.01	&0.69$\pm$0.02	&0.5$\pm$0.01	&0.23$\pm$0.01	&0.05$\pm$0.0	&0.01$\pm$0.0	&\\
				18	&23.95$\pm$0.24	&13.71$\pm$0.14	&-	&-	&-	&2.43$\pm$0.05	&0.87$\pm$0.02	&0.34$\pm$0.01	&0.07$\pm$0.0	&0.02$\pm$0.0	&\\
				19	&-	&-	&-	&-	&-	&4.34$\pm$0.1	&1.5$\pm$0.03	&0.56$\pm$0.01	&0.11$\pm$0.0	&0.03$\pm$0.0	&\\
				20	&2.16$\pm$0.02	&-	&3.33$\pm$0.03	&2.57$\pm$0.03	&1.86$\pm$0.02	&0.62$\pm$0.01	&0.27$\pm$0.01	&0.11$\pm$0.0	&0.02$\pm$0.0	&0.01$\pm$0.0	&\\
				21	&5.34$\pm$0.05	&-	&6.96$\pm$0.07	&4.62$\pm$0.05	&3.01$\pm$0.03	&0.85$\pm$0.02	&0.34$\pm$0.01	&0.13$\pm$0.0	&0.03$\pm$0.0	&0.01$\pm$0.0	&\\
				22	&2.91$\pm$0.03	&2.38$\pm$0.02	&4.01$\pm$0.04	&2.54$\pm$0.03	&1.93$\pm$0.02	&0.54$\pm$0.01	&0.23$\pm$0.01	&0.09$\pm$0.0	&0.02$\pm$0.0	&0.0$\pm$0.0	&\\
				23	&1.33$\pm$0.01	&1.71$\pm$0.02	&2.33$\pm$0.02	&1.88$\pm$0.02	&1.51$\pm$0.02	&0.51$\pm$0.01	&0.25$\pm$0.01	&0.1$\pm$0.0	&0.02$\pm$0.0	&0.01$\pm$0.0	&\\
				24	&3.87$\pm$0.04	&3.35$\pm$0.03	&5.53$\pm$0.06	&3.75$\pm$0.04	&2.69$\pm$0.03	&0.79$\pm$0.02	&0.34$\pm$0.01	&0.13$\pm$0.0	&0.03$\pm$0.0	&0.01$\pm$0.0	&\\
				25	&0.6$\pm$0.01	&0.83$\pm$0.01	&1.09$\pm$0.01	&0.86$\pm$0.01	&0.68$\pm$0.01	&0.23$\pm$0.01	&0.11$\pm$0.0	&0.04$\pm$0.0	&0.01$\pm$0.0	&0.0$\pm$0.0	&\\
				26	&7.93$\pm$0.08	&5.46$\pm$0.05	&-	&6.2$\pm$0.06	&-	&1.12$\pm$0.03	&0.44$\pm$0.01	&0.17$\pm$0.0	&0.03$\pm$0.0	&0.01$\pm$0.0	&\\
				27	&-	&10.79$\pm$0.11	&-	&-	&-	&2.01$\pm$0.05	&0.75$\pm$0.02	&0.29$\pm$0.01	&0.06$\pm$0.0	&0.02$\pm$0.0	&\\
				28	&-	&-	&-	&-	&-	&7.48$\pm$0.2	&2.81$\pm$0.07	&1.08$\pm$0.02	&0.22$\pm$0.0	&0.06$\pm$0.0	&\\
				29	&7.93$\pm$0.08	&5.46$\pm$0.05	&-	&-	&-	&1.1$\pm$0.03	&0.43$\pm$0.01	&0.16$\pm$0.0	&0.03$\pm$0.0	&0.01$\pm$0.0	&\\
				30	&-	&-	&-	&0.74$\pm$0.01	&-	&0.19$\pm$0.0	&0.09$\pm$0.0	&0.04$\pm$0.0	&0.01$\pm$0.0	&0.0$\pm$0.0	&\\
				31	&9.8$\pm$0.1	&7.0$\pm$0.07	&-	&-	&-	&1.53$\pm$0.04	&0.63$\pm$0.02	&0.23$\pm$0.01	&0.05$\pm$0.0	&0.01$\pm$0.0	&\\
				32	&-	&-	&-	&-	&-	&4.83$\pm$0.12	&1.7$\pm$0.04	&0.66$\pm$0.01	&0.13$\pm$0.0	&0.04$\pm$0.0	&\\
				33	&8.62$\pm$0.09	&6.56$\pm$0.07	&-	&-	&-	&1.43$\pm$0.03	&0.58$\pm$0.01	&0.23$\pm$0.0	&0.04$\pm$0.0	&0.01$\pm$0.0	&\\
				34	&4.82$\pm$0.05	&3.99$\pm$0.04	&6.59$\pm$0.07	&4.47$\pm$0.04	&3.36$\pm$0.03	&0.94$\pm$0.02	&0.4$\pm$0.01	&0.16$\pm$0.0	&0.03$\pm$0.0	&0.01$\pm$0.0	&\\
				35	&6.78$\pm$0.07	&4.8$\pm$0.05	&-	&5.5$\pm$0.06	&-	&1.0$\pm$0.02	&0.4$\pm$0.01	&0.16$\pm$0.0	&0.03$\pm$0.0	&0.01$\pm$0.0	&\\
				36	&-	&11.51$\pm$0.12	&-	&-	&-	&2.14$\pm$0.05	&0.8$\pm$0.02	&0.3$\pm$0.01	&0.06$\pm$0.0	&0.02$\pm$0.0	&\\
				37	&9.36$\pm$0.09	&6.5$\pm$0.07	&-	&-	&-	&1.28$\pm$0.03	&0.49$\pm$0.01	&0.19$\pm$0.0	&0.04$\pm$0.0	&0.01$\pm$0.0	&\\
				38	&-	&-	&-	&-	&-	&7.83$\pm$0.17	&2.93$\pm$0.06	&1.11$\pm$0.02	&0.22$\pm$0.0	&0.06$\pm$0.0	&\\
				39	&0.28$\pm$0.0	&-	&0.61$\pm$0.01	&0.56$\pm$0.01	&-	&0.15$\pm$0.0	&0.08$\pm$0.0	&0.03$\pm$0.0	&0.01$\pm$0.0	&0.0$\pm$0.0	&\\
				40	&9.99$\pm$0.1	&6.93$\pm$0.07	&-	&-	&-	&1.33$\pm$0.03	&0.53$\pm$0.01	&0.2$\pm$0.0	&0.04$\pm$0.0	&0.01$\pm$0.0	&\\
				41	&6.91$\pm$0.07	&-	&-	&-	&3.82$\pm$0.04	&1.03$\pm$0.02	&0.4$\pm$0.01	&0.16$\pm$0.0	&0.03$\pm$0.0	&0.01$\pm$0.0	&\\
				42	&-	&-	&-	&-	&-	&24.5$\pm$0.54	&12.87$\pm$0.37	&5.4$\pm$0.13	&1.13$\pm$0.05	&0.29$\pm$0.01	&\\
				43	&8.7$\pm$0.09	&-	&-	&-	&-	&1.12$\pm$0.03	&0.45$\pm$0.01	&0.17$\pm$0.0	&0.03$\pm$0.0	&0.01$\pm$0.0	&\\
				44	&-	&13.84$\pm$0.14	&-	&-	&-	&2.32$\pm$0.06	&0.81$\pm$0.02	&0.32$\pm$0.01	&0.06$\pm$0.0	&0.02$\pm$0.0	&\\
				45	&6.6$\pm$0.07	&-	&-	&-	&3.65$\pm$0.04	&0.98$\pm$0.02	&0.4$\pm$0.01	&0.15$\pm$0.0	&0.03$\pm$0.0	&0.01$\pm$0.0	&\\
				46	&0.02$\pm$0.0	&0.14$\pm$0.0	&0.1$\pm$0.0	&0.14$\pm$0.0	&0.14$\pm$0.0	&0.07$\pm$0.0	&0.04$\pm$0.0	&0.02$\pm$0.0	&0.0$\pm$0.0	&0.0$\pm$0.0	&\\
				47	&-	&-	&-	&-	&-	&3.67$\pm$0.08	&1.36$\pm$0.03	&0.54$\pm$0.01	&0.1$\pm$0.0	&0.03$\pm$0.0	&\\
				48	&10.75$\pm$0.11	&-	&-	&-	&-	&1.72$\pm$0.04	&0.69$\pm$0.02	&0.26$\pm$0.01	&0.05$\pm$0.0	&0.01$\pm$0.0	&\\
				49	&0.05$\pm$0.0	&0.23$\pm$0.0	&0.18$\pm$0.0	&0.23$\pm$0.0	&0.2$\pm$0.0	&0.09$\pm$0.0	&0.05$\pm$0.0	&0.02$\pm$0.0	&0.0$\pm$0.0	&0.0$\pm$0.0	&\\
				50	&15.83$\pm$0.16	&-	&-	&-	&-	&2.23$\pm$0.05	&0.89$\pm$0.02	&0.33$\pm$0.01	&0.06$\pm$0.0	&0.02$\pm$0.0	&\\
				51	&0.05$\pm$0.0	&0.25$\pm$0.0	&0.2$\pm$0.0	&0.25$\pm$0.0	&0.21$\pm$0.0	&0.09$\pm$0.0	&0.05$\pm$0.0	&0.02$\pm$0.0	&0.0$\pm$0.0	&0.0$\pm$0.0	&\\
				52	&0.27$\pm$0.0	&-	&0.61$\pm$0.01	&0.59$\pm$0.01	&0.45$\pm$0.0	&0.17$\pm$0.0	&0.08$\pm$0.0	&0.03$\pm$0.0	&0.01$\pm$0.0	&0.0$\pm$0.0	&\\
				53	&2.8$\pm$0.03	&2.38$\pm$0.02	&3.76$\pm$0.04	&2.66$\pm$0.03	&1.83$\pm$0.02	&0.55$\pm$0.01	&0.23$\pm$0.0	&0.09$\pm$0.0	&0.02$\pm$0.0	&0.0$\pm$0.0	&\\
				54	&11.05$\pm$0.11	&7.6$\pm$0.08	&-	&-	&-	&1.57$\pm$0.04	&0.63$\pm$0.01	&0.25$\pm$0.0	&0.05$\pm$0.0	&0.01$\pm$0.0	&\\
				55	&0.42$\pm$0.0	&0.66$\pm$0.01	&0.84$\pm$0.01	&0.72$\pm$0.01	&0.57$\pm$0.01	&0.2$\pm$0.01	&0.09$\pm$0.0	&0.04$\pm$0.0	&0.01$\pm$0.0	&0.0$\pm$0.0	&\\
				56	&-	&-	&-	&0.41$\pm$0.0	&0.32$\pm$0.0	&0.13$\pm$0.01	&0.1$\pm$0.0	&0.04$\pm$0.0	&0.01$\pm$0.0	&0.0$\pm$0.0	&\\
				57	&1.13$\pm$0.01	&1.23$\pm$0.01	&1.77$\pm$0.02	&1.33$\pm$0.01	&1.03$\pm$0.01	&0.31$\pm$0.01	&0.14$\pm$0.0	&0.05$\pm$0.0	&0.01$\pm$0.0	&0.0$\pm$0.0	&\\
				58	&0.2$\pm$0.0	&0.47$\pm$0.0	&0.49$\pm$0.0	&0.49$\pm$0.0	&0.39$\pm$0.0	&0.14$\pm$0.0	&0.07$\pm$0.0	&0.03$\pm$0.0	&0.01$\pm$0.0	&0.0$\pm$0.0	&\\
				59	&1.83$\pm$0.02	&1.66$\pm$0.02	&2.67$\pm$0.03	&1.74$\pm$0.02	&1.36$\pm$0.01	&0.41$\pm$0.01	&0.18$\pm$0.0	&0.07$\pm$0.0	&0.01$\pm$0.0	&0.0$\pm$0.0	&\\
				60	&12.12$\pm$0.12	&8.57$\pm$0.09	&-	&-	&-	&1.63$\pm$0.03	&0.6$\pm$0.01	&0.24$\pm$0.0	&0.04$\pm$0.0	&0.01$\pm$0.0	&\\
				61	&25.55$\pm$0.26	&-	&-	&-	&-	&24.14$\pm$0.6	&12.2$\pm$0.52	&5.31$\pm$0.15	&1.1$\pm$0.04	&0.28$\pm$0.0	&\\
				62	&0.82$\pm$0.01	&1.04$\pm$0.01	&1.35$\pm$0.01	&1.12$\pm$0.01	&0.81$\pm$0.01	&0.27$\pm$0.01	&0.13$\pm$0.0	&0.05$\pm$0.0	&0.01$\pm$0.0	&0.0$\pm$0.0	&\\ \hline
		\end{tabular}}
	\end{table*}

	\begin{table}
		
		\renewcommand\thetable{D} 
		\caption{Properties of peculiar stars that deviate from standard evolutionary models in the optical CMD. Column 1, 2, 3, 6 \& 7 list the RA, Dec, distance, radial velocity from Gaia DR2, and membership probability of stars, columns 4 \& 5, give the G mag and the G$_{BP}$-G$_{RP}$ colour taken from Gaia DR2 catalogue, and column 8 gives the type of stars depending upon their position in the optical CMD.}
		\label{tab:Peculiar}
		\begin{threeparttable}
			\resizebox{0.6\columnwidth}{!}{\begin{tabular}{ccccccccc}
					\hline
					RA & Dec & Dist & Gmag & BP-RP & RV & PMemb  & Type \\
					($^h$ $'$ $"$)  & ($^\circ$ $'$ $"$) & (pc) & (mag) & (mag) & (km/s) &   & \\ \hline
					8:03:22 & -28:26:43 & 656.0 $\pm$ 16.9 & 9.65 & 0.48  & - & 0.9  & SSG\tnote{1} \\
					8:04:33 & -27:47:13 & 711.3 $\pm$ 19.0 & 10.06 & 0.38 & - & 1.0  & SSG \\
					8:05:09 & -27:56:50 & 665.8 $\pm$ 16.1 & 10.06 & 0.4  & - & 1.0 & SSG \\
					8:04:57 & -28:08:46 & 640.1 $\pm$ 13.6 & 10.09 & 0.32 & - & 1.0 & SSG \\
					8:03:46 & -26:58:56 & 672.5 $\pm$ 65.8 & 17.04 & 1.25 & - & 0.6 & Gap\tnote{2} \\
					7:58:13 & -27:41:5 & 733.3 $\pm$ 56.2 & 15.67 & 0.96  & - & 0.6 & Gap \\
					8:03:42 & -29:21:39 & 604.2 $\pm$ 31.4 & 16.02 & 1.02 & - & 0.6 & Gap \\
					8:06:23 & -28:42:31 & 737.7 $\pm$ 93.8 & 17.91 & 1.33 & - & 0.6 & Gap \\
					8:07:46 & -28:29:59 & 628.7 $\pm$ 21.8 & 16.41 & 1.67 & - & 1.0 & Gap \\
					7:59:46 & -27:06:30 & 682.7 $\pm$ 14.6 & 9.73 & 1.64  & 90 $\pm$ 1& 0.9  & RS\tnote{3} \\ \hline
			\end{tabular}}
		\end{threeparttable}
		\begin{tablenotes}
			\scriptsize
			\item[1] SSG : Sub-subgiant
			\item[2] Gap : Gap stars
			\item[3] RS : Red Straggler
		\end{tablenotes}
		
	\end{table}